\documentclass[a4paper,11pt]{article}
  \expandafter\let\csname equation*\endcsname\relax
  \expandafter\let\csname endequation*\endcsname\relax
\usepackage{graphicx,epsfig, amssymb,amsmath,pst-node,multicol}
\usepackage{url}
\usepackage{fullpage}
\usepackage{authblk}
 \usepackage{enumitem} 
\graphicspath{{.},{./fig/}}
\newcommand{\R}{{\mathbb R}}
\newcommand{\Z}{{\mathbb Z}}

\newcommand{\eop}{\nopagebreak\hfill{\vrule width3pt height8pt depth0pt}}
\def\OO#1{\mathop{{\bf O}(#1)}}
\def\SO#1{\mathop{{\bf SO}(#1)}}
\def\D{\mathop{\bf D}}

\def\ZZ#1{\mathop{{\Z}_#1}}
\newcommand{\OB}{\mathbb{O}} 

\newtheorem{proposition}{Proposition}
\newtheorem{lemma}{Lemma}
\newtheorem{definition}{Definition}
\newtheorem{theorem}{Theorem}

\newcommand{\ttfil}[1]{{\textcolor{black}{#1}}}
\newcommand{\bffil}[1]{{\textcolor{black}{#1}}}

\newcommand{\ttfilr}[1]{{\textcolor{black}{#1}}}

\newcommand{\mylab}[1]{\label{#1}}
\newcommand{\myfig}[1]{Fig.~\ref{fig:#1}}

\newcommand{\myeq}[1]{Eq.~\eqref{eq:#1}}

\title{Onset of intermittent octahedral patterns in spherical B\'enard convection}

\author[1]{Philippe Beltrame}
\author[2]{Pascal Chossat}
\affil[1]{\small Universit\'e d'Avignon - INRA, UMR1114 EMMAH, F-84914 Avignon, FRANCE} 
\affil[2]{\small laboratoire J-A Dieudonn\'e (CNRS \& Universit\'e de Nice Sophia Antipolis) Parc Valrose 06108 Nice, FRANCE}
\begin{document}

\maketitle

\begin{abstract}
The onset of convection for spherically invariant Rayleigh-B\'enard fluid flow is driven by marginal modes associated with spherical harmonics of a certain degree $\ell$, which depends upon the aspect ratio of the spherical shell. 
At certain critical values of the aspect ratio, marginal modes of degrees $\ell$ and $\ell+1$ coexist. Initially motivated by an experiment of electrophoretic convection between two concentric spheres carried in the International Space Station (GeoFlow project), we analyze the occurrence of intermittent dynamics near bifurcation in the case when marginal modes with $\ell=3, 4$ interact. The situation is by far more complex than in the well studied $\ell=1, 2$ mode interaction, however we show that heteroclinic cycles connecting equilibria with octahedral as well as axial symmetry can exist near bifurcation under certain conditions. Numerical simulations and continuation (using the software AUTO) on the center manifold help understanding these scenarios and show that the dynamics in these cases exhibit intermittent behaviour, even though the heteroclinic cycles may not be asymptotically stable in the usual sense.
\end{abstract}

\section{Introduction}
Hydrodynamical systems with a spherical symmetry can undergo complex 
patterns and dynamics near the onset of non uniform flow. This dynamics occurs when critical spherical modes for the 
linearized system, associated with spherical harmonics $Y_\ell^m(\vartheta,\varphi)$ ($\ell$ positive integer, $-\ell\leq 
m\leq +\ell)$, exist with two or more different values of $\ell$, hence "competing" for the instability of the trivial state 
\cite{cho2}. In the case of the onset of thermal convection between two concentric spheres with a central gravity 
force, the value of $\ell$ is an increasing function of the aspect ratio $\eta=R_{inner}/R_{outer}$ where $R_{inner}$ 
(resp. $R_{outer}$) is the radius length of the inner (resp. outer) boundary of the domain of convection \cite{Chos79}. Therefore 
isolated values of $\eta$ exist, at which modes $Y_\ell^m$ ($-\ell\leq m\leq \ell$) and $Y_{\ell+1}^n$ ($-\ell+1\leq n\leq 
\ell+1$) lead to instability {\it simultaneously}. Of course, if $\eta$ is assumed close to the value 1, a larger number of 
modes can become unstable "almost" simultaneously. For values of $\eta$ not too large (typically $\eta< 
0.5$), it is relevant to consider that only two types of spherical modes, those with "degrees" $\ell$ and $\ell+1$, are 
competing for instability. Numerical calculations show that, if one assumes rigid boundary conditions, this competition 
occurs with $\ell=1$ (at a critical value $\eta\simeq 0.25$), $\ell=2$ (at $\eta\simeq 0.4$) or $\ell=3$ (at $\eta\simeq 0.45$).  

The most documented case is when $\ell=1$. Numerical simulations by Friedrich and Haken in 1986 \cite{FriHak86} 
have shown that there existed ranges of parameter values in which the dynamics for the {\it amplitude equations} 
(equations projected onto the critical modes by center manifold reduction, or slaving principle) exhibited an 
intermittent like behaviour. It showed long periods of quasi steady states with {\it axisymmetric pattern} followed by sudden 
excursions to regimes "far" from equilibrium and relamination to another steady state similar the initial one, possibly 
with an axis of symmetry rotated by 90 degrees. This cycle repeats itself forever in an aperiodic manner. This case has 
been analysed by Chossat and Armbruster \cite{ArCh91} who showed that under these conditions a {\it robust} 
heteroclinic cycle was bifurcating from the state of rest. A heteroclinic cycle is a set of equilibria (or other bounded 
solutions of a dynamical system) which are connected to each other by heteroclinic orbits in a cyclic manner. The 
robustness of this invariant set, which normally does not hold, can be forced by the symmetries of the system \cite{clm}.  
If a robust heteroclinic cycle is dynamically an attractor, then the solutions with an initial condition  sufficiently close to it 
exhibit the intermittent behaviour described above. In a further work \cite{ChGuLa99}, Chossat, Guyard and Lauterbach have 
shown that the heteroclinic cycle occurring in this case is a "larger" object than the one described in \cite{ArCh91}, and is an attractor 
when the parameters of the problem belong to the range numerically  explored by \cite{FriHak86}. They called this object 
a {\it generalized} heteroclinic cycle.

Recently some interest has arisen again for onset of convection of a fluid in a spherical shell when the aspect ratio is 
such that spherical modes with $\ell=3$ and $4$ compete for instability. This was in relation with the preparation of the
{\sc GeoFlow} experiment supported by the European Space Agency, which consists of a spherical vessel filled with a 
fluid subjected to an electrophoretic central force \cite{egbersetal}. This device is placed in the International Space 
Station in order to simulate a self-gravitating fluid. For technical reasons the aspect ratio of the spherical shell cannot 
be too small and the first unstable spherical modes are expected to be of degree $\ell=3$ or $4$. Numerical 
simulations near onset, using a center manifold approach in the case of mode interaction, have shown flow regimes 
which are quite suggestive of the presence of an attracting heteroclinic cycle involving steady states with cubic and/or 
tetrahedral symmetries \cite{BTGE06}. Direct simulations show the same behaviour for some time, although for the range 
of parameter values that have been considered, the flow seems to stabilize on a steady-state with tetrahedral 
symmetry after a few switches \cite{gebeeg}.

Our aim in this paper is to investigate this intermittent dynamics by analysing the bifurcation of generalized heteroclinic cycles 
in the case of $\ell=3, 4$ mode interaction, in the same spirit as \cite{ArCh91,ChGuLa99}. 
The main ingredients of the analysis are the spherical symmetries, but also a general property of the Rayleigh-B\'enard 
equations, namely the fact that the equations for the modes with $\ell=4$ on the center manifold undergo a transcritical 
bifurcation of {\it unstable} steady-states (see \cite{clm}), but have a quadratic term which is 
relatively small and allows for a "bending back" of this branch near bifurcation \cite{rgd}. 
It is a well-known fact that this quadratic term vanishes when the gravity force and buoyancy force follow the same law \cite{ChGu96,rgd2}\footnote{This is a general fact when $\ell$ is even. The quadratic term is always 0 when $\ell$ is odd.}. Moreover in this case, it is also known that the signs of coefficients of quadratic terms mixing modes with degrees $\ell$ and $\ell+1$ are strongly constrained and favor the occurrence of robust heteroclinic connections between pure mode steady-states. \\
In the experimental device, the gravity field is replaced by an electrophoretic field which has a radial dependence in $r^{-5}$ instead of $r^{-2}$ for the buoyancy field. Nevertheless, numerical computations of these quadratic coefficients show that in  
a range of physically plausible values of the Prandtl number, the above properties are still satisfied. We subsequently consider two cases: (i) gravity and buoyancy forces both having a radial dependance in $r^{-2}$, (ii) gravity force with radial dependence in $r^{-5}$ with a suitably chosen Prandtl number.

It should also be noted that in \cite{ChGu96}, robust heteroclinic cycles involving steady-states with cubic (octahedral) symmetry were found to exist for values of $\ell= 8$, 9 and 13. In the present case the situation is more complex and the invariant sets which we have found involve axisymmetric as well as cubic and, in some cases, other non axisymmetric patterns.

In Section 2 we introduce the model equations of Rayleigh-B\'enard convection in a spherical shell under gravity or electrophoretic central force field, we analyze the marginal stability of the state of pure conduction, and we compute the equations on the center manifold in the case of a $\ell=3,4$ spherical mode interaction (subsequently called 3-4 mode interaction), with two free parameters: the deviations of Rayleigh number and aspect ratio from their critical values. In Section 3 we first recall basic facts about bifurcation in the presence of symmetry, then we describe the geometry of the 3-4 mode interaction (lattice of isotropy types). In Section 4 we analyze bifurcations and heteroclinic connections in the subspace of pure $\ell=4$ modes. It is indeed an essential feature of this problem that $\ell=4$ modes form an invariant submanifold of the center manifold. Moreover the equations restricted to this submanifold are gradient-like \cite{clm}. Therefore, as shown in Section 5, transverse instabilities in the $\ell=3$ modes play a crucial role for the onset of non trivial dynamics, especially of heteroclinic cycles, even when parameters are set such that the trivial state is linearly stable along these modes. This is the case which we analyze here. The results of Sections 4 and 5 are summarized in Section 6 where the existence of heteroclinic cycles is stated. Section 7 presents a numerical exploration of the dynamics on the center manifold in the parameter range determined in previous sections. This report is concluded by a discussion on the results which have been obtained. Finally, an annex contains some material which is not immediately necessary for the understanding of this work: in Annex A the complete lattice of isotropy subgroups for the 3-4 mode interaction, in Annex B the precise form of the equivariant quadratic and cubic terms in the bifurcation equations, in Annex C the numerical scheme for the computation of the coefficients of the equations on the center manifold, in Annex D a table of eigenvalues computed at the equilibria involved in generalized heteroclinic cycles in three specific cases.

\section{The model equations and their center manifold reduction for the 3-4 mode interaction}
\subsection{Rayleigh-B\'enard convection in a spherical shell}
The Rayleigh-B\'enard convection is studied considering an incompressible Newtonian fluid under the Boussinesq 
approximation \cite{Chand61}. The fluid is confined between two concentric spheres of radii $R_{in}$ and $R_{out}$ 
($R_{in}<R_{out}$).
A radial force field proportional to $g(r)\mathbf{e_r}$ acts on the fluid. When the inner sphere is heated uniformly at 
$T_{in}$ and the outer sphere is cooled uniformly at $T_{out}<T_{in}$ a temperature gradient  $\nabla T_0(r)$ appears. 
For a pure diffusive state, i.e. the fluid being static, the temperature gradient is proportional to  $h(r)=1/r^2$.
Due to the buoyancy force, this state may be unstable beyond a critical temperature difference leading to the 
convection motion. The fluid velocity $\mathbf{u}$ and the temperature perturbation $\Theta=T-T_0$ are governed by 
the Navier-Stokes equation and the heat transport equation. The non-dimensional equations depend on three 
numbers: the aspect ratio $\eta=\frac{R_{in}}{R_{out}}<1$, the Prandtl number $Pr$ (ratio of 
kinematic viscosity to thermal diffusivity) and the Rayleigh number $Ra$ measuring the buoyancy force. 
The resulting equations with no-slip boundary conditions can be found in many references in literature (e.g. 
\cite{BTGE06,BaNi09,PHSG11}) and in our case they read after a suitable choice of scales
\begin{subequations}\label{eq:pde}
\begin{eqnarray}
\frac{\partial\mathbf{u}}{\partial t}&=&-\nabla P + \Delta \mathbf{u} +\lambda g(r)\Theta \mathbf{e_r}- \mathbf{u.\nabla u}, 
\mylab{eq:pde1} \\
\frac{\partial\Theta}{\partial t}&=&Pr^{-1}\left( \Delta \Theta + \lambda h(r)  \mathbf{u.e_r}\right)- \mathbf{u.\nabla} 
\Theta\\
\nabla\mathbf{.u}&=&0,\\
\mathbf{u(r)}&=&\mathbf{0} \text{ for }r=\eta \text{ or }1\\
\Theta(\mathbf{r})&=&0\text{ for }r=\eta \text{ or }1
\end{eqnarray}
\end{subequations}
where $P$ is the pressure and the parameter $\lambda$ is related to the Rayleigh number and aspect ratio by the formula
\begin{equation} \label{def:lambda}
\lambda = \left(\frac{\eta}{1-\eta}\right)^{3/2}\sqrt{Ra}.
\end{equation}
Using   the spherical coordinates $(r,\vartheta,\varphi)$ the system of equations (\ref{eq:pde}) is defined in the domain
\begin{equation}
\Omega=\left\{(r,\theta,\varphi) | \eta\leq r \leq 1\right\}.
\end{equation}
\par
The gravity fields $g(r)$ encountered in the geophysical context are mainly proportional to $r$ for high-density domain 
(Earth's mantle) and to $1/r^2$ for low-density fluid surrounding a high density ball (like the Earth's inner core). In the 
laboratory, the simulated central force field  can be a $1/r^5$-dependent field as for the dielectrophoretic force in the {\sc 
GeoFlow} experiment \cite{Eetal03} or for a magnetic field in the dynamo experiment presented in \cite{Fruh05}.
In this work we focuse on force fields due either to gravity or dielectrophoretic effect. This latter is produced by applying an 
periodic high voltage ($V\simeq 10kV$) between inner and outer sphere on a dielectric fluid (silicon oil).
The force acting on the volume element of the dielectric medium, consists of three parts: Coulomb force $
{\mathrm{\bf{F}}_{c}}=\rho_{fr}\mathrm{\bf{E}}$  ($\rho_{fr}$ free charge density), dielectrophoretic force ${\mathrm{\bf{F}}
_{d}}=-\frac{1}{2}\mathrm{\bf{E}}^2\nabla\epsilon$ and the gradient part $\frac{1}{2}\nabla\Big(\rho\frac{\partial\epsilon}
{\partial\rho}\mathrm{E}^{2} \Big )$. The last term is included in the pressure gradient $\nabla P$ in \myeq{pde1}. The 
period voltage $V$ being much smaller than the relaxation time of free charge, the Coulomb force is neglegible. Finally 
the dielectrophoretic force ${\mathrm{\bf{F}}_{d}}$ varies as $1/r^5$. The general theory is presented in \cite{YFBY84} 
and the derivation for the {\sc GeoFlow} experiment in \cite{TEH03}.
Note that the definition of the Rayleigh number depends on the force type:
\begin{eqnarray}
Ra_g&=&\frac{\alpha g(R_2)}{\nu\kappa}R_2^2(T_{in}-T_{out})\text{ : gravity force} \\ \label{eq:rag}
Ra_e&=&2\frac{\gamma\epsilon_0\epsilon_r}{\rho_0\nu\kappa}V^2(T_{in}-T_{out})\text{ : dielectrophoretic force}
\label{eq:rae}
\end{eqnarray}
The notations are as follows: $\alpha$ is the coefficient of volume expansion,
$\nu$ the viscosity, $\kappa$ the thermal conductivity and $\rho_{0}$ the density.
Furthermore, $\epsilon_{r}$ is the dielectric constant,  ${V}$ the effective voltage and
$\gamma$ the dielectric variability. This last constant is related to the dielectric constant
linear dependence on the temperature: $\epsilon=\epsilon_{0}\epsilon_{r}(1-\gamma({{T}_{1}}-{{T}_{2}}))$.
In the {\sc GeoFlow} experiment, the Rayleigh number $Ra_e$ is tuned by varying the voltage ${V},$ not the 
temperature difference as is usually the case in planar convection experiments.\par
In this paper, we are interested in the dynamics close to the onset of convection as governed by the PDE's system~(\ref{eq:pde}) with $g(r)=r^n
$, $n=-5$ or $-2$. 

\subsection{Linear stability analysis of the rest state} \label{subsec:linearstability}
The linear stability of the pure conduction state $(\mathbf{u},\Theta)=(\mathbf{0},0)$ in spherical symmetry has been well-studied in the case of geophysical flows since the seminal work of Chandrasekhar \cite{Chand61}. Because of the spherical symmetry, the eigenvalue problem is solved in irreducible representation of $\OO{3}$ of degree $\ell$. To be more precise, expressing ${\bf r}=(r,\vartheta,\varphi)$ and expanding the unknown fields of velocity and temperature in suitable series of spherical functions (harmonics) $Y_\ell^m(\vartheta,\varphi)$ ($-\ell\leq m\leq +\ell$, $\ell=0,1,\dots$), the angular dependance is eliminated at each order $\ell$ and the eigenvalue problem reduces to solving differential boundary value problems in the $r$ variable, of the form
\begin{eqnarray*}
{\cal D}_\ell{\cal D}_\ell(ru_\ell)-\sigma u_\ell(r) &=& \ell(\ell+1)\lambda g(r)\theta_\ell(r) \\
({\cal D}_\ell-\sigma Pr)\theta_\ell &=& -\lambda h(r)ru_\ell \\
u_\ell = \frac{d(ru_\ell)}{dr} =\theta_\ell &=& 0 ~\text{ at } r=\eta,~1
\end{eqnarray*}
where $\sigma$ is the eigenvalue, $u_\ell$, $\theta_\ell$ are respectively the component of the radial velocity and the component of the temperature field along the $\ell$-th spherical harmonics, and ${\cal D}_\ell=d^2/dr^2+2r^{-1}d/dr-\ell(\ell+1)r^{-2}$, see \cite{Chand61,Chos79} for details. \\
Note that, the above equations do not depend on the index $m$ of spherical harmonics. This is a first consequence of spherical symmetry. Therefore a given solution of these equations, corresponding to a given value of $\ell$, spans a $2\ell+1$ dimensional space of spherical eigenmodes associated with the same eigenvalue $\sigma$. By construction this space is an irreducible representation of $\OO{3}$ of degree $\ell$.

The eigenvalues $\sigma$ are always real and they are negative when $\lambda$ (or $Ra$) is small enough. For a given aspect ratio, the state of pure thermal conduction becomes unstable when the parameter $\lambda$ exceeds the value $\lambda_c$ at which the rightmost eigenvalue becomes positive. We can equivalently replace $\lambda$ by the Rayleigh number $Ra$ thanks to (\ref{def:lambda}), henceforth defining a critical Rayleigh number $Ra_c$. The neutral stability curve for a fixed $\eta$ is the set of points in the plane $(\ell,Ra)$ at which non trivial solutions of the eigenvalue problem exist with $\sigma=0$. It has been shown in \cite{Chos79} when $g(r)=h(r)=r$ or $1/r^2$ that these curves are strictly convex. Therefore there exists a value $\ell_c$ at which $Ra$ is minimal, which in turn defines the critical value $Ra_c$. Note that, these critical values do not depend on $Pr$. Numerical simulations corroborate this behavior for $g(r)=1/r^5$  \cite{BEH03}. It follows that for generic values of $\eta$ the critical eigenspace corresponds to an irreducible representation of degree $\ell_c$, therefore it has dimension $2\ell_c+1$. However $\ell_c$ depends on the aspect ratio and tends to $\infty$ when $\eta$ tends to 1.
Therefore $\ell_c$ is a step function of $\eta$ and at the boundaries of the intervals so defined, two different critical degrees $\ell_c$ and $\ell_c+1$ coexist. These values $\eta_c$ define in the $(\eta,Ra)$ plane, codimension 2 bifurcation points with $\ell_c,\ell_c+1$ mode interaction. 

This is illustrated by the figure \ref{fig:rac-etac} below, which shows numerical results on the computation of the neutral stability curve in the plane $(\eta,Ra)$ for different values of $\ell$. The critical $Ra_c$ and $\ell_c$ are defined by the lower enveloppe of the curves and the codimension 2 bifurcation points are the points on the enveloppe at which two curves intersect. \\
 In the rest of this paper we concentrate on the $3,4$ mode interaction, which in both cases $g(r)=1/r^2$ and $g(r)=1/r^5$ correspond to a critical $\eta_c\sim 0.45$. The corresponding critical eigenspace $V$ has therefore dimension $16$. \\
 In the following we define $V^{k}$ to be the space of spherical harmonics of degree $\ell=k$, so that $V=V^{3}\oplus V^{4}$.

\begin{figure}[ht!]
\begin{center}
\includegraphics[height=6.8cm]{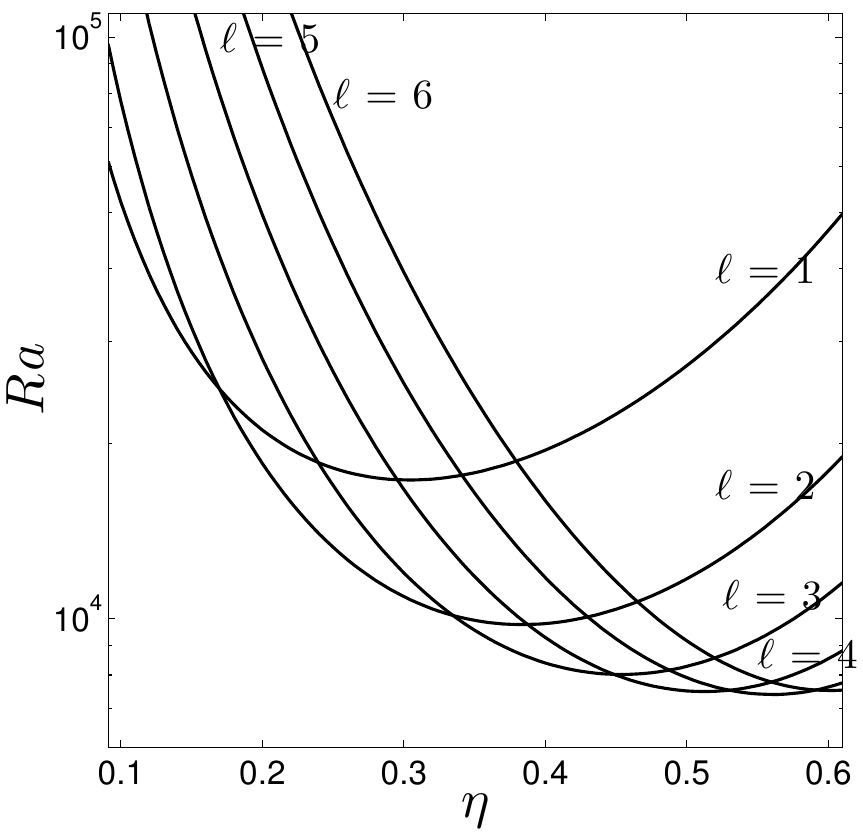} \hskip 1cm
\includegraphics[height=7cm]{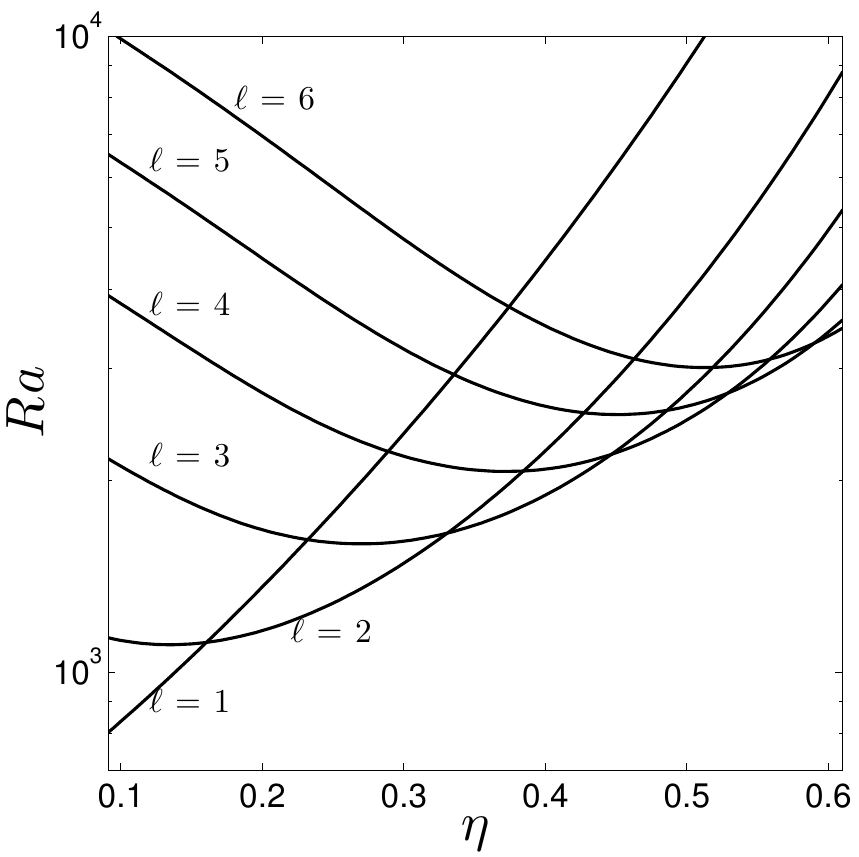}
\caption{Neutral stability curves in the cases $g(r)=1/r^2$ (left) and $g(r)=1/r^5$ (right) for values of $\ell$ up to 6. For a given $\eta$, the lowest point gives the critical values of $Ra$ and $\ell$.} \label{tab:rac-etac}
\end{center}
\label{fig:rac-etac}
\end{figure}

\subsection{Center manifold reduction for the 3-4 mode interaction} \label{sec:centermanifold}
\begin{figure}[ht!]
\begin{center}
\includegraphics[width=0.49\hsize]{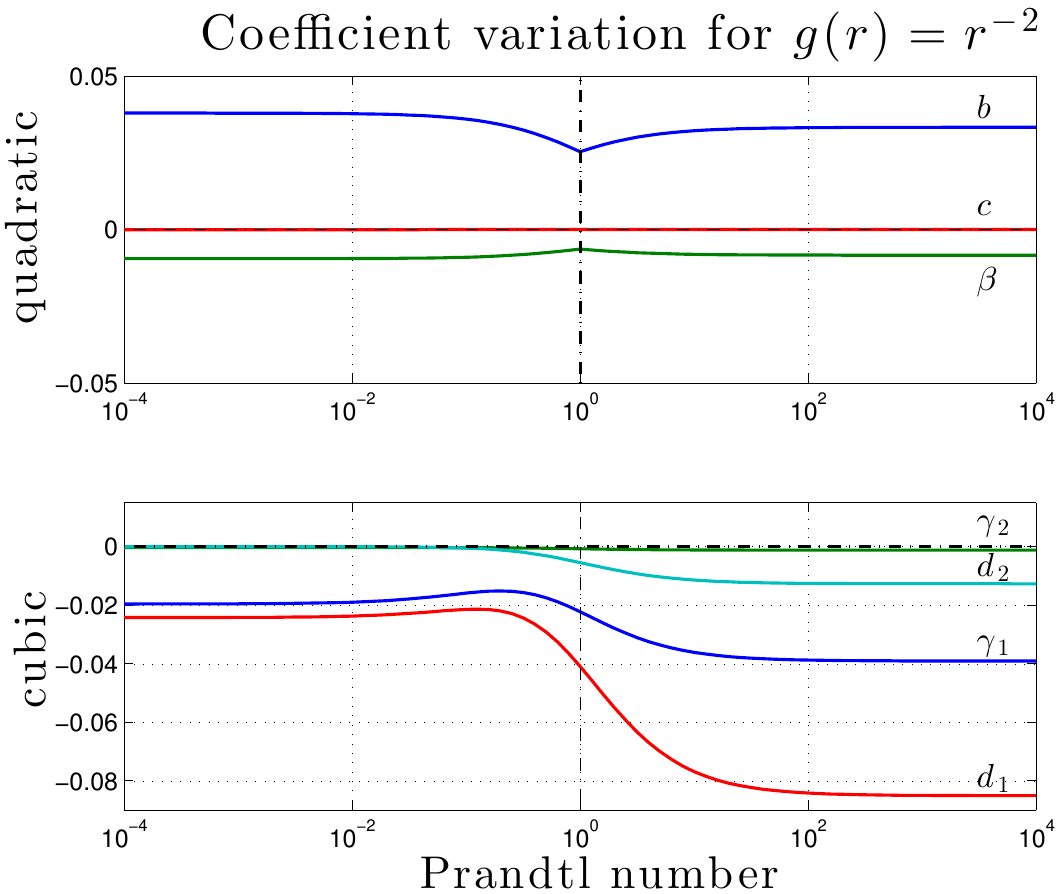} \hfill
\includegraphics[width=0.49\hsize]{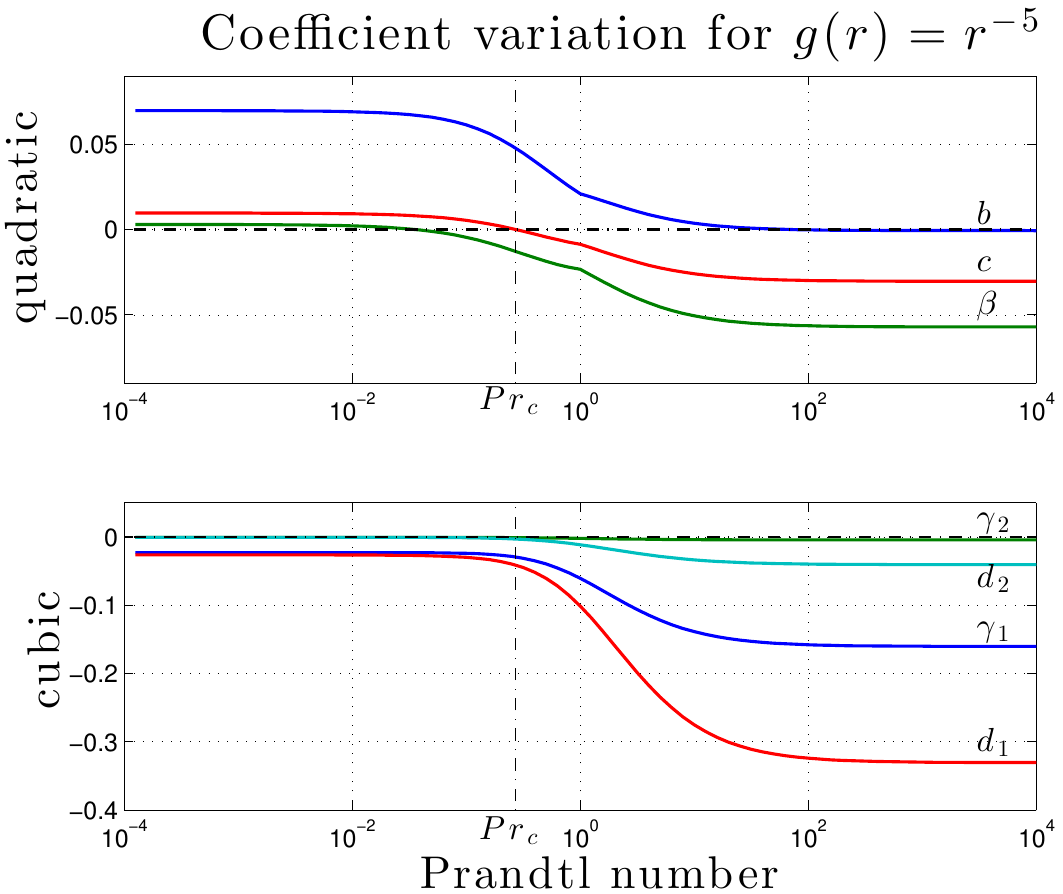}
\caption{Quadractic and cubic coefficients in Equations (\ref{eq:ode}) as functions of the Prandtl number. Left: case $g(r)=1/r^2$, right: case $g(r)=1/r^5$. $Pr_c$ indicates the critical value vanishing $c$ for $g(r)=r^{-5}$.} \label{fig:coef-versus-pr}
\end{center}

\end{figure}
\ttfilr{When $\lambda=\lambda_c$ and $\eta=\eta_c$ the marginal modes of the linear problem span the space $V$ of eigenmodes with $\ell=3$ and $4$. We shall therefore consider the 2-parameter bifurcation problem with parameters $\tilde\lambda=\lambda-\lambda_c$ and $\tilde\eta=\eta-\eta_c$. It is a well-established fact that the problem (1a)-(1e) satisfies the hypothesis of the center manifold reduction theorem, which allows to reducing the system of PDE's to an ODE defined in the space $V$ \cite{VI92, HaIo}. The idea is that near bifurcation the solutions can be expressed in terms of the marginal modes, the other modes behaving like "slave variables". Performing Taylor series expansion of the solution with respect to $X\in V$ and parameters $\tilde\lambda$, $\tilde\eta$, and identifying terms of same order in the equations, it is then possible to solve the resulting systems step by step from lowest order (see appendix \ref{app:coef}), to get an approximate polynomial ODE}
\begin{equation}\label{eq:taylor}
\dot{X} = \sum_{pqr} \tilde\lambda^q \tilde\eta^r R_{qr}^p(X,...,X),
\end{equation}
where $R_{qr}^p$ are $p$-linear and symmetric operators.\\
By taking the coordinates of $X$ along the spherical harmonics $Y_3^j$ and $Y_4^k$, we can set $X=\left[(x_{-3},\cdots, x_3),(y_{-4},\cdots, y_4)\right]$. Note that, since $X$ is real, the coordinates verify $x_{-m} = (-1)^m \overline{x}_m$ and $y_{-n} = (-1)^n \overline{y}_n$.
It is essential to know that the operators $R_{qr}^p$ keeps the $O(3)$ symmetry. In other words if $T_g$ denotes the action of $g\in O(3)$ in $V$, then $R_{qr}^p(T_gX,...,T_gX)=T_gR_{qr}^p(X,...,X)$. This property leads to a substantial simplification of the expression of these terms, many of them being actually identically equal to 0. Calculations which we do not detail here (see \cite{ChGu96}) show that the leading part of the equations finally read
\begin{subequations}\label{eq:ode}
\begin{eqnarray}
\dot x_j &=& \mu_1 x_j + \beta Q^{(1)}_j(x_{-3},\cdots,x_3,y_{-4},\cdots,y_4) + \gamma_1\|(x_{-3},\cdots,x_3)\|^2x_j  \label{eq_l=3} \nonumber \\ 
& & ~~~~~~~~~~~~~~~~~~~~~~~~~~~~~~~~~~~~~~~~~~~~~~~~~~  + \gamma_2C^{(1)}_j(x_{-3},\cdots,x_3) \\
\dot y_k &=& \mu_2 y_k +b Q^{(2)}_k(x_{-3},\cdots,x_3) + c Q^{(3)}_k(y_{-4},\cdots,y_4) + d_1\|(y_{-4},\cdots,y_4)\|^2y_k \nonumber 
\label{eq_l=4} \\
& &  ~~~~~~~~~~~~~~~~~~~~~~~~~~~~~~~~~~~~~~~~~~~~~~~~~ + d_2 C^{(2)}_k(y_{-4},\cdots,y_4)
\end{eqnarray}
\end{subequations}
\noindent where the quadratic $Q$'s and cubic $C$'s are expressed in the Annex \ref{app:equivariants}. \\
The norm of $X\in V$ is defined as follows: 
$$\|X\|^2=\sum_{-3}^3{(-1)^kx_kx_{-k}} + \sum_{-4}^4{(-1)^my_my_{-m}}.$$ 
The two sums correspond respectively to the square of the norm in the $\ell=3$ modes and to the square of the norm in the $\ell+1=4$ modes. 
Note that the subspace of "pure" 4 modes is flow-invariant (this is a consequence of properties of isotropy subgroups for the action of $O(3)$ in $V$, which we shall recall in the next section). \\
The coefficients in front of these polynomial terms depend on the physical parameters and in particular, on the Prandtl number. Using the inductive algorithm presented in \cite{ChGu96} the coefficients are expressed in Annex \ref{app:coef} and computed order by order for various Prandtl numbers.  \\
The results for $g=1/r^5$ can be found in \cite{BeEg04}. They are gathered in \myfig{coef-versus-pr} for both force fields. According to \cite{FriHak86,BeEg04} when  $Pr>1$, time and unknowns are changed as follows: time is multiplied by $Pr$ and $X$ is divided by $\sqrt{Pr}$. It results that the quadratic coefficient are divided by $\sqrt{Pr}$. 

In the case $g(r)=1/r^2$, the results agree with the following properties, which were proved in a more general context in \cite{cho2}, independantly of the value of $Pr$: 
\begin{enumerate}
\item $c=0$
\item $b=-2\beta$
\item $\gamma_1<0$ and $d_2<0$.
\end{enumerate}
The proof is based on the fact that $L_0$ is a self-adjoint operator when $g(r)=1/r^2$. \\
Moreover we also see from \myfig{coef-versus-pr} that $d_1<d_2<0$ for all $Pr$. \\
In the case $g(r)=1/r^5$, equalities (i) and (ii) are nearly satisfied if $Pr$ is equal to a critical value $Pr_c\simeq 0.27$, despite the fact that $L_0$ is no longer self-adjoint. Moreover all the above inequalities are also true in this case. These properties of the coefficients will play an important role in our bifurcation and stability analysis. \\
The condition $c=0$ introduces an additional degeneracy in the bifurcation problem, as we shall see in Section \ref{sec:l=4}. Assuming either that the force field $g(r)$ is slightly perturbed from $g(r)=1/r^2$, or in the case $g(r)=1/r^5$, that $Pr$ is close to $Pr_c$, we can consider the coefficient $c$ to be an additional parameter close to 0.  

Remark that $\beta<0$ when $g(r)=1/r^2$ and also when $g(r)=1/r^5$, near the critical $Pr_c$. Changing the time scale we can always assume $\beta=-1$, which we shall do in the subsequent analysis.

\section{Basic facts about bifurcation with symmetry and the isotropy lattice}
\subsection{Basic facts about equivariant bifurcation theory} \label{equivbif_theory}
Because symmetry is responsible for the high dimension of the center manifold while its geometry is by no way simple, the use of concepts and techniques of Equivariant Bifurcation Theory \cite{gost2,cl} will be of crucial importance in the forthcoming bifurcation analysis. We introduce first some general and basic facts, then in the next subsection we describe the geometry of the action of $\OO{3}$ on the center manifold, and in the last subsection we apply these informations to describe the elementary bifurcations which occur in this problem.  

\ttfilr{We denote by $g\mapsto T(g)$, $g\in G$, the representation of the group $G=\OO{3}$ in the space $V$. In other words $T$ is a homomorphism from $G$ to $GL(V)$ (the group of invertible linear maps in $V$).  }

\begin{definition}
(i) Given $x\in V$, the set $G_x=\{g\in G~/~ T(g)x=x\}$ is the {\it isotropy subgroup} of $x$. \\
(ii) Given an isotropy subgroup $H$, the linear space $Fix(H)=\{x\in X~/~T(H)x=x\}$ is the {\it fixed-point subspace} associated with $H$. If $\dim Fix(H)=1$, one often call it an {\it axis of symmetry} for the action of $G$ in $V$. \\
(iii) Given a point $x\in V$, the set ${\cal O}_x=\{T(g)x~/~g\in G\}$ is the $G$-orbit of $x$. 
\end{definition}
The three following statements are elementary: (i) If $H\subset H'$ are two isotropy subgroups, then $Fix(H)\supset Fix(H')$. \\
(ii) The largest subgroup of $G$, which leaves $Fix(H)$ invariant is its normalizer $N(H)$\footnote{The normalizer of $H$ is the group of elements $g$ in $G$ such that $gHg^{-1}=H$.}. \\
(iii) Two points in the same $G$-orbit have conjugated isotropy subgroups. \\
In fact the $G$-orbit of a point is a compact manifold whose dimension is given by
\begin{equation} \label{eq:dimorbite}
\dim O_x=\dim G - \dim G_x
\end{equation}
Since here $\dim G=3$ and $\dim G_x$ can be either 0 (case of a finite isotropy subgroup), 1 or 3, $\dim {\cal O}_x$ can be either 3, 2 or 0. We can classify $G$-orbits by conjugacy classes of isotropy subgroups. These conjugacy classes are called {\it isotropy types} or {\it orbit types} and they are partially ordered by group inclusion. This ordering is called the {\it lattice} of isotropies of the group action. 

The knowledge of the lattice of isotropy types is useful in the study of dynamical systems with symmetry for the following reason \cite{cl}. Consider a $G$-equivariant differential equation in a space $V$, like Eq. (\ref{eq:ode}). Then any fixed-point subspace is invariant by this equation. As a consequence, if the initial condition belongs to a subspace $Fix(H)$, then the whole trajectory lies in $Fix(H)$. In fact, the subset of points which have exactly isotropy $H$ is flow-invariant. This subset is open in $Fix(H)$ and its boundary is filled with points with higher isotropy. \\ 
Symmetry has another general consequence. Suppose that the action of $G$ in $V$ does not contain the trivial representation, as in our $3-4$ mode interaction. Then $x=0$ is a solution for all parameter values. Indeed writing the bifurcation equation $F(x,\mu)=0$, we have by equivariance $T(g)F(0,\mu)=F(0,\mu)$, but since $V$ does not contain the trivial representation only $0$ is fixed by $T(g)$ for all $g$. \\
These two remarks put together allow to state a general bifurcation result in systems with symmetry, known as the Equivariant Branching Lemma \cite{gost2}. 
We give here a version of the lemma suitable to our purpose. Note that, if $Fix(H)$ is an axis of symmetry, $N(H)$ can act on $Fix(H)$ either trivially or by $x\mapsto\pm x$. As above we assume the action of $G$ in $V$ does not contain the trivial representation.
\begin{lemma} \label{lemma:equivariantbranching}
Let $H$ be an isotropy subgroup with $\dim Fix(H)=1$. 
Let $f(x,\mu)=0$ be the bifurcation equation restricted to $Fix(H)$, $f\in {\cal C}^k(\R\times\R^m,\R)$, $k>2$, such that $f'_x(0,0)=0$ (bifurcation condition). Then \\
(i) if $N(H)$ acts trivially on $Fix(H)$, $f''_{x\mu}(0,0)\neq 0$ and $f''_{x^2}(0,0)\neq 0$, a transcritical branch of solutions with isotropy $H$ bifurcates from 0; \\
(ii) if $N(H)$ acts by $\pm Id$ in $Fix(H)$, $f''_{x\mu}(0,0)\neq 0$ and $f'''_{x^3}(0,0)\neq 0$, a pitchfork branch of solutions bifurcates from 0.
\end{lemma}
Note that, the conditions on the derivatives of $f$ are "generic" in each of cases (i) and (ii), which loosely means that they should be only exceptionally wrong. However in case (i), as we have already seen in Section \ref{sec:centermanifold}, the condition $f''_{x^2}(0,0)\neq 0$ can be unsatisfied in certain cases in Rayleigh-B\'enard convection. This point will be of crucial importance in all our bifurcation analysis. \\   
Of course bifurcated solutions with lower isotropy may also exist, but there is no such simple statement as the above lemma. \\ 
As we shall see, symmetry affects not only the bifurcation of steady-states but also the type of dynamics that can exist. 

\subsection{The isotropy lattice and fixed-point subspaces}
We have listed all isotropy types and determined representative fixed-point subspaces by exploiting the informations provided in \cite{cl} (Appendix A).
The lattice of isotropies is a quite complicated graph, which we show in \ref{annex:lattice}. Table \ref{tableFix(H)} contains the relevant informations for isotropy types with fixed-point subspace of dimension lower that 6. These are the most relevant for our purpose. As above (Section \ref{sec:centermanifold}) we note $(x_{-3},\cdots, x_3)$ for the component along the $\ell=3$ spherical harmonics $Y_3^j$, and $(y_{-4}, \cdots, y_4)$ for the component along the $\ell=4$ spherical harmonics $Y_4^k$. Note that, $Y_\ell^{-m}=(-1)^m \overline Y_\ell^m$ and therefore the same holds for the coordinates (real space). \\
Also note that the antipodal symmetry $S:~{\bf r}\mapsto -{\bf r}$ acts as $-Id$ on spherical harmonics with odd degree $\ell$, while it acts trivially on spherical harmonics with even degree. It follows that the group $\Z_2^c=\{Id,S\}$ is an isotropy subgroup whose fixed-point subspace is 9 dimensional and consists of all the spherical modes with $\ell=4$ (irreducible representation of degree 4). We shall note $V^{4}=Fix(\Z_2^c)$.
\begin{table}[h] \label{table:isotropies}
\centering 
\begin{tabular}{|c|c||c|c|}
\hline
Isotropy $H$ & $Fix(H)$ & Isotropy $H$ & $Fix(H)$ \\ \hline
$\OO2\oplus\Z_2^c$ & $y_0$ & $\D_2\oplus\Z_2^c$ & $(y_0,y_{2r},y_{4r})$ \\
$\OB\oplus\Z_2^c$ & $y_{4r}=\pm\eta y_0$ & $\D_4^d$ & $(x_{2i},y_0,y_{4r})$ \\ 
$\OO2^-$ & $(x_0,y_0)$  & $\D_3$ & $(x_{3i},y_0,y_{3r})$ \\
$\D_4\oplus\Z_2^c$ & $(y_0,y_{4r})$ & $\D_2$ & $(x_{2i},y_0,y_{2r},y_{4r})$ \\ 
$\D_6^d$ & $(x_{3r},y_0)$ & $\D_3^z$ & $(x_0,x_{3r},y_0,y_{3r})$ \\
$\D_3\oplus\Z_2^c$ & $(y_0,y_{3r})$ & $\D_2^z$ & $(x_0,x_{2r},y_0,y_{2r},y_{4r})$ \\
$\OB^-$ & $(x_{2i},y_{4r}=\eta y_0)$ & $\Z_2\oplus\Z_2^c$ & $(y_0,y_2,\bar y_2, y_4, \bar y_4)$  \\
$\D_4^z$ & $(x_0,y_0,y_{4r})$ & $\ZZ4^-$ & $(x_2, \bar x_2, y_0, y_4, \bar y_4)$ \\
\hline
\end{tabular}
\caption{\footnotesize Fixed-point subspaces for representatives of the isotropy types such that $\dim Fix(H)\leq 5$. We 
have noted $\eta=\sqrt{\frac{5}{14}}$. Another useful representation of $Fix(\OB\oplus\Z_2^c)$ is $y_{3r}=\nu y_0$ with 
$\nu=\sqrt{\frac{10}{7}}$.}
\label{tableFix(H)}
\end{table}
The notations of groups in table \ref{tableFix(H)}  are taken from \cite{gost2}. These groups are defined as follows (up to conjugacy). \\
- $\D_n$ is the group generated by the $n$-fold rotation $R_n$ about an axis $\delta$ (which we choose as the vertical $z$-axis) and rotation by $\pi$ around  an axis perpendicular to $\delta$ (noted $\chi$). This group is isomorphic to the dihedral group of order $2n$. Note that $\chi$ acts in $V$ as follows: $\chi\cdot x_j=-\bar x_j$, $\chi\cdot y_k=\bar y_k$. $\D_n\oplus\Z_2^c$ is then the full group of symmetry of a prism with regular $n$-gon basis (the notation $\oplus$ means "direct product"). Similarly, $\OO2\oplus\Z_2^c$ is the full symmetry group of a cylinder. \\
- $\OB$ is the group of direct (rotational) symmetries of a cube or octahedron. Hence $\OB\oplus\Z_2^c$ is the full symmetry group of a cube. \\
- Let $K$ be the reflection through a plane containing $\delta$. Then $\D_n^z$ is the group generated by $R_n$ and $K$. This group is isomorphic to $\D_n$. Similarly, $\D_{2n}^d$ is the group generated by $\D_n$ and $K$. It contains $4n$ elements. \\ 
- $\OB^-$ is the group generated by the tetrahedral group $\mathbb{T}$ (group of direct symmetries of a tetrahedron) and by $K$ (here $\delta$ is the axis of a 3-fold rotation in $\mathbb{T}$). This group contains 24 elements and is isomorphic to the octahedral group $\OB$. \\
- Let $K_\perp$ be the reflection through the plane orthogonal to $\delta$. Then $\Z_{2n}^-$ ($n>1$) is the group generated by the transformation $KR_{2n}$ (it is cyclic of order $2n$). $\Z_2^-$ is the 2-element group generated by any reflection through a plane. The reflection $\kappa$ through $K_\perp$ acts as follows in $V$: $\kappa\cdot x_j=(-1)^{j+1}x_j$, $\kappa\cdot y_k=(-1)^ky_k$.

Note that, all isotropy subgroups which do not contain $\Z_2^c$ have fixed-point subspaces which contain $\ell=3$ modes. We call these subspaces {\it mixed mode subspaces}.

\section{Bifurcation with pure $\ell=4$ modes} \label{sec:l=4}
We have seen that the subspace $V^{4}$ of pure $\ell=4$ modes is also the subspace of points which are fixed by the antipodal symmetry, therefore it is flow-invariant. The bifurcation problem with $\ell=4$ has been completely resolved in the "generic", codimension 1 case in \cite{clm}. \\
By Table \ref{table:isotropies}, two isotropy types have axes of symmetry: $\OO2\oplus\Z_2^c$ and $\OB\oplus\Z_2^c$. Solutions with the former isotropy are axisymmetric and those with the latter isotropy have octahedral (cubic) symmetry. Both solutions are also invariant by reflections through the equatorial plane. \\ 
According to Lemma \ref{lemma:equivariantbranching}, the bifurcated branches should be transcritical since $N(H\oplus\Z_2^c)=H\oplus\Z_2^c$ when $H=\OO2$ and $\OB$. However we have also seen in Section \ref{sec:centermanifold} that the quadratic terms of the equations restricted to the pure $\ell=4$ modes vanish when $g(r)=r^{-2}$, or, in the case when $g(r)=r^{-5}$, if the Prandtl number is close to a critical value $Pr_c$. This leads to a codimension 2 bifurcation problem, which has been studied from a singularity theory point of view in \cite{rgd}, then applied to the spherical B\'enard problem in \cite{rgd2} (with different force field and boundary conditions). \\
The subsequent analysis can be put into this perspective: the coefficient $c$ of the "pure" quadratic terms is now considered a {\it free parameter close to 0}, and we further assume $c>0$. Indeed according to \cite{rgd}, the case $c<0$ leads to more complicated bifurcation diagrams in $V^4$ and hardly identifiable heteroclinic cycles. \\
Next we provide useful material for the further dynamical analysis. First we compute the primary branches of steady-states and their stability, then we study the bifurcation diagrams in the invariant plane $Fix(\D_4\oplus\Z_2^c)$, finally we state a proposition about the bifurcation diagram in $V^4$.

\subsection{The axisymmetric equilibria.} 
On the axis $Fix(\OO2\oplus\Z_2^c)$ the (scalar) bifurcation equation reads
$$0 = \mu_2y_0 + 9c y_0^2 +(d_1+d_2)y_0^3. $$
where $\mu_2$ is the bifurcation parameter. It follows that the bifurcated equilibria satisfy the relation
\begin{equation} \label{mu2axisymmetric}
\mu_2 = -9c y_0-(d_1+d_2)y_0^2 
\end{equation}
A turning point exists at $y_0=-\frac{9c}{2(d_1+d_2)}$ and since $d_1+d_2<0$ (for any value of $Pr$) the parabola is always oriented towards $y_0>0$. \\
Even in the limit $c=0$, the two branches of the parabola correspond to symmetrically distinct states.
We denote by $\alpha_\pm$ these branches. According to (\ref{eq:dimorbite}), the corresponding $\OO3$-orbits of equilibria have dimension 2. \\
In order to determine their stable and unstable manifolds in $V$ we need to compute the eigenvalues of the Jacobian matrix $L_{\alpha_\pm}$ of the vector field linearized at $\alpha_\pm$, hence to determine first the isotypic decomposition of the action of $\OO2\oplus\Z_2^c$ in the $3,4$ representation of $\OO3$ \cite{cl}. A straightforward (and classical) analysis shows that $L_{\alpha_\pm}$ decomposes into two $1\times 1$ and 
seven $2\times 2$ diagonal blocks along the coordinates $x_0$, $y_0$, $(x_j,\bar x_j)$ ($j=1,2,3$) and  $(y_m,\bar 
y_m)$ ($m=1,2,3,4)$ respectively. The block in the $(y_1,\bar y_1)$ subspace is the 0 matrix, corresponding to the fact 
that this plane is tangent to the $\OO3$ orbit of $\alpha_\pm$ \cite{clm}. The other eigenvalues are listed in the 
following table \ref{vp_O(2)} where $y_0$ is the coordinate of $\alpha_\pm$.

\begin{table}[h] \footnotesize
\centering
\begin{tabular}{|l|c|c|}
\hline
Eigenvalues & Multiplicity & Eigenspaces \\ \hline
$\sigma_0^{\alpha}=\mu_1-6 y_0$  & 1 & $x_0$ \\
$\sigma_1^{\alpha}=\mu_1- y_0$ & 2 & $\{x_1,\bar x_1\}$ \\
$\sigma_2^{\alpha}=\mu_1+7 y_0$  & 2 & $\{x_2,\bar x_2\}$ \\
$\sigma_3^{\alpha}=\mu_1-3 y_0$  & 2 & $\{x_3,\bar x_3\}$ \\
\hline
$\lambda_0^{\alpha}=9c y_0+(2d_1+ d_2)y_0^2$  & 1 & $y_0$ \\
$\lambda_1^{\alpha}=0$ & 2 & $\{y_1,\bar y_1\}$ \\
$\lambda_2^{\alpha}=-20c y_0+5/2d_2 y_0^2$ & 2 & $\{y_2,\bar y_2\}$ \\
$\lambda_3^{\alpha}=-30c y_0-45/28d_2 y_0^2$ & 2 & $\{y_3,\bar y_3\}$ \\
$\lambda_4^{\alpha}=5c y_0-20/7d_2 y_0^2$ & 2 & $\{y_4,\bar y_4\}$ \\
\hline
\end{tabular}
\caption{\small Eigenvalues of the linearized vector field at $\alpha=\alpha_\pm$}
\label{vp_O(2)}
\end{table}

\subsection{The octahedral equilibria.}
We now turn to the bifurcations along the axis $Fix(\OB\oplus\ZZ2^c)$ defined by the relation $y_{4r}=\sqrt{5/14}y_0$. The bifurcation 
equation along this axis, which we parametrize with $y_0$, reads
$$0 = \mu_2y_0 +14 c y_0^2 +(\frac{12}{7}d_1-\frac{16}{49}d_2)y_0^3$$
and therefore the bifurcated branches are given by
\begin{equation} \label{mu2cubic}
\mu_2 = -14c y_0-Dy_0^2  ~~\text{where}~D=12/7d_1-16/49d_2
\end{equation}
Since $d_1<d_2<0$ (Fig. \ref{fig:coef-versus-pr}), $D<0$ and the parabola is oriented towards $y_0>0$. Here again the two branches correspond to symmetrically distinct states.

We note these solutions $\beta_+$ and $\beta_-$. They generate three-dimensional $\OO3$-orbits of equilibria. 
The eigenvalues of the Jacobian matrix $L_{\beta_\pm}$ of the vector field linearized at $\beta_\pm$ are listed in the 
following table \ref{vp_O}, together with the corresponding eigenspaces. Here again the multiplicity and eigenspaces 
result from the isotypic decomposition of the action of $\OB\oplus\ZZ2^c$ in the $\ell=3,4$ representation of $\OO3$. 
The convention  is that coordinates which do not appear in the definition of eigenspaces must be set equal to 0. The $
\beta_\pm$ solutions are parametrized by the coordinate $y_0$.

\begin{table}[h] \footnotesize
\centering
\begin{tabular}{|l|c|c|}
\hline
Eigenvalues & Multiplicity & Eigenspaces \\ \hline
$\sigma_0^{\beta}=\mu_1+12 y_0$  & 1 & $\{x_{2i}\}$ \\
$\sigma_1^{\beta}=\mu_1-6 y_0$ & 3 & $\{x_0, x_{3}=-\frac{\sqrt{15}}{3}\bar x_{1}\}$ \\
$\sigma_2^{\beta}=\mu_1+2 y_0$  & 3 & $\{x_{2r}, x_{3}=\frac{\sqrt{15}}{5}\bar x_1\}$ \\
\hline
$\lambda_0^{\beta}=14c y_0+(24/7d_1- 32/49d_2)y_0^2$  & 1 & $\{y_{4r}=\sqrt{\frac{5}{14}}y_0)$ \\
$\lambda_1^{\beta}=0$ & 3 & $(y_{4i}, y_{3}=-\frac{\sqrt{7}}{7}\bar y_1)\}$ \\
$\lambda_2^{\beta}=-10c y_0+40/7d_2 y_0^2$ & 2 & $\{y_{2r}, y_{4r}=-\sqrt{\frac{7}{10}}y_0\}$ \\
$\lambda_3^{\beta}=-40c y_0+10/7d_2 y_0^2$ & 3 & $\{y_{2i}, y_{3}=\sqrt{7}\bar y_1\}$ \\
\hline
\end{tabular}
\caption{\small Eigenvalues of the linearized vector field at $\beta=\beta_\pm$}
\label{vp_O}
\end{table}

\subsection{Bifurcation in the invariant planes.} \label{subsec:P}
There are two types of fixed-point planes in $V^{4}$,with representatives 
$$\left.\begin{array}{c}
P=Fix(\D_4\oplus\ZZ2^c)=\{y_0,y_{4r}\} \\ P_1=Fix(\D_3\oplus\ZZ2^c)=\{y_0,y_{3r}\}.
\end{array}\right.$$
Each of them contains the axis $Fix(\OO2\oplus\ZZ2^c)=\{y_0\}$ and two copies of the axes of cubic isotropy $\OB\oplus\ZZ2^c$: namely $\{y_{4r}=\pm\sqrt{5/14} y_0\}$ in $P$, and $\{y_{3r}=\pm\sqrt{10/7} y_0\}$ in $P_1$.

\begin{figure}[h]
\begin{center}
\includegraphics[width=0.49\hsize]{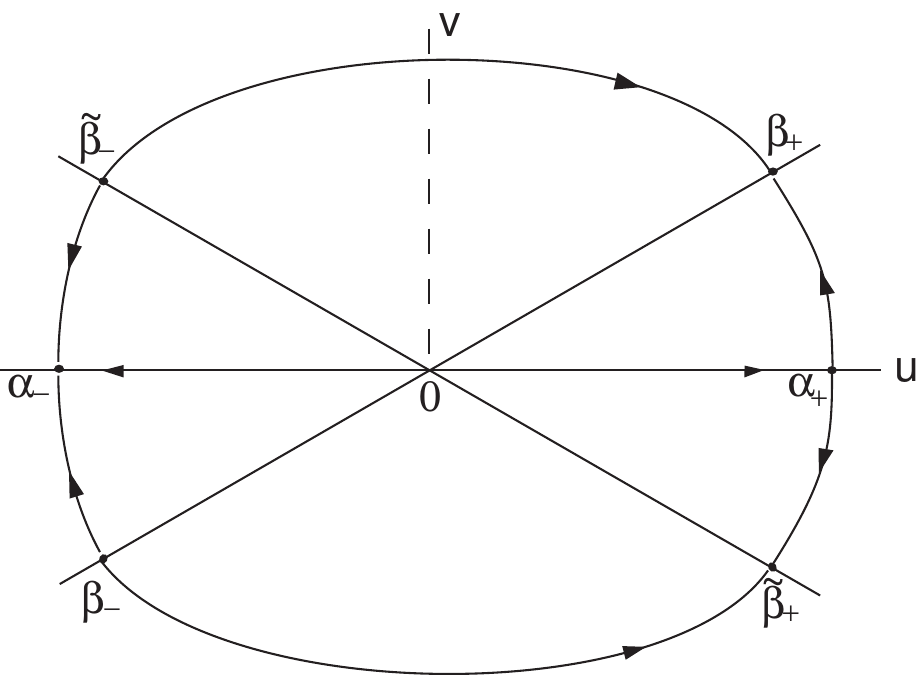} \hfill
\includegraphics[width=0.49\hsize]{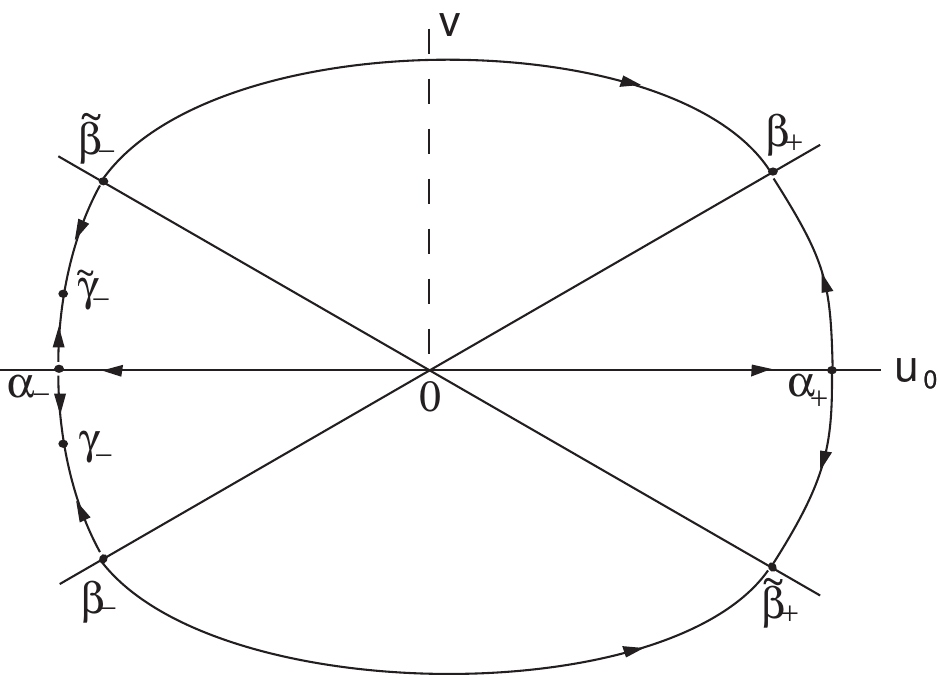}
\caption{\footnotesize Sketch of the phase portrait in $P=Fix(\D_4\oplus\ZZ2^c)$. Left: $0<\mu_2<\mu_2^\alpha$, right: $\mu_2^\alpha<\mu_2<\mu_2^\beta$.}
\label{phasediag_D4}
\end{center}
\end{figure}

We focus on the plane $P$. The case of $P_1$ would be treated similarly and we just indicate the result. Writing $u=y_0$ and $v=y_{4r}$ to simplify notations, the equations in $P$ read
$$\left.\begin{array}{ccc}
\dot u &=& \mu_2 u + c (9u^2+14v^2)+d_1u(u^2+2v^2)+d_2u(u^2-\frac{26}{7}v^2) \\
\dot v &=& \mu_2 v +14c uv +d_1v(u^2+2v^2)+d_2v(-\frac{13}{7}u^2+\frac{30}{7}v^2)
\end{array}\right.$$ 
We note $\beta_\pm$, resp. $\tilde\beta_\pm$, the equilibria satisfying $y_{4r}=\eta y_0$, resp. $y_{4r}=-\eta y_0$. $\beta_+$ and $\tilde\beta_+$ are exchanged by rotation $\phi=\pi/4$, idem for $\beta_-$ and $\tilde\beta_-$.
Equilibria off the invariant axes must satisfy the conditions
$u = \frac{7c}{4d_2}$ and  $\mu_2 = -\frac{7c^2}{16d_2^2} \left(7d_1+43d_2\right) -(2d_1 +\frac{30}{7}d_2)v^2$. 
We see that a secondary bifurcation off the $\alpha$ branch occurs at 
\begin{equation} \label{mu2alpha}
\mu_2^\alpha=-(7d_1+43d_2)\frac{7c^2}{16d_2^2}
\end{equation}
This branch is a supercritical {\it pitchfork}. If $c>0$ the branching occurs from an $\alpha_-$ equilibrium. It corresponds to a change of sign of the eigenvalue $\lambda_4^\alpha$ (Table \ref{vp_O(2)}). \\
Let us denote by $\gamma$, $\tilde\gamma$ these equilibria with isotropy $\D_4\oplus\ZZ2^c$. As $\mu_2$ is increased, they move closer to the 
invariant axes with cubic isotropy until they cross them at the value 
\begin{equation} \label{mu2beta}
\mu_2^\beta=-(21d_1+94d_2)\frac{c^2}{4d_2^2}.
\end{equation}
Observe that one always have $\mu_2^\alpha<\mu_2^\beta$. 
This value $\mu_2^\beta$ corresponds to a {\it transcritical} bifurcation from $\beta$ and $\tilde\beta$ solutions.
When $c>0$, these secondary bifurcations correspond to the change of sign of the eigenvalue 
$\lambda_2^{\beta_-}$ (see Table \ref{vp_O}). \\
The phase diagram in $P$ when $0<\mu_2<\mu_2^\beta$ looks like in Fig. \ref{phasediag_D4}. 
Observe that as long as $\mu_2<\mu_2^\alpha$, saddle-sink heteroclinic orbits connect the $\alpha_-$ to the $\beta_-$ equilibria in $P$. \\
When $\mu>\mu_2^\beta$, $\gamma$ passes on the other side of the $\beta$ axis and the connections between $\alpha_-$ and $\beta_-$ equilibria are restored but the arrows are reversed. 

In the plane $P_1$, same calculations show that as long as $\mu_2$ is smaller than $\hat\mu_2^\alpha=-8(168d_1+17d_2)c^2/d_2$ the phase diagram looks like Fig. \ref{phasediag_D4} (left), except that the $\beta$ and $\tilde\beta$ axes are reversed and $\alpha_+$ is stable while $\alpha_-$ is unstable. Note that, $\hat\mu_2^\alpha>\mu_2^\beta$.


\subsection{Bifurcation off the invariant planes in $V^{4}$}
A complete description of the bifurcations with parameters $\mu_2$ and $c$ in $V^4$ was made in \cite{rgd} and we rely on their results. The system is 3-determined and the equations restricted to order 3 are gradient, which implies that no non trivial dynamics exists in a neighborhood of the bifurcation. We need however to making precise statements about heteroclinic connections between equilibria in $V^4$.  

\begin{proposition} \label{prop:Z2c}
Suppose $\mu_2>0$ and $c>0$ is close enough to 0. Then 
\begin{itemize}
\item[(i)] A flow-invariant and attracting topological sphere exists in $V^4$, which contains all the bifurcated equilibria.
\item[(ii)] The $\OO3$ orbit of octahedral equilibria $\beta_+$ is attracting. The other branches of equilibria have types $\beta_-$ and $\alpha_\pm$. If $\mu_2>\mu_2^\alpha$, a secondary branch $\gamma$ with isotropy $\D_4\oplus\ZZ2^c$ bifurcates from $\alpha_-$ and crosses the branch $\beta_-$ when $\mu_2$ crosses $\mu_2^\beta$. There are no other bifurcated equilibria when $\mu_2$ is smaller than a critical value which is greater that $\mu_2^\beta$.
\item[(iii)] The robust heteroclinic connections between these orbits of equilibria are sketched in Fig. \ref{connexionsV4}. The numbers indicate the dimensions of the manifolds of connections in the unstable manifolds of the equilibria at the base of the arrows. Remaining dimensions (recall that $\dim V^4=9$) correspond to stable submanifolds of the equilibria. 

\begin{figure}[h]
\centering
\includegraphics[width=0.8\hsize]{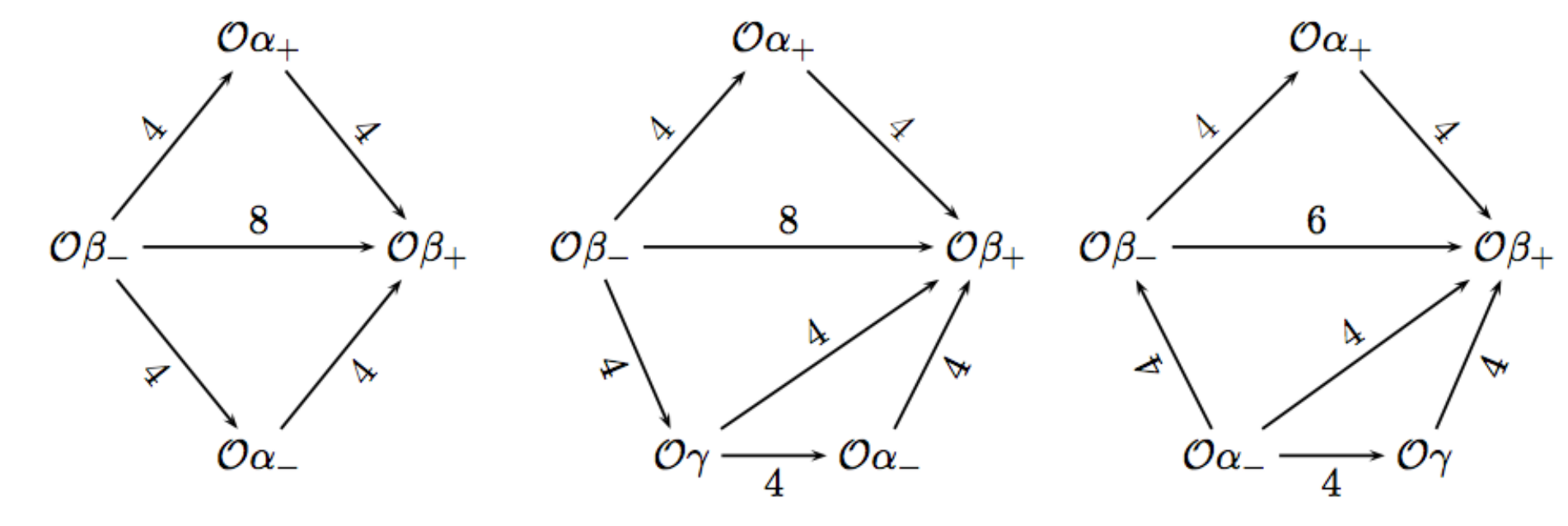}
\caption{\footnotesize Heteroclinic connections in $V^4$.
Left: $\mu_2<\mu_2^\alpha$. Center: $\mu_2^\alpha<\mu_2<\mu_2^\beta$. Right: $\mu_2^\beta<\mu_2$.}
\label{connexionsV4}
\end{figure}

\end{itemize} 
\end{proposition}
{\bf Proof.} The existence of the invariant sphere is a consequence of a theorem by M. Field \cite{Field89}, which we can apply because (i) $c>0$ is assumed small enough, and (ii) the quartic form $d_1\|Y\|^4+d_2Y\cdot C^{(2)}(Y)$ is negative, where $C^{(2)}(Y)$ is the cubic map defined in Table \ref{mapC^(2)} (straightforward proof if $d_2$ is small). \\
The (orbital) stability of $\beta_+$ is easy to check under the hypotheses of the theorem. We rely on \cite{rgd} to assert that when $\mu_2<\mu_2^\beta$, there is no secondary branches of equilibria than $\gamma$.  
Finally point (iii) follows from a careful examination of the phase portrait in the invariant subspaces, which itself relies on the sign of eigenvalues in Tables \ref{vp_O(2)} and \ref{vp_O}.
\eop

\begin{figure}[h]
\begin{center}
\epsfig{figure=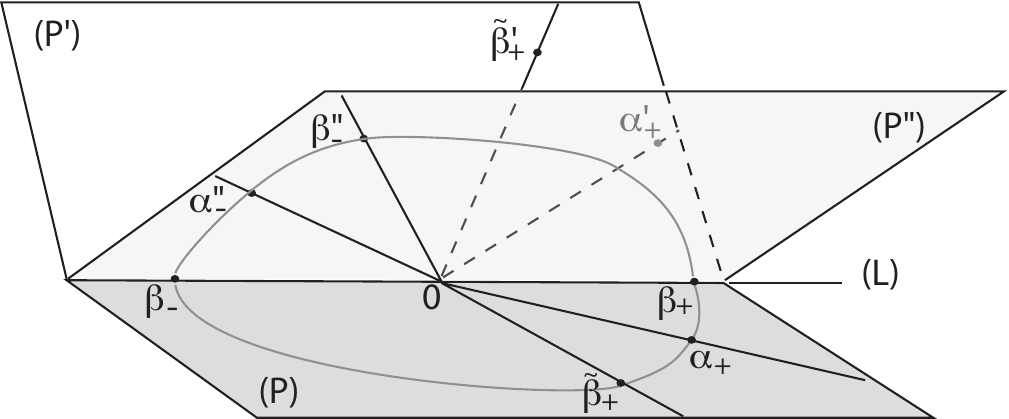,height=4cm}
\caption{\footnotesize The subspace $Fix(\D_2\oplus\ZZ2^c)$.}
\label{Fix(D2)} 
\end{center}
\end{figure}

\begin{figure}[h]
\begin{center}
\includegraphics[height=4.5cm]{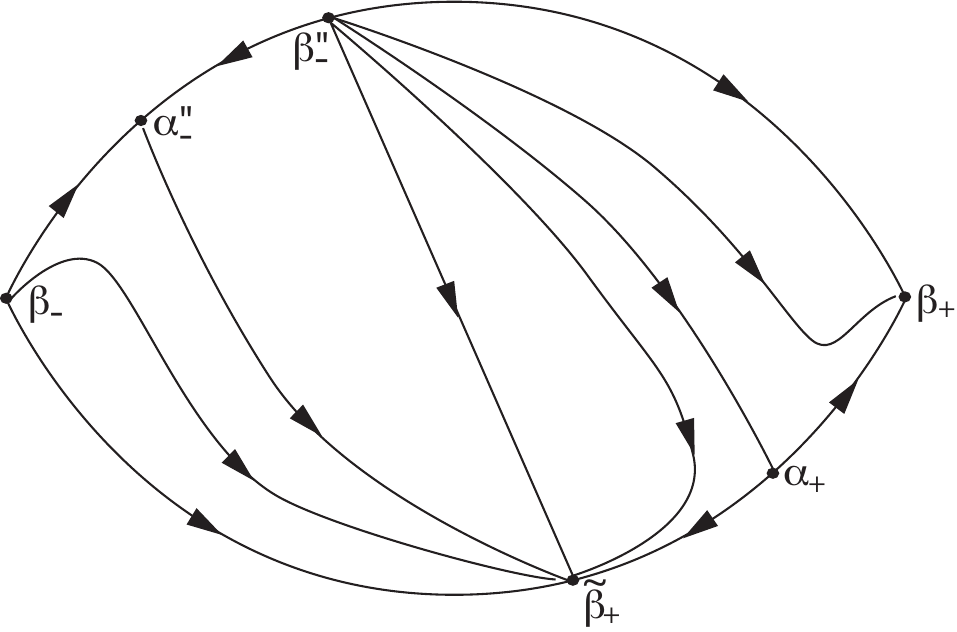} \hskip 0.5cm
\includegraphics[height=4.5cm]{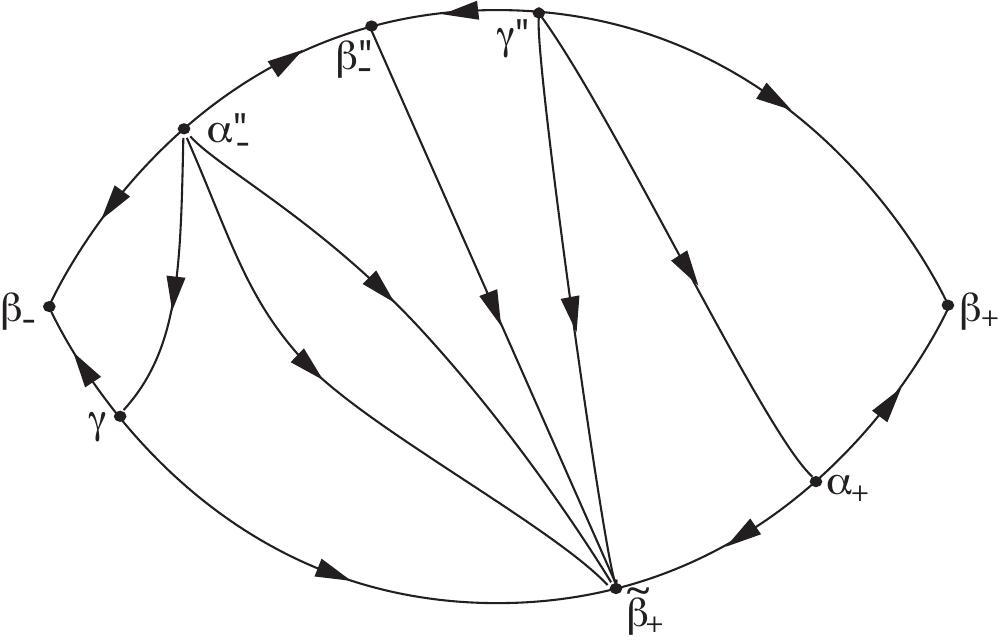}
\caption{\footnotesize $Fix(\D_2\oplus\ZZ2^c)$: the phase portrait on the portion of invariant sphere bounded by $P$ and $P"$ with $y_0>0$. {\bf Left:} $\mu_2<\mu_2^\alpha$. {\bf Right:} $\mu_2>\mu_2^\beta$.}
\label{connexionsD2Z2c}
\end{center}
\end{figure}

In order to illustrate the phase portrait in higher dimension in the case $\mu_2<\mu_2^\alpha$, we consider the 3 dimensional subspace $Fix(\D_2\oplus\ZZ2^c)$. This space contains 
three copies $P$, $P'$ and $P"$ of $Fix(\D_4\oplus\ZZ2^c)$, which intersect all three on the same axis $L=Fix(\OB
\oplus\ZZ2^c)$. Using coordinates $(y_0,y_{2r},y_{4r})$ (see Table \ref{tableFix(H)}), these planes have equations 
$y_{2r}=0$ and $y_{2r}=\pm(\sqrt{10/4}y_0-\sqrt{7/2}y_{4r})$. We note $L=\{y_{2r}=0,y_{4r}=\sqrt{5/14}y_0\}$. 
The three planes are exchanged by the action of $N(\D_2)/\D_2\simeq D_3$ (we note $N(\D_2)$ the normalizer of $
\D_2$ in $\SO3$, which is known to be $\OB$ \cite{gost2}).  This structure is sketched in Fig. \ref{Fix(D2)}. \\
Assuming the existence of the flow-invariant sphere, the bounded dynamics restricts to that sphere and can 
be projected on a 2D picture, see Fig. \ref{connexionsD2Z2c} for cases $\mu_2<\mu_2^\alpha$ and $\mu_2>\mu_2^\beta$. We see that $\beta_+$ and $\tilde\beta_+$ are sinks, 
$\beta_-$ and $\beta_-''$ have robust connections to $\alpha_-$ (in $Fix(\D_4\oplus\ZZ2^c)$) which itself has a robust 
connection to $\tilde\beta_+$. This combined with the phase portrait in $Fix(\D_3\oplus\ZZ2^c)$ gives a good description of the various connections in $Fix(\ZZ2^c)$.

\section{Bifurcations involving the modes with $\ell=3$} \label{sec:bif_l=3}
Even under the hypotheses of Proposition \ref{prop:Z2c}, the high dimensionality of the phase space $V$ makes it very likely that the bifurcation diagram and dynamics are so complicated that their complete description is either intractable or extremely laborious. We shall therefore restrict our analysis to subdomains in parameter space, in which spatio-temporally interesting dynamics is likely to bifurcate.

To be more precise, we now assume $\mu_1<0$. This implies that the modes with $\ell=3$ are (weakly) stable at bifurcation. As we shall see, this assumption helps isolating complex spatio-temporal patterns that can be analyzed on the center manifold using the tools of equivariant bifurcation and dynamics. Actually, when $\mu_1>0$ various kinds of steady-state, periodic, quasiperiodic or chaotic dynamics can be observed numerically, as it can be expected from the signs of eigenvalues of the Jacobian matrix at the pure $\ell=4$ equilibria. Our aim is at isolating regions in parameter space at which as simple as possible recurrent dynamics can be observed. Such conditions are defined in this section, and the resulting invariant objects (generalized heteroclinic cycles) will be discussed in the next section.  

\subsection{Dynamics in the mixed mode plane $\Pi=Fix(\OB^-)$} \label{subsec:fix(O-)}
\begin{figure}[h]
\begin{center}
\epsfig{figure=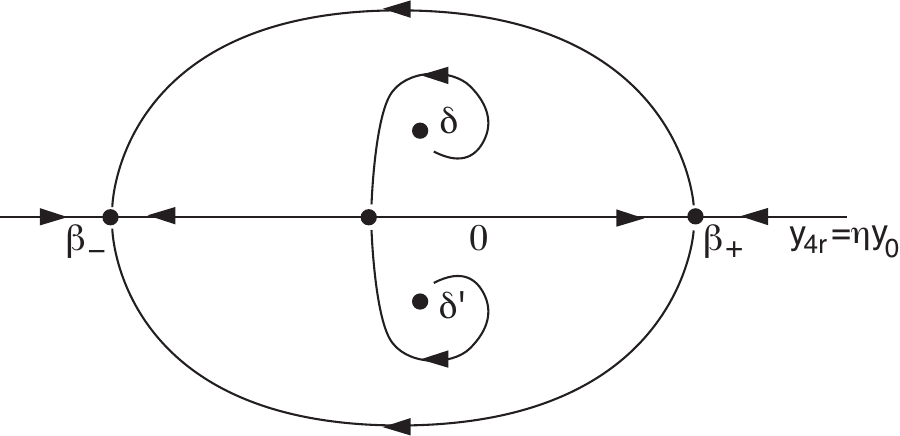,height=5cm} 
\caption{\footnotesize Sketch of the phase portrait in $\Pi$ when $\hat\mu_1<\mu_1<0$.}
\label{fig:Pi} 
\end{center}
\end{figure}

We assume the hypotheses of Proposition \ref{prop:Z2c} are fulfilled. Hence the $O(3)$-orbit of the octahedral steady-state $\beta_+$ is an attractor in the space $V^4$. We now look at the stability of $\beta_+$ in the invariant plane $\Pi$, the coordinates of which are $x_{2i}$ and $y_{4r}=\eta y_0$ ($\eta=\sqrt{5/14}$), and which is the only invariant plane containing $\beta_\pm$ equilibria. \\
$\Pi$ contains only the axis of symmetry $\{x_{2i}=0\}$. Using $y_0$ as a parameter for this axis, the equations for the flow on $\Pi$ read
\begin{eqnarray}
\dot x_{2i} &=& \mu_1 x_{2i} + 12 x_{2i}y_0+(2\gamma_1+20\gamma_2) x_{2i}^3 \\
\dot y_0 &=& \mu_2y_0 +14c y_0^2 -14 x_{2i}^2 +(12/7d_1-16/49d_2)y_0^3
\end{eqnarray}   
The bifurcation scenario for this kind of planar system is classical and we don't go into details of the calculations \cite{ArCh91}.  Let us assume first that $c=0$ and fix $\mu_2>0$. It follows from the above equations that a (secondary) bifurcation occurs from $\beta_+$ when $\mu_1$ reaches the value
\begin{equation} \label{tildemu1}
\tilde\mu_1=-12\sqrt{\frac{\mu_2}{-D}}
\end{equation}
where $D=12/7d_1-16/49d_2<0$ (see (\ref{mu2cubic}).
This corresponds to the value of $\mu_1$ at which the eigenvalue $\sigma_0^{\beta_+}=0$ (see Table \ref{vp_O}). \\
As $\mu_1$ is varied the following sequence of events occurs: \\
- For $\mu_1<\tilde\mu_1$, there exist no equilibria off the invariant axis and $\beta_\pm$ are both sinks in $\Pi$. \\
- At $\mu_1=\tilde\mu_1$ a pitchfork bifurcation occurs from $\beta_+$ off the axis. We note $\delta$ and $\tilde\delta$ 
the bifurcated equilibria with isotropy $\OB^-$ (which are exchanged by taking $x_{2i}$ to $-x_{2i}$). \\
- As $\mu_1$ is increased further, a Hopf bifurcation occurs from $\delta$ (and $\tilde\delta$). We call $C_\Pi$ and $\tilde C_\Pi$ the bifurcated periodic orbits.\\
- The two limit cycles grow until they collide with the stable manifold of the origin. Then they disappear and two {\it 
saddle-sink connections} from $\beta_+$ to $\beta_-$ are established in $\Pi$ at a value $\mu_1=\hat\mu_1$ which satisfies $
\tilde\mu_1<\hat\mu_1<0$, see Fig. \ref{fig:Pi}. \\
- The equilibria $\delta$, $\tilde\delta$ die off at the origin in a "reverse" bifurcation at $\mu_1=0$. \\
- The heteroclinic orbit from $\beta_+$ to $\beta_-$ persists until a new pitchfork bifurcation occurs from $\beta_-$ at a 
positive value of $\mu_1$. \\
These properties persist when $c\neq 0$ is small enough.

\subsection{Dynamics of mixed modes off the plane $\Pi$} \label{sec:offPi}
We now want to find conditions on parameters such that the heteroclinic connections found above, not only exist but also drive the dynamics along the $\ell=3$ modes. Then the construction of the generalized heteroclinic cycles will follow. 

Let us denote by $y_0^{\alpha_\pm}$ (resp. $y_0^{\beta_\pm}$) the value of $y_0$ on the branches of equilibria $\alpha_\pm$ (resp. $\beta_\pm$) at a given parameter value $\mu_2>0$. 
\begin{definition}
We set $\mu_1'=-7y_0^{\alpha_+}$ (value at which $\sigma_2^{\alpha_+}=0$)  and $\mu_1''=-2y_0^{\beta_+}$ (value at which $\sigma_2^{\beta_+}=0$).
\end{definition}
Note that $\mu_1'<0$ and $\mu_1''<0$. The following lemma gives conditions on the parameters such that the eigenvalues of the Jacobian matrix of $\alpha_\pm$, $\beta_\pm$ along the modes $\ell=3$ are mostly negative.

\begin{lemma} \label{lemma1}
We suppose $\mu_1<0$.Then: \\
(i) $\sigma_j^{\alpha_+}<0$ ( $j=0,1,3$), $\sigma_2^{\alpha_-}<0$, $\sigma_1^{\beta_+}<0$, $\sigma_0^{\beta_-}<0$ and $\sigma_2^{\beta_-}<0$. \\
(ii) if $\mu_1<\min(\mu_1',\mu_1'')$, then $\sigma_j^{\alpha_-}<0$ ( $j=0,1,3$), $\sigma_2^{\alpha_+}<0$, $\sigma_1^{\beta_-}<0$ and    $\sigma_2^{\beta_+}<0$; \\
(iii) If $\mu_1<\min(\mu_1',6\mu_1'')$, then in addition to inequalities (ii), we also have $\sigma_0^{\beta_+}<0$. 
\end{lemma}
{\bf Proof.} Point (i) follows straightforwardly from Tables \ref{vp_O(2)} and \ref{vp_O}. It also follows from these tables that
the signs of the eigenvalues in point (ii) are true if
$$\mu_1<6y_0^{\alpha_-}~~, \mu_1<-7y_0^{\alpha_+},~~ \mu_1<6y_0^{\beta_-},~~ \mu_1<-2y_0^{\beta_+}.$$
Solving (\ref{mu2cubic}) and (\ref{mu2axisymmetric}) for $y_0$, we find that
\begin{eqnarray*} 
y_0^{\alpha_\pm} &=& \frac{1}{2(d_1+d_2)}\left( -9c\mp\sqrt{81c^2-4(d_1+d_2)\mu_2}\right) \\
y_0^{\beta_\pm} &=& \frac{1}{D}\left( -7c\mp\sqrt{49c^2-D\mu_2}\right)
\end{eqnarray*}
Recall that $d_1+d_2<0$ and $D<0$ (see \ref{mu2cubic}). Replacing these formulas in the above inequalities, point (ii) follows.
Also note that $6\mu_1''=\tilde\mu_1$, from which (iii) follows.
\eop 

\begin{figure}[h]
\begin{center}
\epsfig{figure=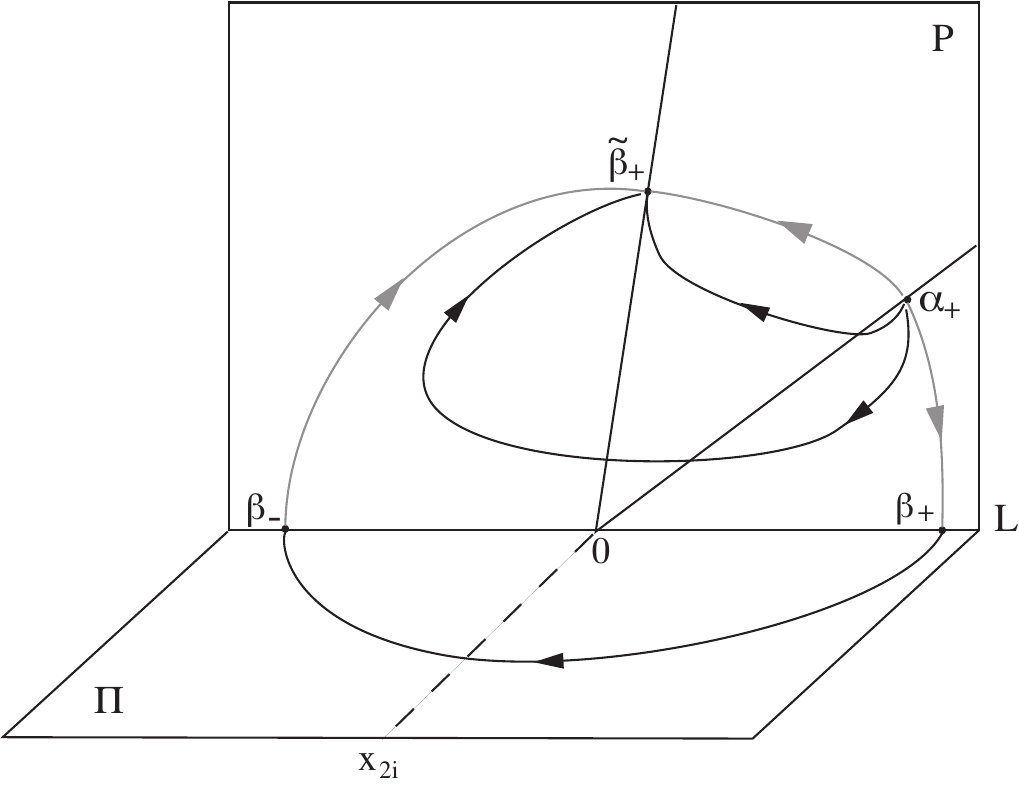,height=6cm}
\caption{\footnotesize The phase portrait in $\Delta=Fix(\D_4^d)$ (in grey the heteroclinic connections in $P$).}
\label{Fix(D4d)}
\end{center}
\end{figure}

By point (iii) of this lemma, if $\mu_1<\min(\mu_1',6\mu_1'')$, all the eigenvalues of the Jacobian matrix at $\beta_+$ in eigendirections transverse to the $\OO3$-orbit are negative, therefore $\beta_+$ is an orbitally stable equilibrium (i.e. its orbit ${\cal O}_{\beta_+}$ is an asymptotically stable object). However for the values of the coefficients $d_1$ and $d_2$ shown in Fig. \ref{fig:coef-versus-pr}, and with $c$ close to 0, one can check the inequalities
\begin{equation*}
6\mu_1''<\mu_1'<\mu_1''
\end{equation*}
Moreover, although it is cumbersome to determine the value $\hat\mu_1$ at which the saddle-sink connection $\beta_+\rightarrow\beta_-$ is established in $\Pi$, numerical simulations in \ref{subsec:stability} show evidence that $\hat\mu_1<\mu_1''$. \\
To understand the change of dynamics when $\mu_1$ varies in the interval $(6\mu_1'',\mu_1'')$ it is convenient to consider the 3-dimensional invariant space $\Delta=Fix(\D_4^d)$. Note that $\Delta=P\oplus\Pi$ (see Table \ref{tableFix(H)}). The eigenvalues $\sigma_2^{\alpha_+}$, resp. $\sigma_2^{\beta_+}$, correspond to the eigendirection at $\alpha_+$, resp. at $\tilde\beta_+$, orthogonal to $P$ in $\Pi$, see Fig. \ref{Fix(D4d)}. When $\mu_1$ increases to cross $\mu_1'$, $\sigma_2^{\alpha_+}$ changes sign from $<0$ to $>0$, which corresponds to the {\it reverse} pitchfork bifurcation of a mixed-mode equilibrium in $\Pi$. It can be shown (see \cite{ChGuLa99} for a similar case) that in this case a saddle-sink connection is established in $\Pi$ between $\alpha_+$ and $\tilde\beta_+$. \\    
The dynamics in $\Pi$ when $\mu_2<\mu_2^\alpha$ and $\max(\hat\mu_1,\mu_1')<\mu_1<\mu_1''$ is sketched in Fig. \ref{Fix(D4d)}.
 
In fact, as in the Proposition 4.6 of \cite{ChGuLa99}, one can even show the following result in the 5-dimensional subspace $Z=Fix(\ZZ4^-)$. We refer to \cite{ChGuLa99} for the proof.
\begin{proposition} \label{prop:Z4-}
If $c>0$, $\mu_2>0$ and $\hat\mu_1<\mu_1''$, there exists an open interval of negative values of $\mu_1$ such that the full (4-dimensional) unstable manifold of $\alpha_+$ is included in the stable manifold of the group orbit of $\beta_+$ in $Z$. 
\end{proposition}

\section{Existence of heteroclinic cycles \label{section_genhetcyc}}
\subsection{Generalized heteroclinic cycles}
We summarize the results of sections \ref{sec:l=4} and \ref{sec:bif_l=3} in the following theorem:
\begin{theorem} \label{thm:generalized hetcycles}
For the system (\ref{eq:ode}) with coefficient values as in Fig. \ref{fig:coef-versus-pr}, assume $c>0$ close to 0 and $\max(\hat\mu_1,\mu_1')<\mu_1<\mu_1''$. Then
\begin{itemize}
\item[(i)] If $0<\mu_2<\mu_2^\alpha$, the following sequence of robust heteroclinic connections (saddle-sink connections in flow-invariant subspaces) exists:
\begin{center}
\includegraphics[width=0.66\hsize]{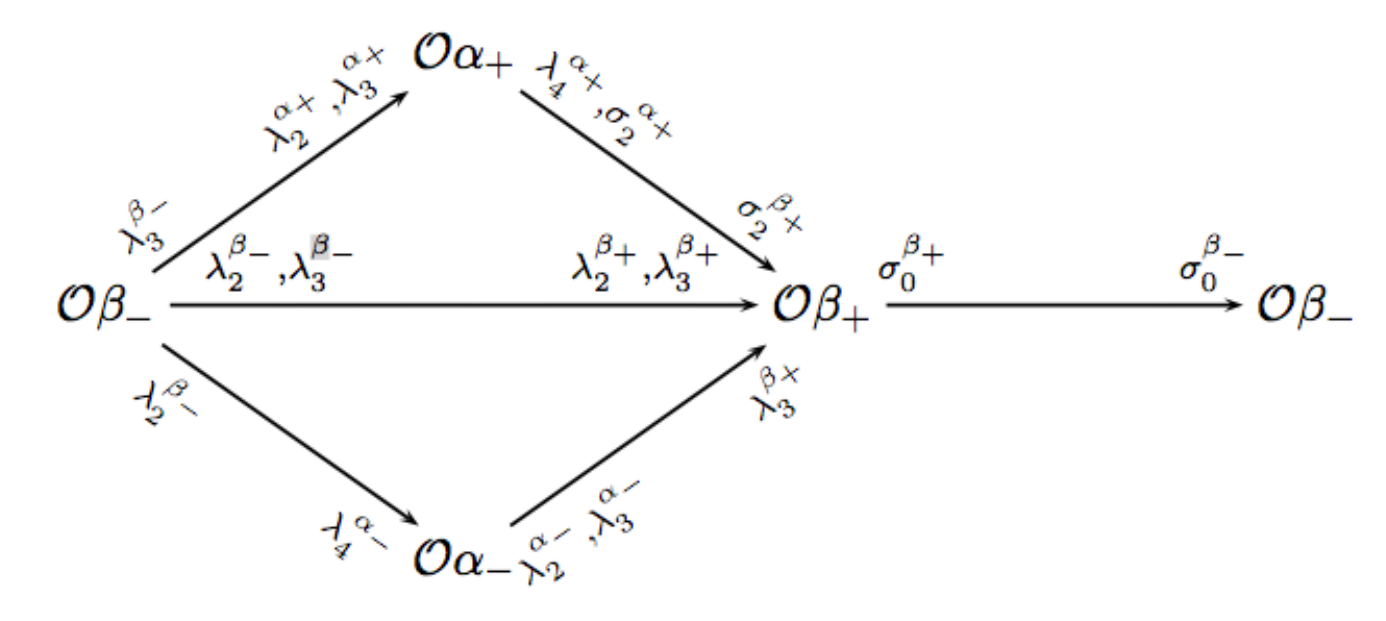}
\end{center}
\item[(ii)] If $\mu_2^\alpha<\mu_2<\mu_2^\beta$, a similar object as in (i) exists, with an additional orbit of equilibria $\gamma$ intercalated between ${\cal O}\beta_-$ and ${\cal O}\alpha_-$, see Fig. \ref{connexionsV4} (center).
\newpage
\item[(iii)] If $\mu_2$ is slightly larger than $\mu_2^\beta$, the sequence of connections is simpler ($\alpha_-$ not involved in the sequence):
\begin{center}
\includegraphics[width=0.7\hsize]{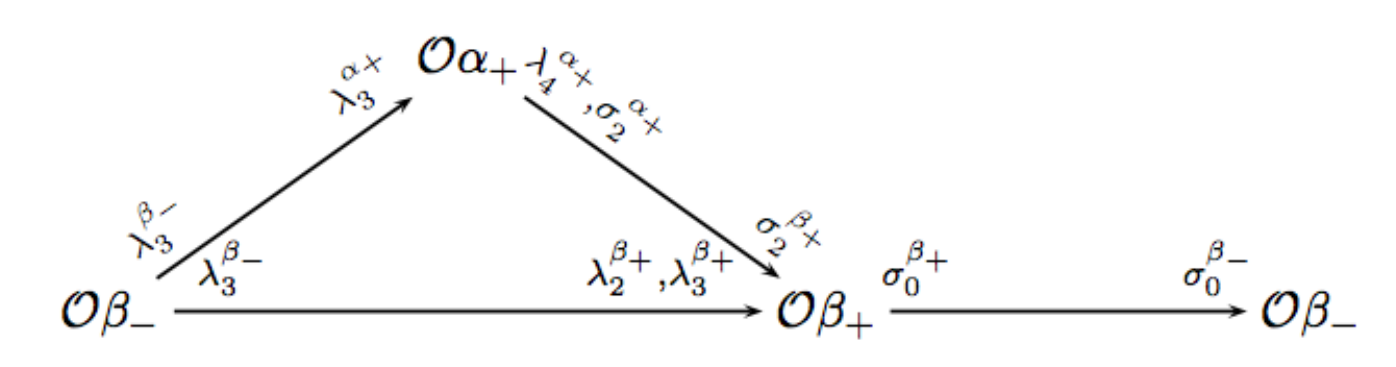}
\end{center}
%
\end{itemize}  
We indicate in the diagrams the eigenvalues for the corresponding stable or unstable parts of the heteroclinic connections. Note that the unstable manifold of each equilibrium is filled with heteroclinic orbits connecting to one or several other equilibria.. \\
The above sequences realize {\it generalized heteroclinic cycles} of types I(i), I(ii) and I(iii) respectively.
\end{theorem}

{\bf Proof.}  By inspection of the diagrams in Fig. \ref{connexionsV4} and Proposition \ref{prop:Z4-}. 
\eop

\subsection{A simple heteroclinic cycle involving mixed-mode equilibria \label{simplehetcyc}}
We have seen in Section \ref{subsec:fix(O-)} that when $\mu_1<\hat\mu_1<0$, the presence of a stable fixed point 
$\delta$ in the plane $\Pi$ prevented the existence of the saddle-sink connections in 
that plane. However if we now look at the 3-dimensional subspace $\Delta=P\oplus\Pi$, a saddle-sink connection 
can exist in that subspace between $\delta$ (in $\Pi$) and $\tilde\beta_+$ (in $P$). A similar situation holds in the $\ell=1, 2$ interaction \cite{ArCh91}. Here the process by which this saddle-sink connection is established is more complex than in the $\ell=1, 2$ case. It results from a sequence of bifurcations of steady-states and/or periodic orbits off the invariant planes and is numerically analyzed in Section \ref{difdiagDelta}. 

Once the saddle-sink heteroclinic connection exists, a heteroclinic cycle is set since $\tilde\beta_+$ belongs to the 
group orbit ${\cal O}\beta_+$ (see the diagram \ref{schemacycle3} below).
\begin{equation}
{\cal O}\beta_+ \longrightarrow  {\cal O}\delta  \longrightarrow  {\cal O}\beta_+
\label{schemacycle3}
\end{equation}
Moreover this cycle is {\it simple}, in the sense that it involves unstable manifolds of dimension 1 only.
Following \cite{ArCh91} we call it a type II heteroclinic cycle. \\
Remark that the full cycle is included in the 5 dimensional subspace $Z=Fix(\Z_4^-)$. Indeed, $\delta$ is connected in $P$ to $\beta_+$ and $\tilde\beta_+$. But $\tilde\beta_+$ is connected to $\tilde\delta$ inside the image $\tilde\Pi$ of $\Pi$ by the 8 fold rotation with vertical axis, which is itself included in $\tilde\Delta=\{x_{2r},y_0,y_{4r}\}$. Now note that $Z=\Delta+\tilde\Delta$. \\
Variants of this cycle, involving periodic orbits in $\Pi$, can also exist for suitable parameter values.
%
%
\section{The numerical simulation of the dynamics on the center manifold and its interpretation \label{sec:numerics}}
The results of the previous section do not guarantee the observability of intermittent dynamics induced by the presence of heteroclinic cycles, if they exist. This depends on the behavior as time increases, of trajectories starting close to the cycle. Obviously if the (generalized) heteroclinic cycle is an attractor, then the nearby dynamics will show aperiodic switching between steady-states with increasing periods of time passed in their vicinity. Even if a heteroclinic cycle is not asymptotically stable, a nearby attractor can exist, so that the intermittent behavior can be observable. 

Stability conditions have been derived in \cite{krumel} for heteroclinic cycles which are called "simple". These conditions are easily applicable to cycles of type II, Section \ref{simplehetcyc}. As a general rule, stability depends upon the relative strength of contracting v.s. expanding eigenvalues at the steady-states in the heteroclinic cycle. The case of generalized heteroclinic cycles, like those of type I (section \ref{simplehetcyc}) is more difficult to analyze. Already in spherical Rayleigh-B\'enard convection with $\ell=1$ and $2$ mode interaction, the characterization of the asymptotic stability of generalized heteroclinic cycles led to non obvious formulas \cite{ChGuLa99}. In the present situation the analysis is even more complicated due to the complex topology of these cycles. \\
In this section we therefore explore the dynamics on the center manifold by time integrating the equations (\ref{eq:ode}). We first determine the "stability curves" which, in the $(\mu_1,\mu_2)$ plane, correspond to the critical values which have been defined in Sections \ref{sec:l=4} and \ref{sec:bif_l=3}. Then the bifurcation diagrams when $\mu_1$ is varied are computed, and the dynamics is explored for selected values of $\mu_1$, $\mu_2$ and $c$. Results are compared with eigenvalues computed for the same parameter values, which correspond to the rates of contraction and expansion at relevant steady-states.

\subsection{Stability diagrams}\label{subsec:stability}
\begin{figure}[h]
\begin{center}
\includegraphics[width=0.49\hsize]{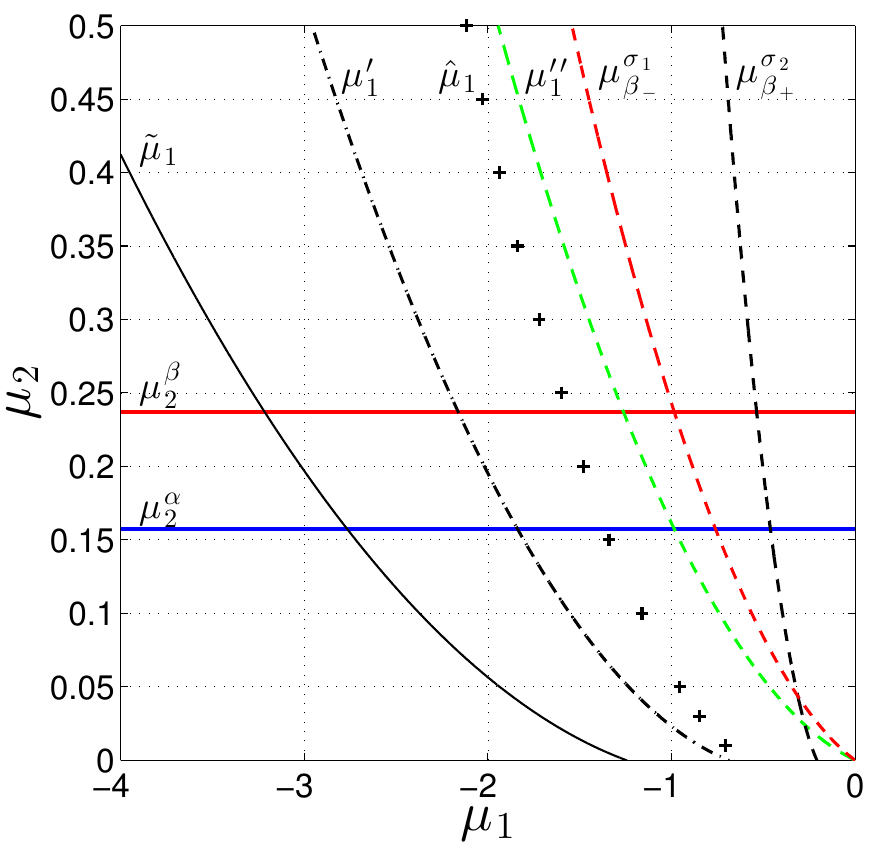}\hspace{0.1cm}
\includegraphics[width=0.49\hsize]{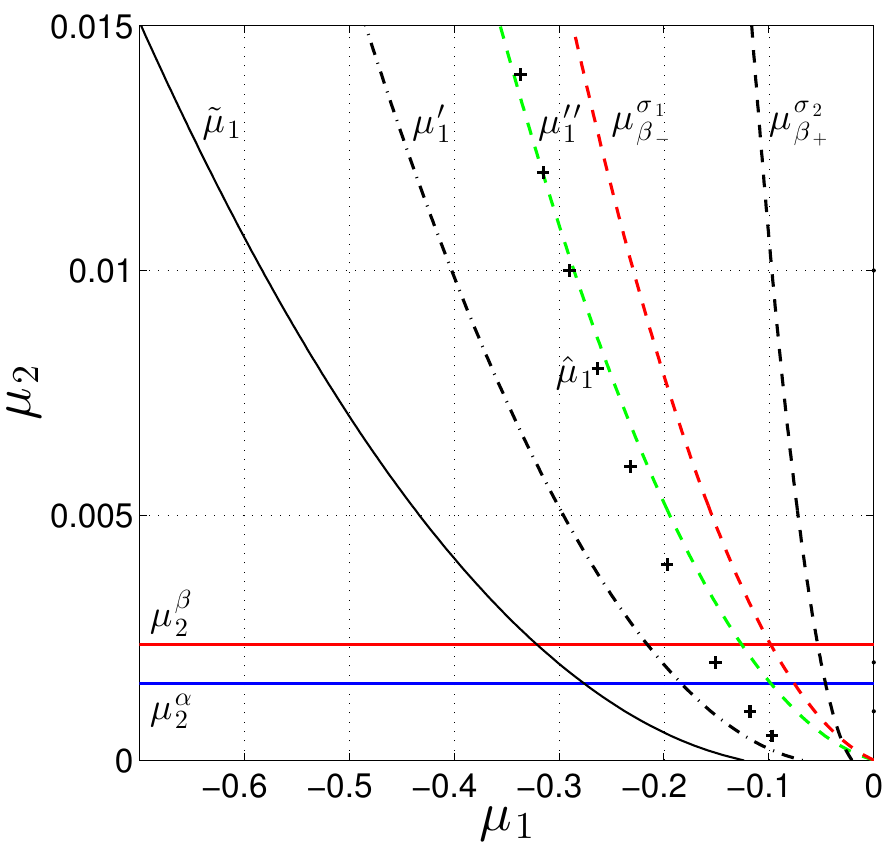}
\caption{Stability diagrams of equilibria $\alpha_\pm$ and $\beta_\pm$ for $g(r)=1/r^2$ and two values of the coefficient $c$: (left) $c=0.04$ and (right) $c=0.004$. For $\mu_2$ fixed, $\mu_{eq}^{\sigma}$ denotes the $\mu_1$ value at which the eigenvalue value $\sigma$ of the equilibrium $eq.$ changes sign. See Tables \ref{vp_O(2)} and \ref{vp_O} for the other critical values notations.}
\label{fig:stabdiag1}
\end{center}
\end{figure}

\begin{figure}[h]
\begin{center}
\includegraphics[width=0.49\hsize]{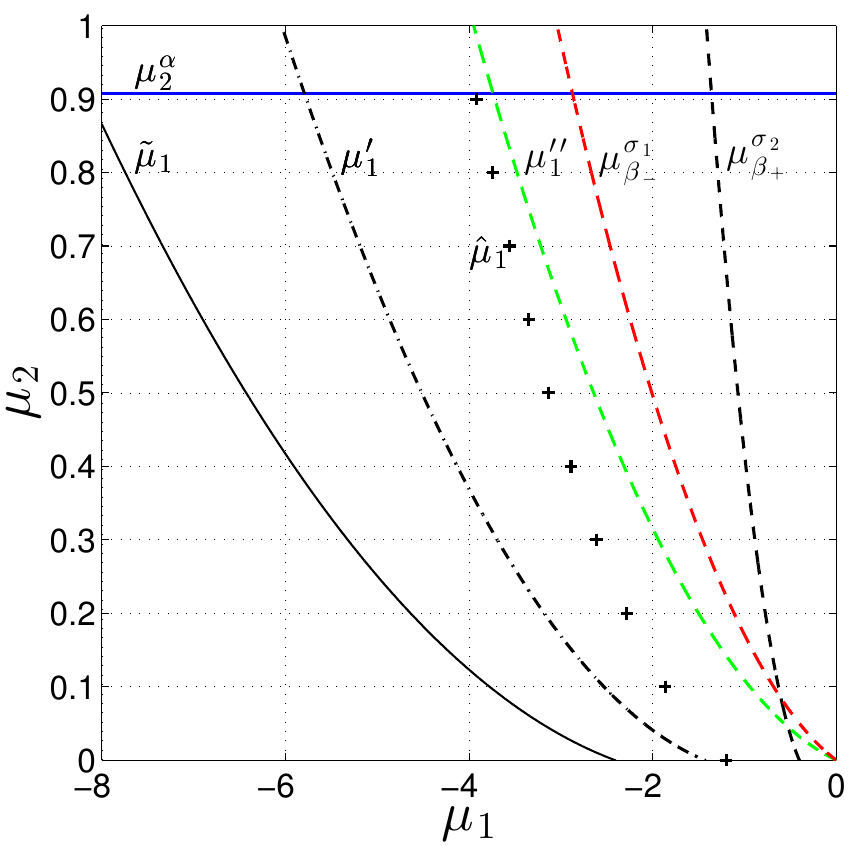}\hspace{0.1cm}
\includegraphics[width=0.49\hsize]{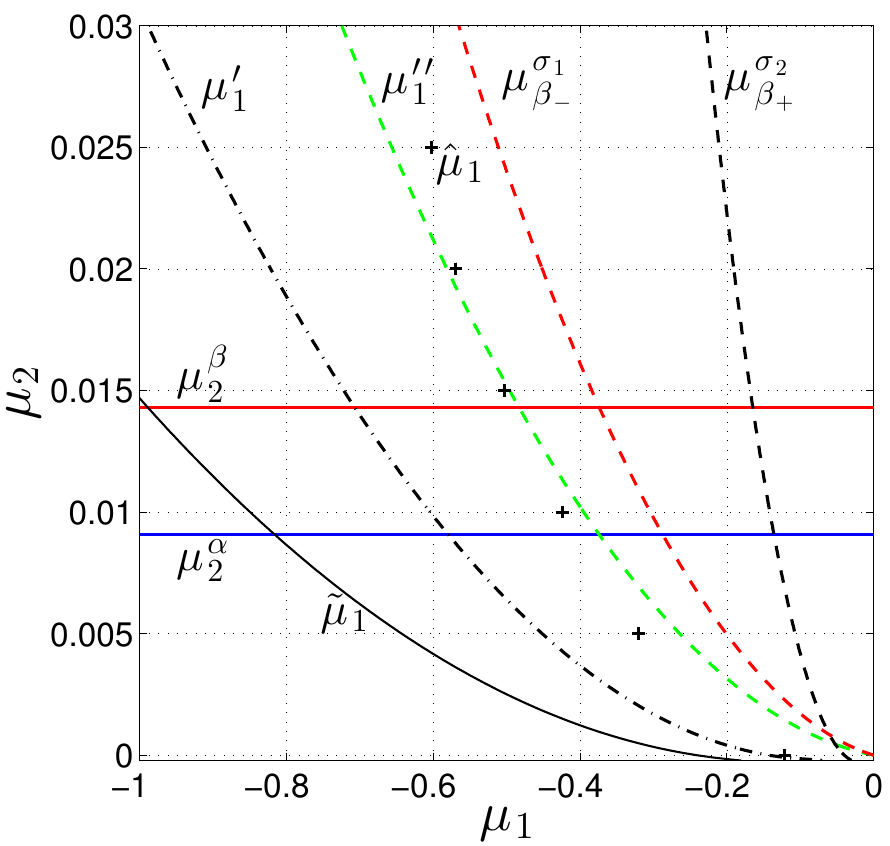}
\caption{Stability diagrams of equilibria $\alpha_\pm$ and $\beta_\pm$ for $g(r)=1/r^5$ and two values of the coefficient $c$: (left) $c=0.04$ and (right) $c=0.004$. Notations are as in Fig. \ref{fig:stabdiag1}}
\label{fig:stabdiag2}
\end{center}
\end{figure}

We focus on the parameter domain delimited by $\mu_1<0$ and $\mu_2>0$. This domain is subdivided into regions bounded by "neutral stability curves" on which the eigenvalues of the $\alpha_\pm$ and $\beta_\pm$ equilibria change sign (Figs. \ref{fig:stabdiag1} and \ref{fig:stabdiag2}). The names of the corresponding 'critical' values of $\mu_1$ have been defined in prevoius sections. See Tables \ref{vp_O(2)} and \ref{vp_O} for an expression of the leading part of these eigenvalues.
The eigenvalue becomes positive when the associated stability curve is crossed by increasing $\mu_1$. 
These domains differ only slightly between both force fields in $1/r^2$ and $1/r^5$. In contrast, the bifurcation parameter $c$ has a strong influence on the region size for each force field. However in any case, the following inequalities hold when $\mu_2>0$ is small enough:
\begin{equation}
\tilde{\mu}_1<\mu_1'<\hat{\mu}_1< min(\mu_1'',\mu_{\beta_-}^{\sigma_1},\mu_{\beta_+}^{\sigma_2}) \label{eq:sortm1}
\end{equation}
where $\mu_{eq}^{\sigma}$ denotes the value of $\mu_1$ at which the eigenvalue $\sigma$ of the equilibrium $eq.$ changes sign (hence becomes positive when $\mu_1>\mu_{eq}^{\sigma}$).
In consequence, according to Theorem \ref{thm:generalized hetcycles}, when $\mu_2$ is small enough and $\hat{\mu}_1<\mu_1<min(\mu_1'',\mu_{\beta_-}^{\sigma_1},\mu_{\beta_+}^{\sigma_2})$, heteroclinic cycles exist. \\
Note that for $c=0.004$ and when $\mu_2$ is large enough, $\hat{\mu}_1$ becomes larger than $\mu_1''$  (Figs. \ref{fig:stabdiag1}b and \ref{fig:stabdiag2}b). In this case the heteroclinic cycles are necessarily unstable along transverse directions. Numerical simulations show a seemingly unstructured dynamics with no intermittent behavior, we therefore do not consider this case. \\
To study the bifurcation scenarios which lead to the establishment of heteroclinic cycles, we keep $\mu_2$ fixed while $\mu_1$ is increased from $\tilde{\mu}_1$ up to the value at which intermittent behavior is lost.
A priori we have to distinguish three cases depending on the relative position of $\mu_2$ with respect to $\mu_2^\alpha$ and $\mu_2^\beta$. However it is observed that the numerical results display similar dynamics in the different cases.  \\
We focus on the central force field $g(r)=1/r^2$. The case $g(r)=1/r^5$ shows very similar results.
\subsection{Steady-state and Hopf bifurcations in $\Delta=Fix(D^d_4)$ \label{difdiagDelta}}
\begin{figure}[ht]
\begin{center}
\includegraphics[width=.49\hsize]{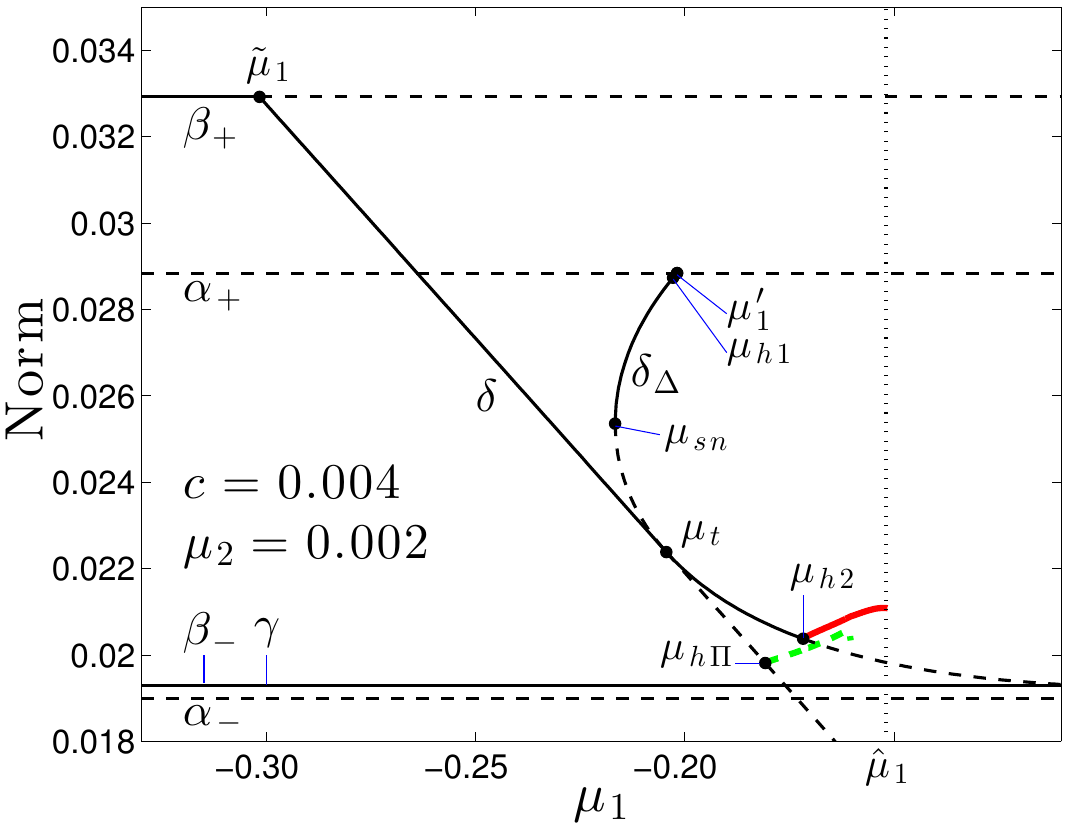}
\includegraphics[width=.49\hsize]{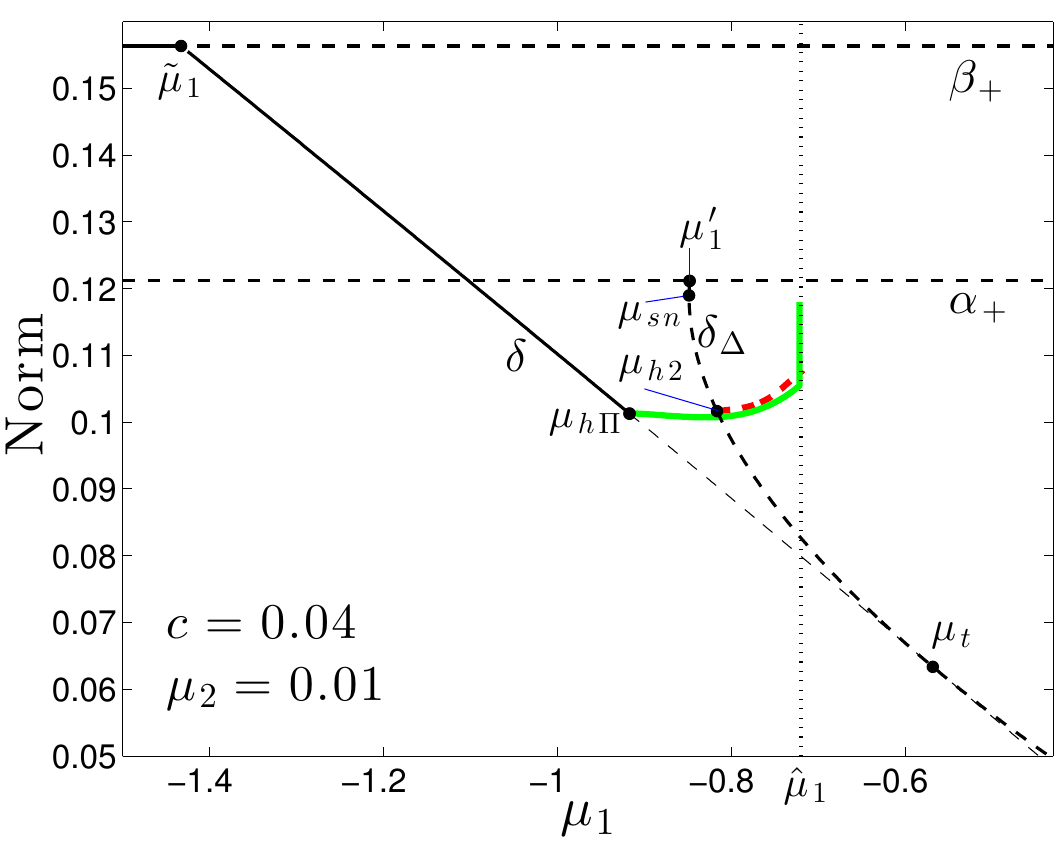}
\includegraphics[width=.49\hsize]{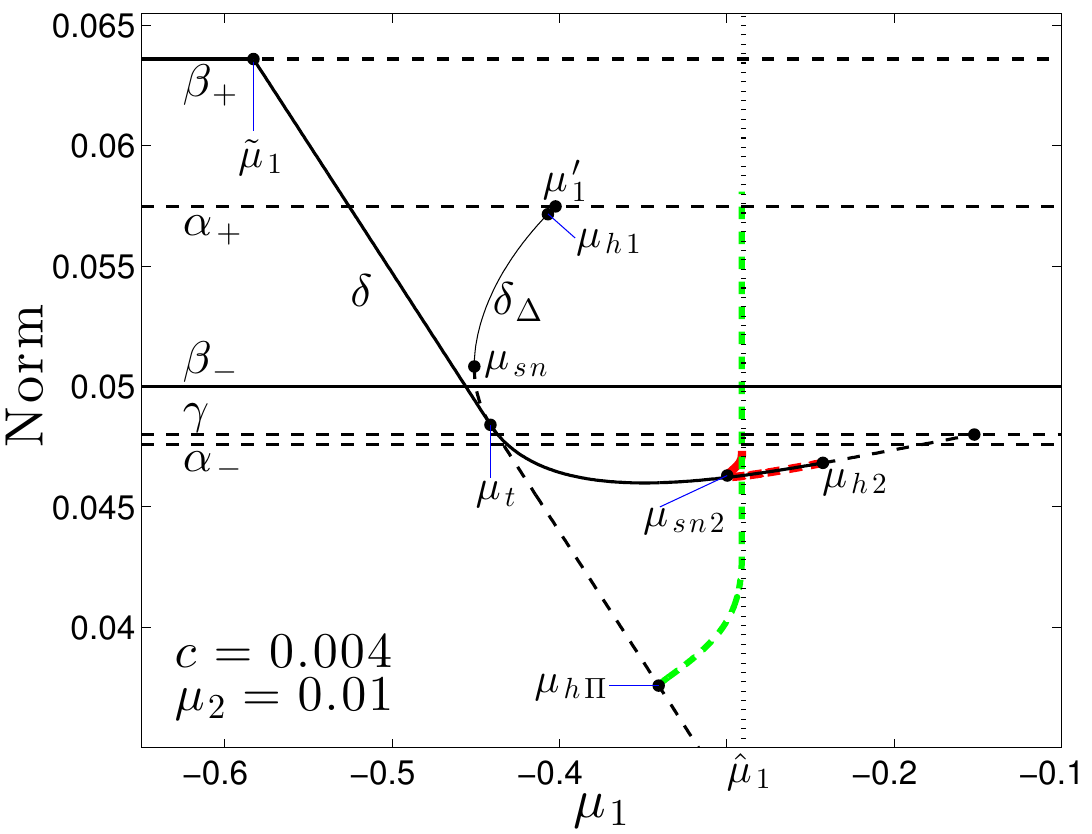}
\includegraphics[width=.49\hsize]{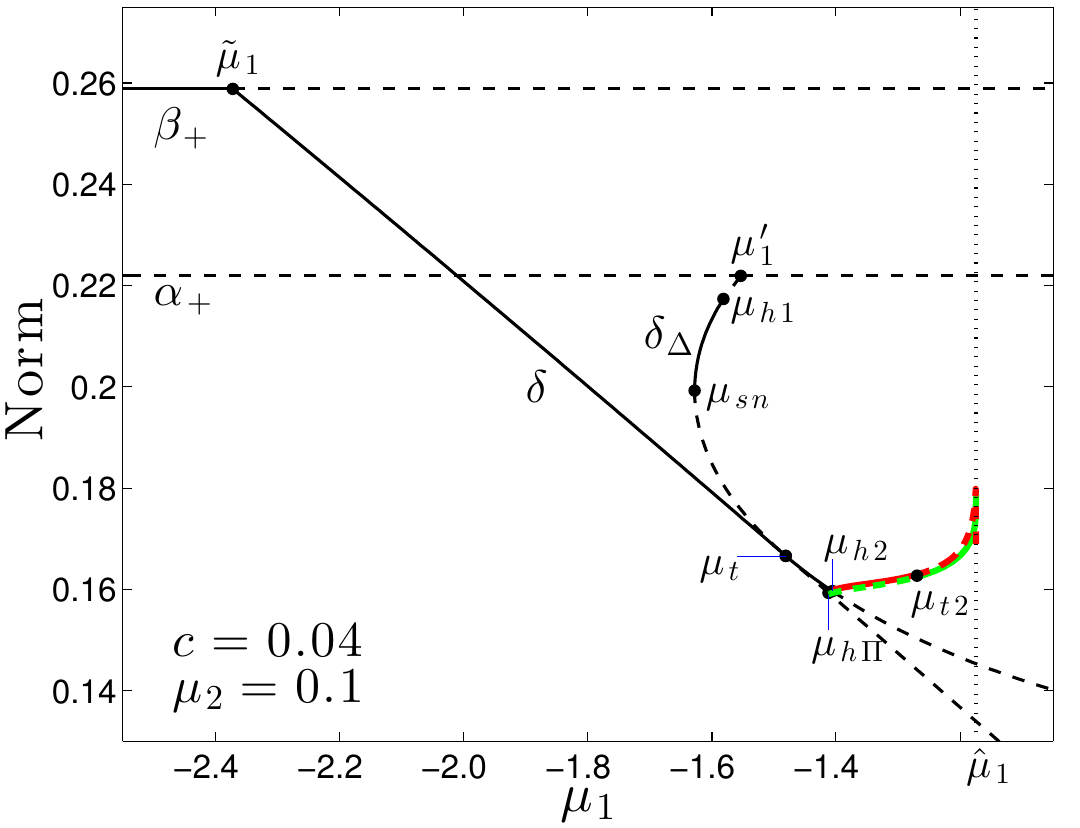}
\caption{Bifurcation diagrams in $\Delta$ with respect to $\mu_1$ for four different parameter couples $(\mu_2,c)$. [Dashed] Plain line indicates [unstable] stable branch. The dotted vertical line corresponds to the value $\hat{\mu}_1$.}
\label{fig:bifdiagD}
\end{center}
\end{figure}
We noticed in Section \ref{section_genhetcyc} that a key condition for the existence of heteroclinic cycles is the establishment of saddle-sink heteroclinic connections between steady-states in the three-dimensional space $\Delta=P\oplus\Pi$. Figure \ref{Fix(D4d)} shows the phase portrait in the flow-invariant domain $y_{4r}<\eta y_0$, in the dynamically simple case $\mu_1\in ]max(\hat{\mu}_1,\mu_1'),\mu_1''[$. 
Here we study in more details, with the help of the continuation software AUTO~\cite{AUTO00}, the bifurcation diagrams and the scenarios which lead to these connections in $\Delta$ for $c=0.004$ and $c=0.04$. 

The results are displayed in \myfig{bifdiagD} for $\mu_2=0.002$ and $0.01$ (case $c=0.004$), and $\mu_2=0.01$ and $0.1$ (case $c=0.04$). These values illustrate the cases described in Theorem \ref{thm:generalized hetcycles}. The presence of a "submaximal" equilibrium $\gamma$ in $P$ does not introduce significant differences in the bifurcation diagrams. \\
We indicate in each case the value $\hat\mu_1$ at which the saddle-sink connection $\beta_+\rightarrow\beta_-$ is established in $\Pi$. This happens when the stable manifold of the periodic orbit which bifurcates from $\delta$ in $\Pi$ at $\mu_1=\mu_{1\Pi}$, meets the stable manifold of the origin (Section \ref{subsec:fix(O-)}). \\
Horizontal lines correspond to equilibria $\alpha_\pm$, $\beta_\pm$ and $\gamma$ in $P$, which do not depend on $\mu_1$. The $\alpha_-$ and $\beta_-$ equilibria do not appear in right diagrams (Fig. \ref{fig:bifdiagD}) in order to focus on the relevant bifurcations. The straight line with negative slope corresponds to the branch $\delta$ equilibrium in $\Pi$, which bifurcates from $\beta_+$ at $\tilde\mu_1$. It ends up at the origin for values of $\mu_1$ which are not shown here. The following list shows the typical sequence of bifurcations for equilibria and periodic orbits emerging off the invariant planes in $\Delta$.

\begin{itemize}
\item At $\mu_1=\mu_1'$ an equilibrium $\delta_\Delta$ bifurcates from $\alpha_+$ (pitchfork bifurcation). The bifurcated branch is subcritical and unstable. 
\item At $\mu_1=\mu_{h1}$ a supercritical Hopf bifurcation occurs along this branch, after which $\delta_\Delta$ becomes stable. Note that $\mu_{h1}$ is very close to $\mu_1'$ in all cases.
\item At $\mu_1=\mu_{sn}$ the branch $\delta_\Delta$ bends back and looses stability (saddle-node bifurcation).
\item This unstable branch crosses the branch of $\delta$ equilibria in the plane $\Pi$ at $\mu_1=\mu_t$ and becomes stable again (transcritical bifurcation). 
\item At $\mu_1=\mu_{h2}$, another Hopf bifurcation occurs on the $\delta_\Delta$ branch. We call $C_2$ these periodic orbits.
\item Finally the equilibria $\delta_\Delta$ merge with the $\beta_-$ or $\gamma$ branch (depending on whether $\mu_2\leq\mu_2^\beta$ or $\mu_2>\mu_2^\beta$), via a reverse pitchfork bifurcation.
\end{itemize}
Note that $\mu_{h2}$ and  $\mu_{h\Pi}$ have the same position relative to $\mu_t$.
\subsection{Numerical time integration in $V$}
We take on the bifurcation scenarios in $\Delta$, Fig. \ref{fig:bifdiagD}, to describe the dynamics that arise in $V$ as $\mu_1$ is varied in the corresponding range of values.

\subsubsection{The case $c=0.004$.}
\begin{figure}[h]
\begin{center}
\includegraphics[width=0.99\hsize]{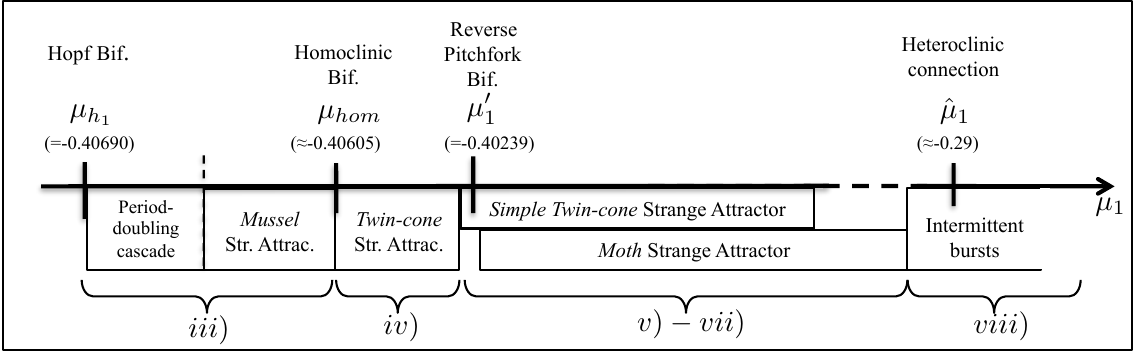}\\
\caption{Stable non-trivial attractors observed by simulating the bifurcation equations (\ref{eq:ode}) for increasing $\mu_1$ between $\mu_{h1}$ and $\hat{\mu}_1$ with $\mu_2=0.01$ and $c=0.004$. The roman numbers correspond to the description items in the text.}
\label{fig:scenarios}
\end{center}
\end{figure}

We fix $\mu_2=0.01$, a value larger than $\mu_2^\beta$. Similar behavior was found when $\mu_2=0.002\in]\mu_2^\alpha,\mu_2^\beta[$ or when $\mu_2=0.001<\mu_2^\alpha$, although in these cases an additional equilibrium $\gamma$ exists in $P$. In fact, in these ranges of parameter values, most of the dynamics in $V$ is driven by the dynamics in $\Delta$ or $Z$.\\
The simulations are performed by increasing $\mu_1$. By stability of an equilibrium in $V$ we mean that its $O(3)$ orbit is an attractor (orbital stability).
\ttfil{%
The attractors observed during the simulation are summarized in \myfig{scenarios}.
Let us detail the different scenarios and transitions.}%
\begin{enumerate}[label=\roman*)]
\item For $\mu_1<\tilde{\mu}_1$, time integration confirms that $\beta_+$ (i.e. its $O(3)$ orbit) is the unique sink in~$V$. 
\item As $\mu_1$ crosses $\tilde{\mu}_1$, bifurcated equilibria $\delta$ in $\Pi$ become the unique sinks untill $\mu_1$ reaches $\mu_{sn}=-0.45094$, value at which the equilibria of type $\delta_\Delta$ become stable too via the saddle node bifurcation in $\Delta$. This bistability persists until $\mu_1$ crosses $\mu_t$. Then $\delta_\Delta$ are the only stable equilibria.
\item At $\mu_1=\mu_{h1}$, $\delta_\Delta$ becomes unstable through Hopf bifurcation followed by a period doubling cascade in a narrow range (\myfig{scenarios}).
Then a strange attractor appears in the vicinity of $\alpha_+$ in $D$, called the 'mussel' because of its shape (\myfig{strat}).  
\begin{figure}[h]
\begin{center}
\includegraphics[height=5cm]{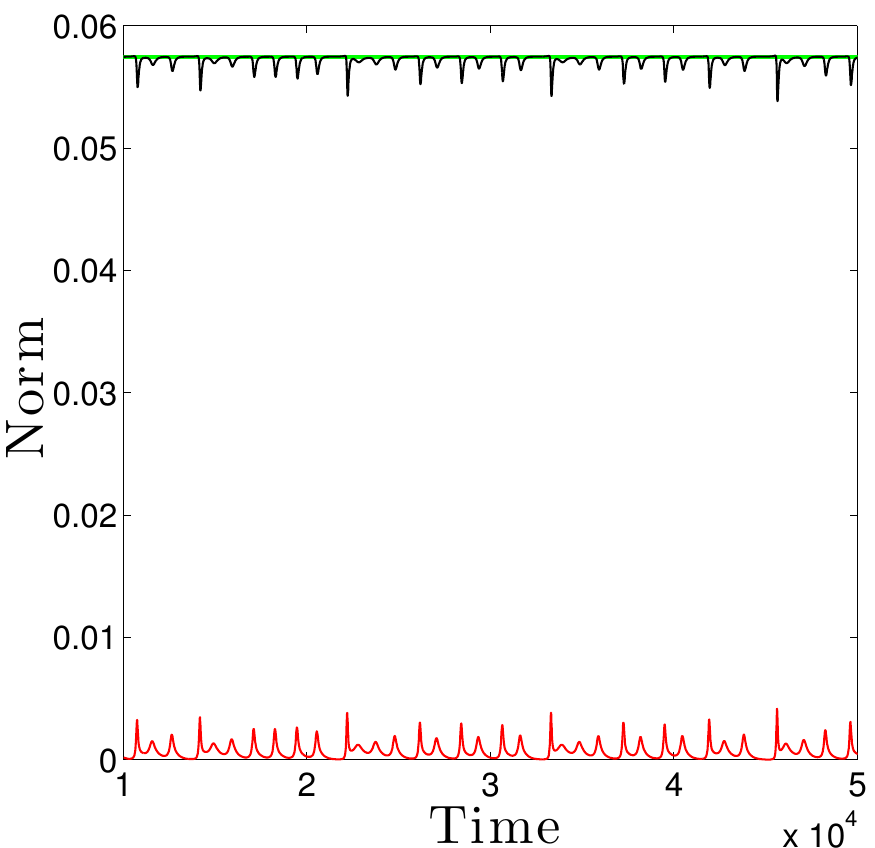}
\includegraphics[height=5cm]{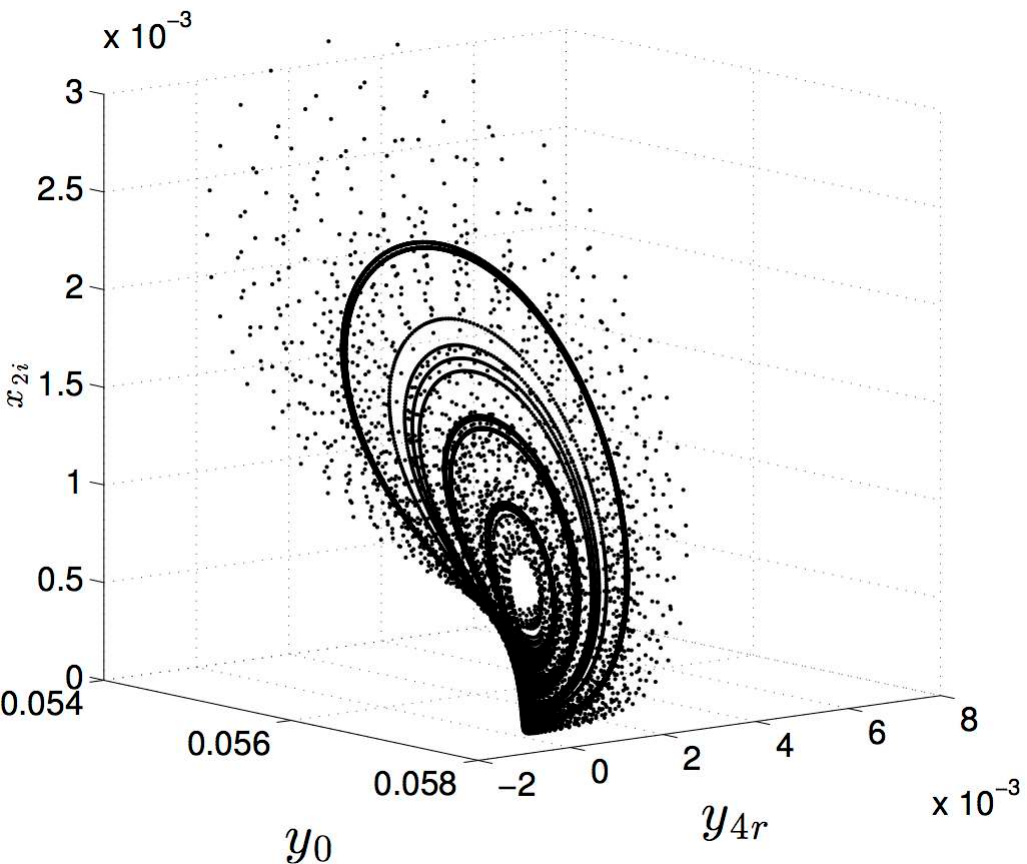}\\
(a)\hspace{4cm}(b)
\caption{The 'mussel' attractor at $\mu_1=-0.4061$, $\mu_2=0.01$, $c=0.004$. (a) Time series of 'energies' (norms) of the $\ell=3$ modes (red) and $4$ modes (black). Green line: norm of $\alpha_+$. (b) Phase portrait in $\Delta$.}
\label{fig:strat}
\end{center}
\end{figure}
\item The periodic orbit disappears at $\mu_{hom}<\mu_1'$ through a homoclinic bifurcation with $\alpha_+$.  Then a heteroclinic connection from $\delta_\Delta$ to $\alpha_+$ is established (\myfig{sketchD}).
\begin{figure}[h]
\begin{center}
\includegraphics[width=8cm]{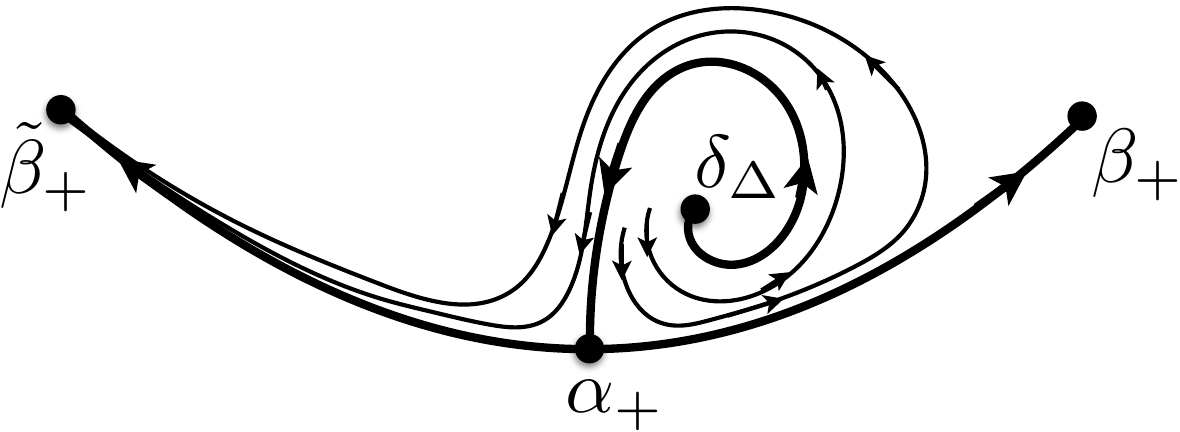}
\caption{Sketch of the phase portrait in $\Delta$ near $\alpha_+$ when $\mu_1$ slightly lower than $\mu_1'$.}
\label{fig:sketchD}
\end{center}
\end{figure}
Moreover  the connection $\delta$ to $\delta_\Delta$ is established at  $\mu_1=\mu_t<\mu_{hom}$. Thus in the range $]\mu_{hom},\mu_1'[$ the following sequence of heteroclinic orbits exists in $\Delta$: $\beta_+\rightarrow\delta\rightarrow\delta_\Delta\rightarrow \alpha_+\rightarrow\beta_+$ or $\tilde\beta_+$. But from $\tilde\beta_+$ a similar sequence of connections in the space $\tilde\Delta=\{x_{2r},y_0,y_{4r}\}$, which is the image of $\Delta$ by the 8-fold rotation around vertical axis. This heteroclinic cycle belongs to $Z$ for the reason given in Section \ref{simplehetcyc}. In fact this is a (non simple) variant of the heteroclinic cycle of type II described there. \\
This object is unstable.
\ttfil{
Instead, it is observed a strange attractor whose dynamics first skirts the connections in $P$ between $\alpha_+$ and $\beta_+$ or $\tilde{\beta}_+$ and  then approaches $\alpha_+$ with a spiraling trajectory (see phase portrait in  \myfig{strcone}). Then, a dynamics similar to the 'mussel' attractor takes place. After a while  the trajectory escapes from the $\alpha_+$ vicinity and skirts again the connection $\alpha_+$ to the cubic steady-state. The alternation between both kinds of dynamics (spiraling trajectory/mussel-like dynamics) arises in an irregular manner leading to an intermittent behavior. The resulting strange attractor is called the {\it twin-cone} attractor.
}%

\begin{figure}[h]
\begin{center}
\includegraphics[width=0.8\hsize,height=6cm]{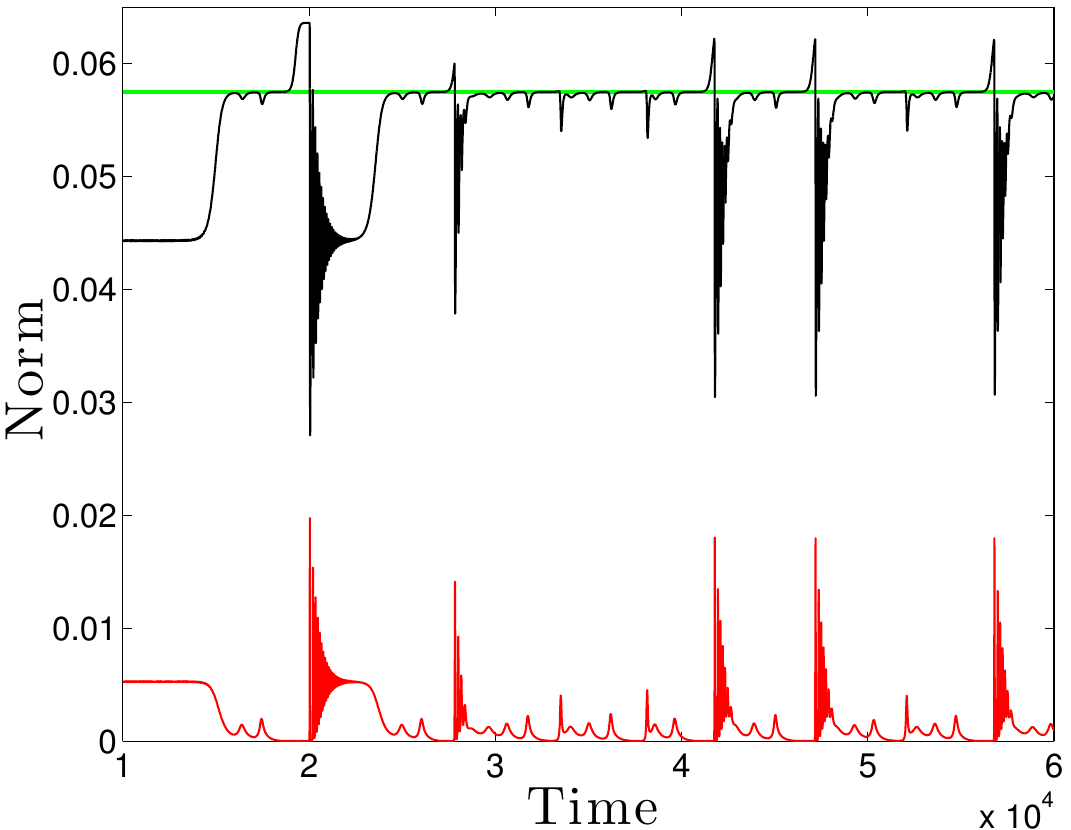}\\
\vspace{0.5cm}
\hspace{1cm}\includegraphics[width=0.4\hsize]{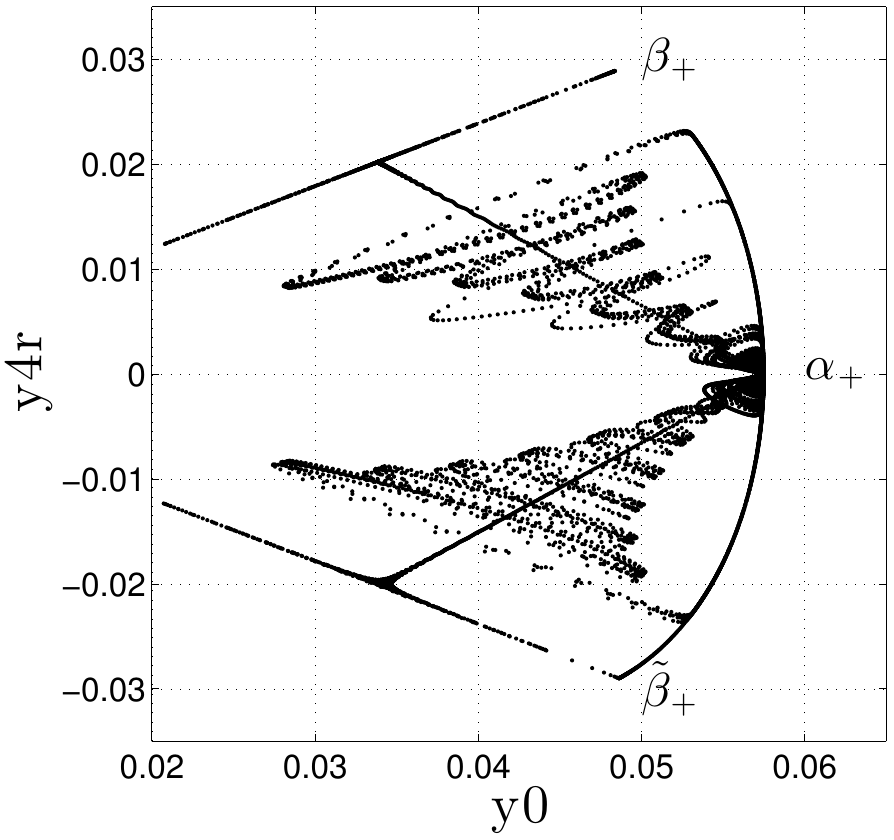}\hfill
\includegraphics[width=0.35\hsize]{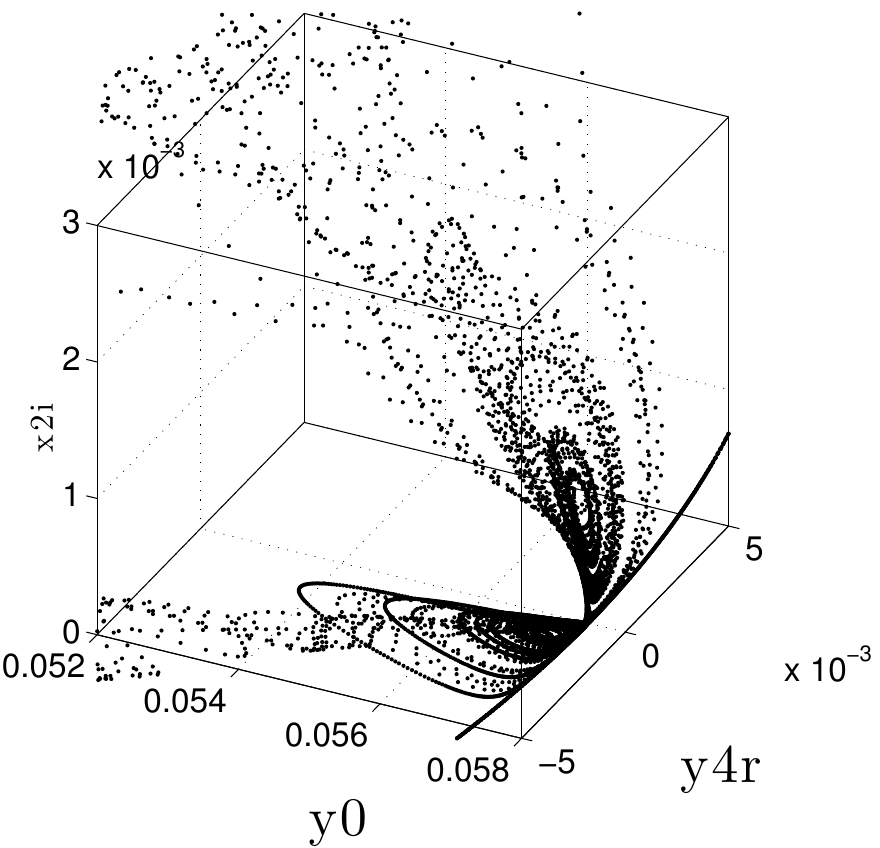}
\hspace{1cm}
\caption{Dynamics in $Z$ at $\mu_1=-0.40605$, $\mu_2=0.01$, $c=0.004$. Above: time series (same color code as in Fig. \ref{fig:strat}). Below left: projection of the phase portrait on $P$. The initial condition is close to the type II heteroclinic cycle. Skew line in the left panel correspond to the $C_\Pi$ periodic branch or its copy in $\tilde\Pi$. After one cycle the dynamics converges to the 'twin-cone' strange attractor. Below right: magnification near $\alpha_+$ displaying a dynamics similar to \myfig{strat}.}
\label{fig:strcone}
\end{center}
\end{figure}
The right panel in \myfig{strcone} shows that the dynamics near the mussel attractor is nearly contained in $\Delta$ or in $\tilde{\Delta}$:  the component $x_{2i}$ is almost zero when $y_{4r}<0$. This dynamics is  stable and it is an attractor in the whole space $V$.
%
\item 
\ttfil{Near $\mu_1'$, the twin-cone attractor becomes more regular and an almost periodic dynamics takes place (\myfig{strats}). We call this attractor the 'simple twin-cone' attractor. Moreover, another strange attractor emerges in the vicinity of $\alpha_+$ called 'moth' attractor. It has a similar structure as the 'simple twin-cone' attractor with two components  almost included in $\Delta$ or in its copy $\tilde{\Delta}$. The alternation between both components is  nearly periodic  (\myfig{strats}).}
\begin{figure}[h]
\begin{center}
\includegraphics[width=.8\hsize,height=5cm]{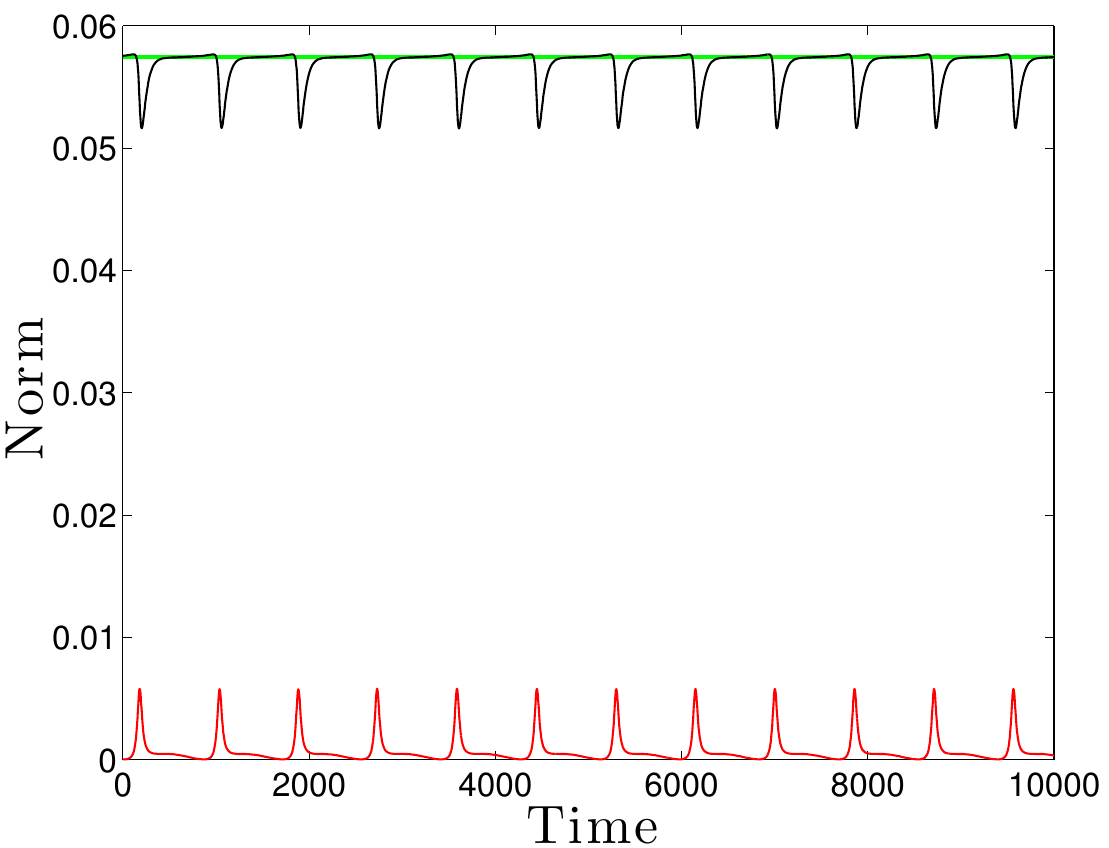}\\
\includegraphics[width=.8\hsize,height=5cm]{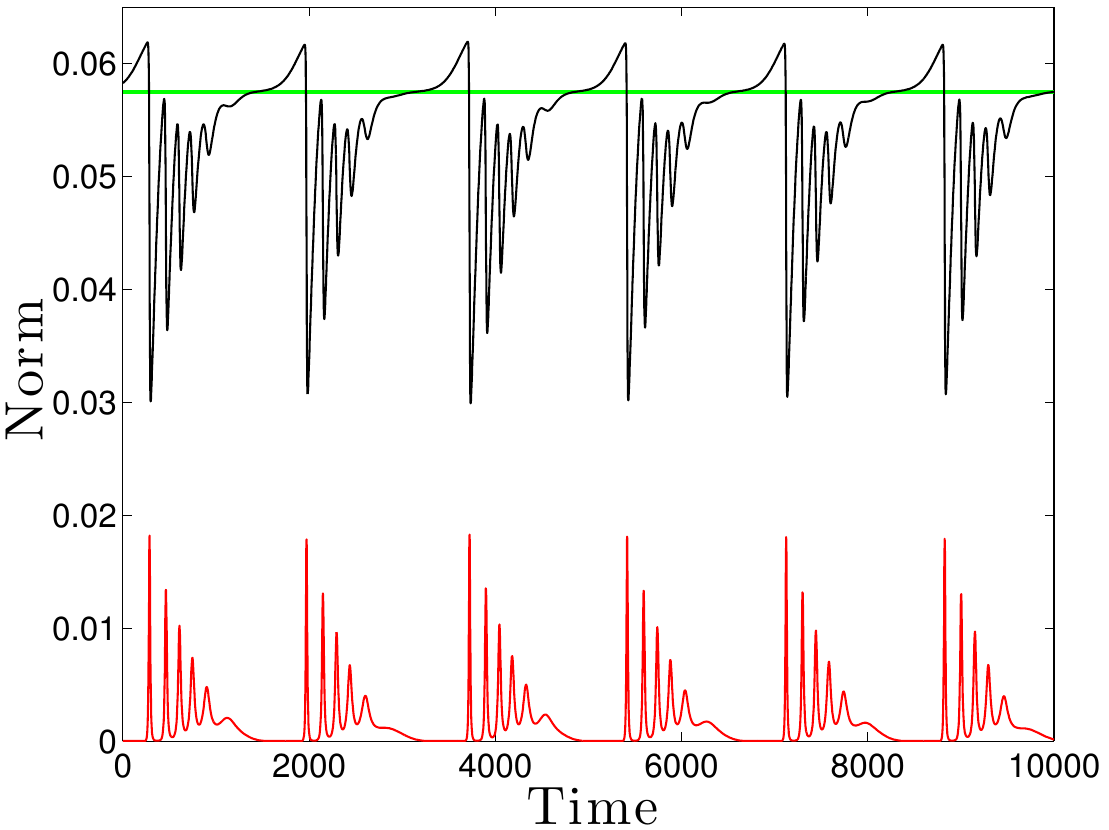}
\caption{Time series of the dynamics following the 'moth' (up) and the 'simple twin-cone' (down) close to $\alpha_+$ for $\mu_1=-0.4>\mu_1'$ and $\mu_2=0.01$, $c=0.004$ (same color code as in Fig. \ref{fig:strat}).}
\label{fig:strats}
\end{center}
\end{figure}
%
\item As $\mu_1$ increases further, the number of loops of the simple twin-cone attractor decreases (three at $\mu_1=-0.37$, two at $\mu_1=-0.36$) as $\mu_1$ increases further, till the attractor disappears at $\mu_1\sim -0.35$.
In contrast  the amplitude of the \ttfil{moth} attractor increases till the attractor boundary comes near the invariant planes $\Pi$ and its copy $\tilde{\Pi}$ in $\tilde\Delta$ (\myfig{nearhc}, left panel).
Indeed the dynamics passes near the limit cycle $C_\Pi$ (bifurcated from $\delta$ at $\mu_{h\Pi}=-0.3407$) or near its copy in $\tilde{D}$. 
\item At $\mu_1=-0.2967$ the moth attractor becomes asymmetric as shown in \myfig{nearhc} right panel.
\ttfil{The simulation in the whole space~$V$ instead of the subspace $Z$, displays a slightly modified attractor: The  amplitude is smaller and the period shorter.}
 Note that in this range of values of $\mu_1$, the limit cycles $C_2$ bifurcated from $\delta_\Delta$ at $\mu_{h2}=-0.2997$ exists but is unstable in $V$.
\begin{figure}[hb!]
\begin{center}
\includegraphics[height=6cm]{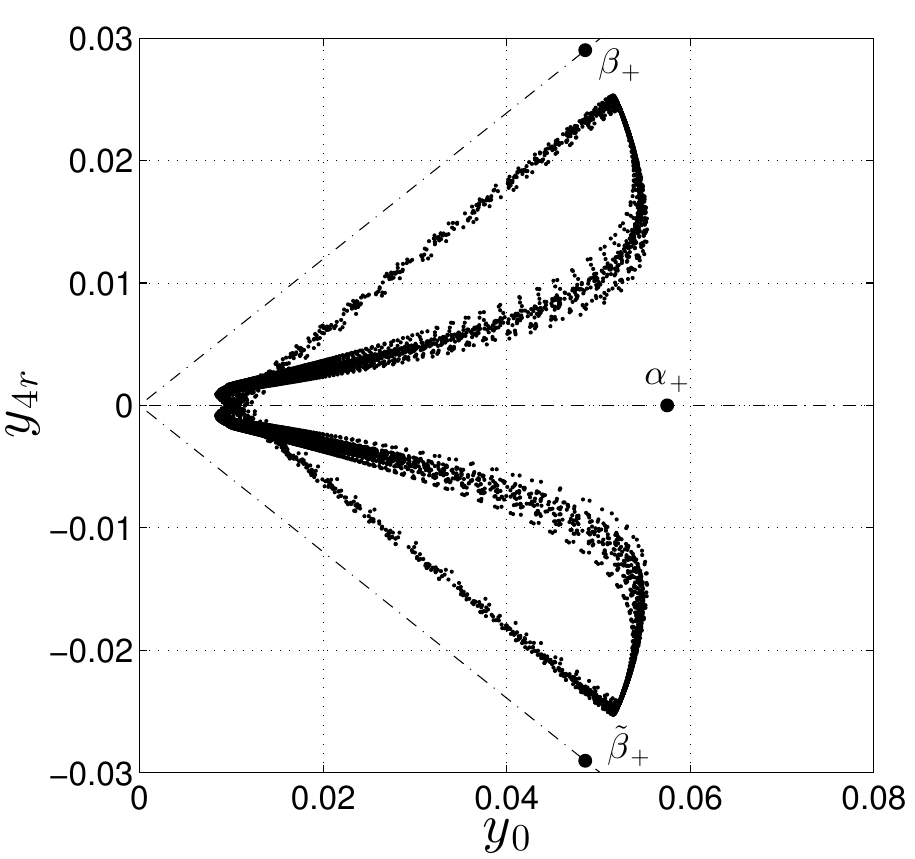}
\hfill
\includegraphics[height=6cm]{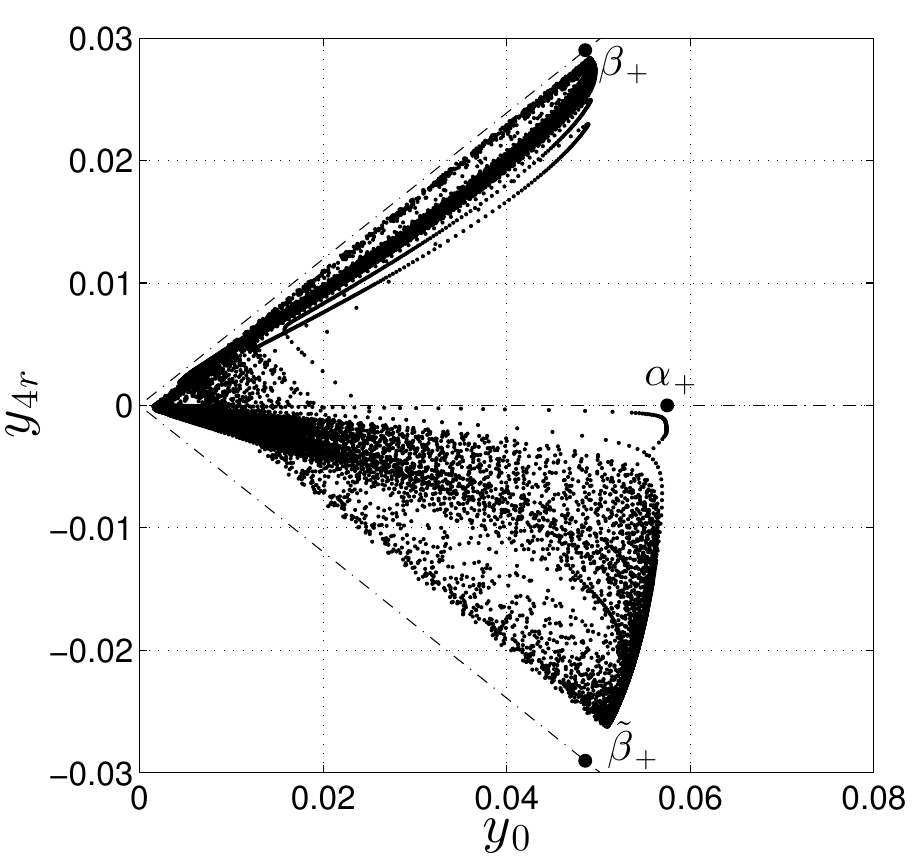}
\caption{Projection on $P$ of the phase portrait of the nearly periodic 'moth' strange attractor for (left) $\mu_1=-0.31$ and (right) $\mu_1=-0.295$. Other parameters:  $\mu_2=0.01$, $c=0.004$. Dot dashed lines indicate the intersections with planes $\Pi$ and $\tilde\Pi$.}
\label{fig:nearhc}
\end{center}
\end{figure}
%
 %
\begin{figure}[h]
\begin{center}
\includegraphics[height=5cm,width=10cm]{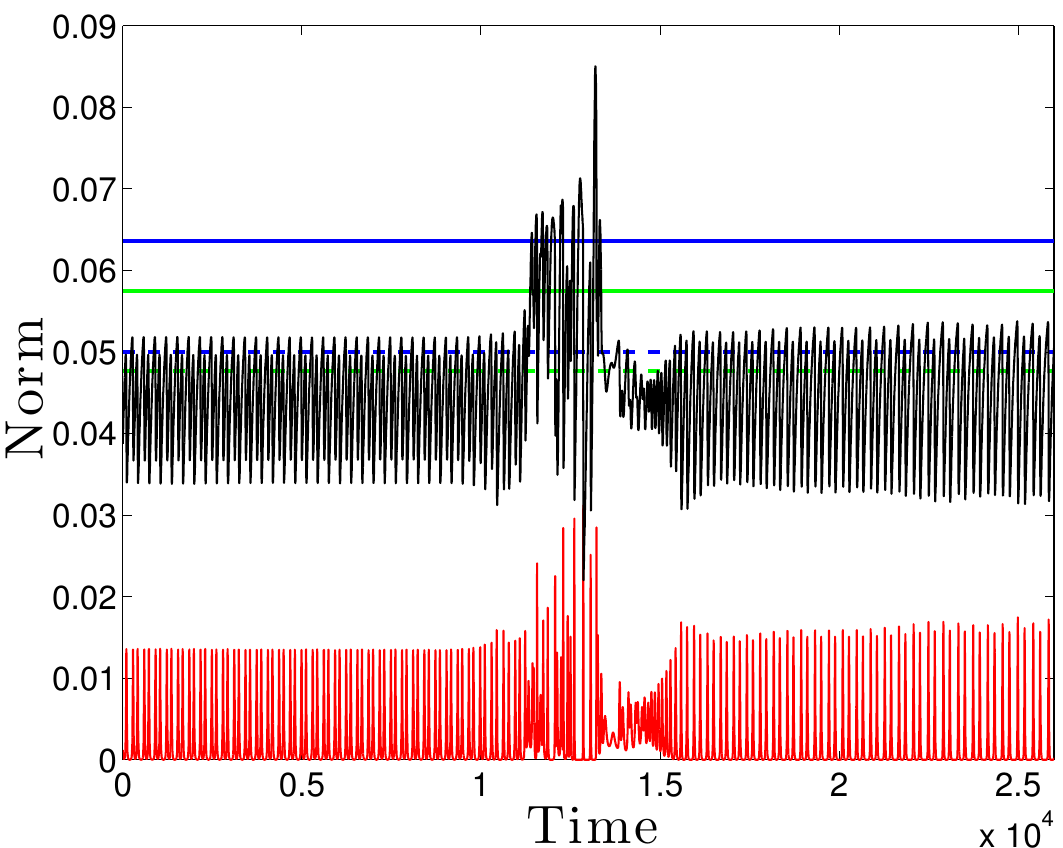}
\caption{Interspersed 2-periodic dynamics when $\mu_1=-0.28,\mu_2=0.01$, $c=0.004$. Same color code as in Fig. \ref{fig:strat}. Green line: norm of $\alpha_+$, blue line: norm of $\beta_+$, green dotted line: norm of $\alpha_-$, blue dotted line: norm of $\beta_-$.}
\label{fig:dpun}
\end{center}
\end{figure}
\begin{figure}[h]
\begin{center}
\includegraphics[width=0.8\hsize,height=5cm]{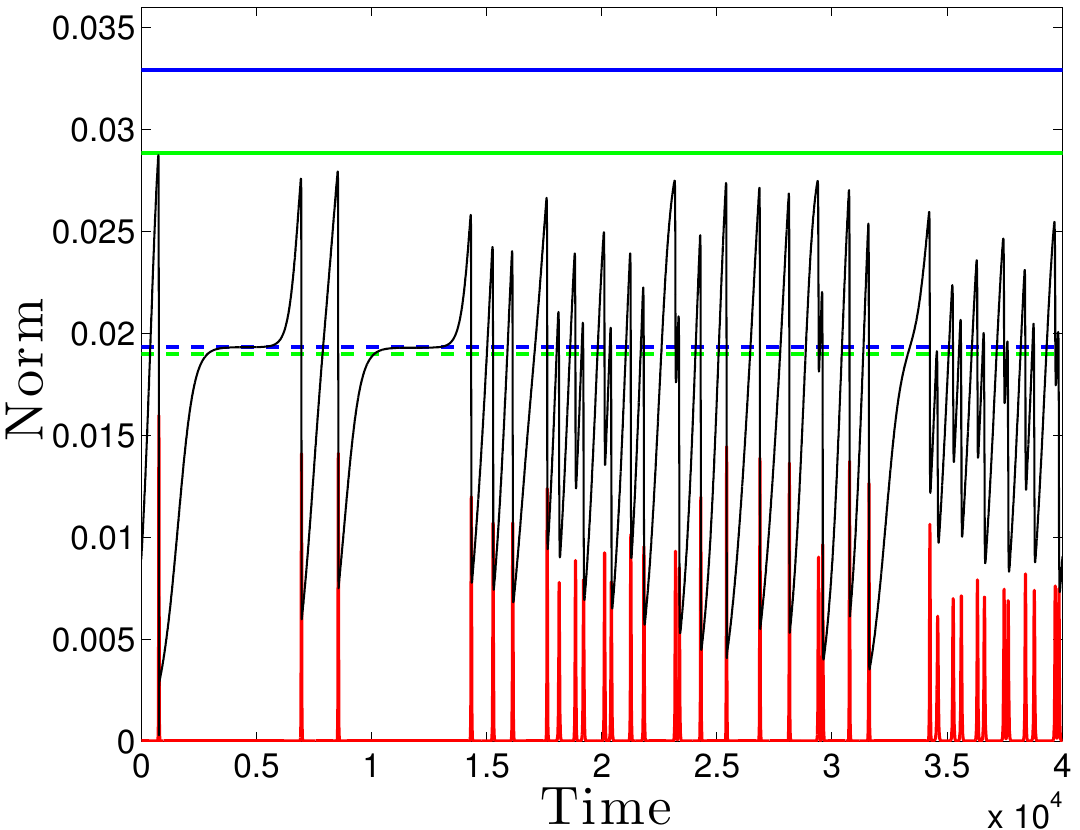}
\caption{Time series for $\mu_1=-0.12,\mu_2=0.002$,  $c=0.004$ (same color code as in Fig. \ref{fig:dpun}).  The plateaux near the energy of $\beta_-$ correspond to a part of the periodic cycle $C_2$ which passes near the equilibrium $\gamma$.}
\label{fig:hcun0,002}
\end{center}
\end{figure}
\item Close to the emergence of the heteroclinic connection $\beta_+$ to $\beta_-$ ($\hat{\mu}_1=-0.29$), the previous strange attractor loses its stability and the dynamics restricted to $Z$ tends to $\beta_-$ or its copy $\tilde\beta_-$. In contrast in~$V$ the dynamics is time-dependent. This dynamics is  similar to the previous one during 100 to 1000 periods but  interspersions by chaotic events with large bursts arise (\myfig{dpun}).  
\item When $\mu_1>\hat{\mu}_1$ the generalized heteroclinic cycle is established. However when $\mu_2=0.01$ we do not observe intermittent behavior related to the cycle, rather periodic or aperiodic fluctuations around either the octahedral pattern $\beta_+$ or the mixed mode, "tetrahedral-like" pattern $\delta$. Nevertheless the simulation for $\mu_2=0.002$ and $\mu_1$ close to $\hat{\mu}_1$ shows a dynamics which roughly follows parts of the generalized heteroclinic cycle, see (\myfig{hcun0,002}). In Annex \ref{ap:ev} the eigenvalues giving the rates of expansion and contraction along the heteroclinic connections are displayed in this case. It shows that the overall contraction rate (product of the absolute value of negative eigenvalues) is larger than the expansion one, which one would expect to give asymptotic stability. However the numerical simulations indicate this is not the case.
\item As $\mu_1$ gets larger than $min(\mu_1'',\mu_{\beta_-}^{\sigma_1},\mu_{\beta_+}^{\sigma_2})$, additional instabilities lead to more complex and diffuse dynamics:  small amplitude dynamics are interspersed by bursts whose occurrence and amplitude increase with $\mu_1$ (see \myfig{chaotic}).
\begin{figure}[h!]
\begin{center}
\includegraphics[width=0.8\hsize, height=6cm]{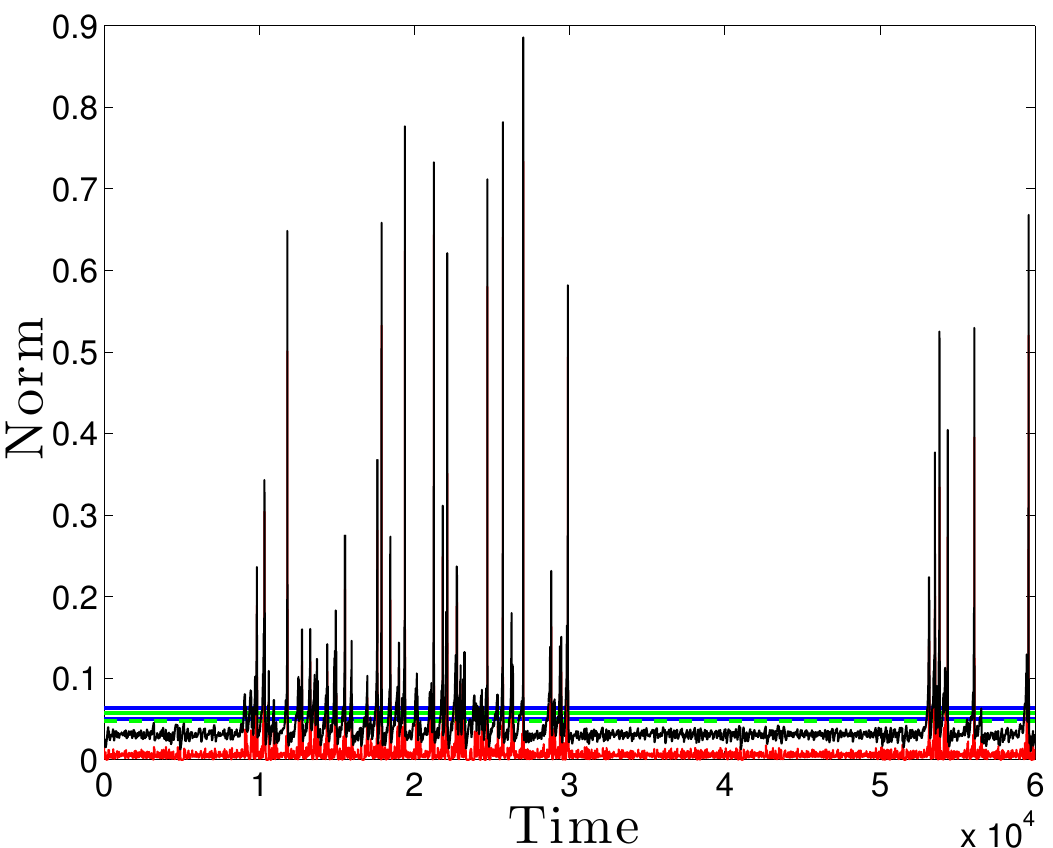}
\caption{Time series for $\mu_1=-0.15,\mu_2=0.01$,  $c=0.004$ (same color code as in Fig. \ref{fig:dpun}).}
\label{fig:chaotic}
\end{center}
\end{figure}
\end{enumerate}
{\it In summary:} the predicted heteroclinic cycles are unstable. Intermittent-like dynamics related to the presence of heteroclinic cycles are observed in narrow ranges of $\mu_1$ (trajectories come near the axisymmetric state $\alpha_+$ in this case). When $\mu_2$ is small enough and $\mu_1>\hat\mu_1$ (close to $\hat\mu_1$) the trajectories explore the vicinity of the octahedral patterns $\beta_+$ and $\beta_-$ in an irregular manner but don't show a clear intermittent behavior.  
%
\subsubsection{The case $c=0.04$}

The values of $\mu_2$ that we chose for the simulations ($\mu_2=0.01$ and $0.1$) are smaller than $\mu_2^\alpha$, in order to keep the study within the range of validity of the center manifold approximation, see \myfig{stabdiag1}. 
Then we observe that the phase portrait in the pure mode subspace $V^4$ corresponds to Fig. \ref{connexionsD2Z2c} (left): the ($O(3)$-orbit of) $\beta_+$ is the unique attractor in $V^4$. \\
Even though the bifurcation diagram close to $\alpha_+$ presents similarity with the case $c=0.004$, the time integration does not show the same attractors. The dynamics converges to either the equilibrium $\delta$ or the periodic orbit $C_\Pi$ in $\Pi$ plane when $\mu_1<\hat\mu_1$. We now list our observations as $\mu_1$ is increased.
\begin{enumerate}[label=\roman*)]
\item When $\mu_1$ crosses $\hat\mu_1$, generalized heteroclinic cycles are established. The simulation displays a dynamics which is compatible with cycle I(i) of Theorem \ref{thm:generalized hetcycles}: trajectories visit the vicinity of equilibria $\beta_+$ and $\beta_-$ although $\beta_-$ may be missed during several cycles, figures \ref{fig:hcpom20,1} and \ref{fig:hcpom20,01}. The axisymmetric steady-state $\alpha_-$ is also visited occasionally. The fact that equilibria of type $\beta_-$ are "missed" on a recurrent basis indicates that the heteroclinic cycle itself is not asymptotically stable, but rather that an attractor exists in a neighborhood of it. We suggest the following interpretation: when $\mu_1$ is close to $\hat\mu_1$ the connection $\beta_+\rightarrow\beta_-$ in $\Pi$ is strongly attracted toward the axis $L=Fix(\OB\oplus\Z_2^c)$ before it joins $\beta_-$. Indeed this is close to the 'homoclinization' of the periodic orbit $C_\Pi$ on the origin and $\beta_+$. Nearby trajectories may then follow the stable manifold of $\beta_-$ in $\Pi$, but from the neighborhood of 0 it may instead follow an orbit along the stable manifold of $\tilde\beta_+$ or $\alpha_-$, which are both sinks in $P$. When the trajectory joins directly $\beta_+$, this dynamics follows the periodic cycle $C_2$ which exist (but is unstable) for a slightly larger value than $\hat{\mu}_1$ as we saw in Section \ref{difdiagDelta}. \\
\item After $\mu_1$ is slightly increased this cycle vanishes through a heteroclinization on the equilibria $\beta_+$, $0$ and $\alpha_+$. Then the time series show the presence of two plateaux at equilibria of types $\beta_-$ and $\beta_+$, at each cycle (Figs. \ref{fig:hcm20,1} and \ref{fig:hcm20,01}). Note that for $\mu_2=0.01$, the plateau near $\beta_+$ is  not really visible (\myfig{hcm20,01}), it means that the trajectory does not spend long time close to the $\beta_+$ equilibria. It is a consequence of the fact that $|\mu_2|\ll|\mu_1|$ so that $\beta_+$ has small contracting eigenvalues (order of $\mu_2$) compared of its unstable direction (order of $|\mu_1|$). When $\mu_2$ is larger, equal to $0.1$, in contrast both plateaux have a similar duration (\myfig{hcm20,1}). \\
\item Even though the generalized heteroclinic cycle may not be stable, the simulations show that a nearby attractor exists. However, by increasing further $\mu_1$, the heteroclinic cycle is gradually more unstable: one or both equilibria are not reached anymore. Thus, for a value of $\mu_1$ which is still smaller than $min(\mu_1'',\mu_{\beta_-}^{\sigma_1},\mu_{\beta_+}^{\sigma_2})$ the dynamics is no more a heteroclinic cycle. Nevertheless, this attractor gets the main characteristic of the $\beta$-heteroclinic cycle: the trajectory spends a long time in $V^4$ following by a short transition in the mixed mode space $D$ (\myfig{hcdm20,1}).\\
\item For larger values of $\mu_1$ we observe isolated burst of large amplitudes as in the case $c=0.004$ (\myfig{hcm1-0,5m20,1}). The frequency of burst occurrence and their amplitude increases when $\mu_1$ approaching zero. Moreover the component of mode 3 does not vanish anymore.\par
\end{enumerate}
{\it In summary:}  type I generalized heteroclinic cycles involving octahedral and axisymmetric equilibria exist in the parameter range predicted in Section \ref{section_genhetcyc}. Numerical simulations show that the dynamics closely follow these invariant sets when $\mu_1$ is larger than but close enough to $\hat\mu_1$. However, as we already noticed in the case $c=0.004$, it does not converge to it. This indicates that the generalized heteroclinic cycles are not  attractors, although the rates of contraction and expansion of the eigenvalues along the heteroclinic connections would suggest they are, see Annex  \ref{ap:ev}. 
\par
\begin{figure}[h]
\begin{center}
\includegraphics[width=0.8\hsize,height=5cm]{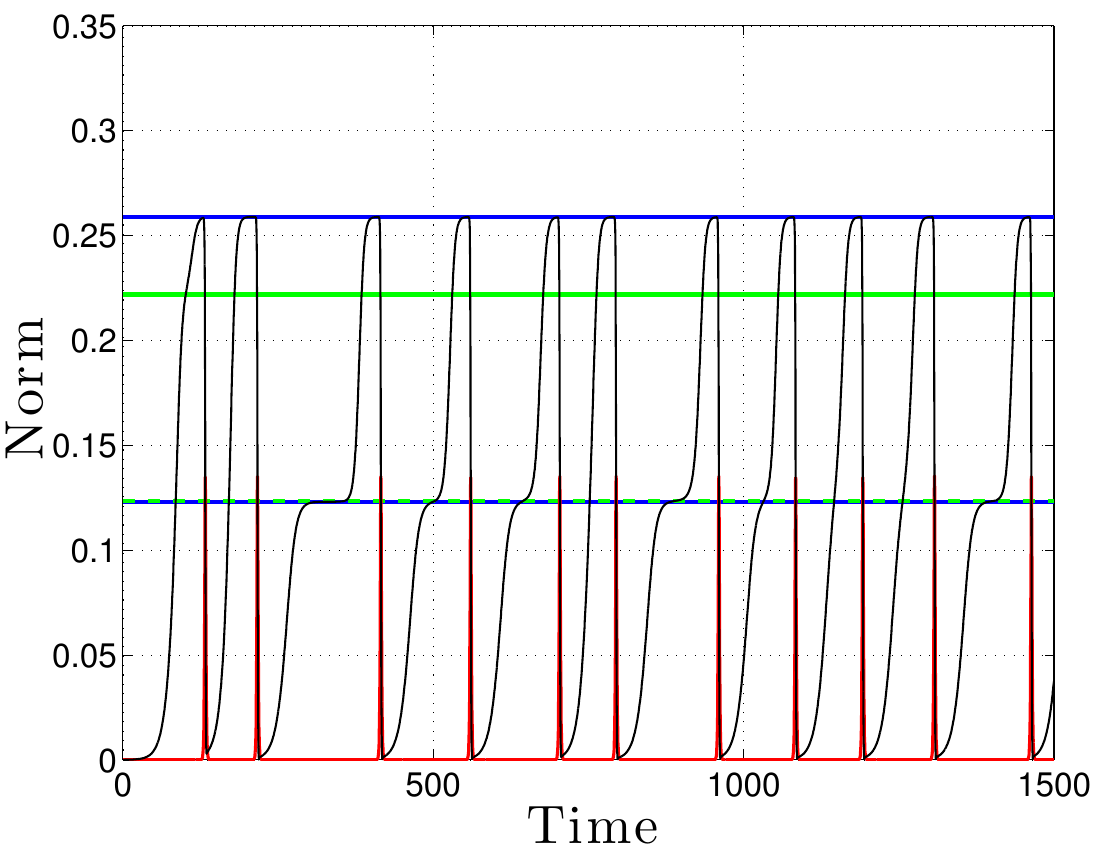}\\
\caption{Time series for $\mu_1=-1.17,\mu_2=0.1$, $c=0.04$. Same color code as in Fig.  \ref{fig:dpun}.}
\label{fig:hcpom20,1}
\end{center}
\end{figure}
\begin{figure}[h]
\begin{center}
\includegraphics[width=0.8\hsize,height=5cm]{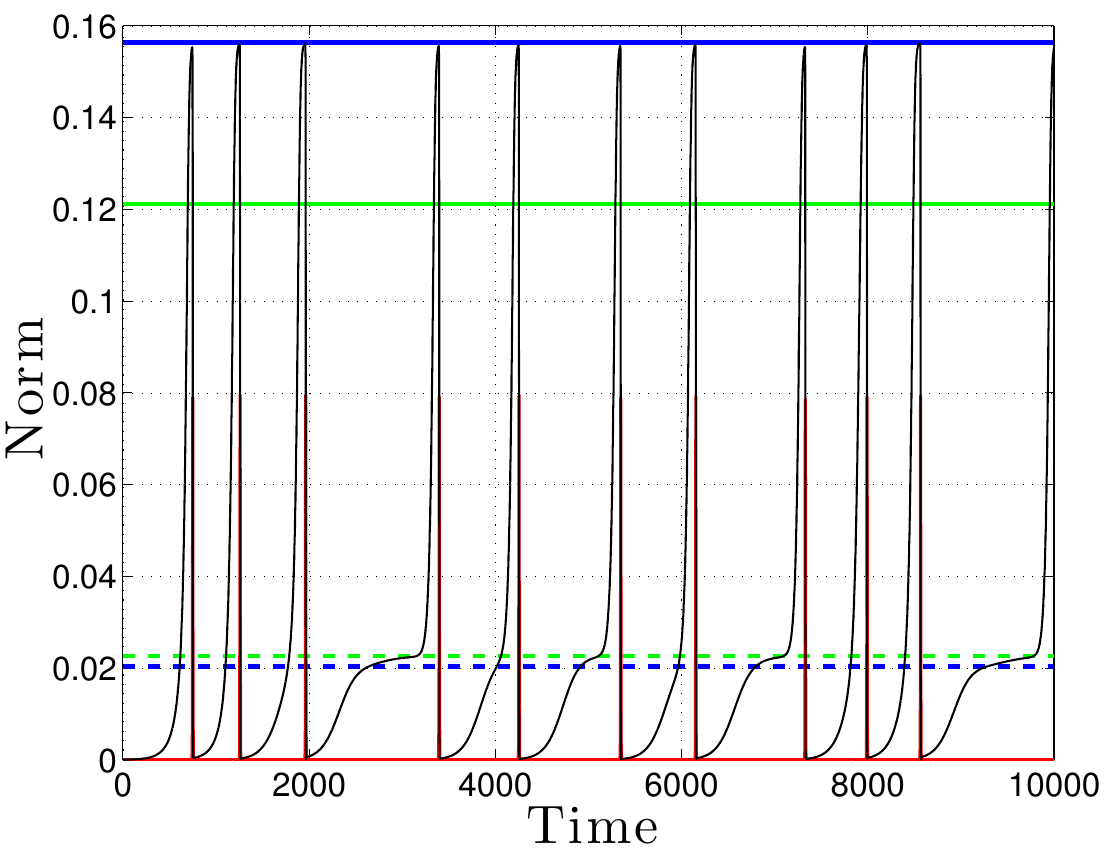}\\
\caption{Time series for $\mu_1=-0.712,\mu_2=0.01$, $c=0.04$. Same color code as in Fig.  \ref{fig:dpun}.}
\label{fig:hcpom20,01}
\end{center}
\end{figure}
\begin{figure}[h!]
\begin{center}
\includegraphics[width=0.8\hsize,height=5cm]{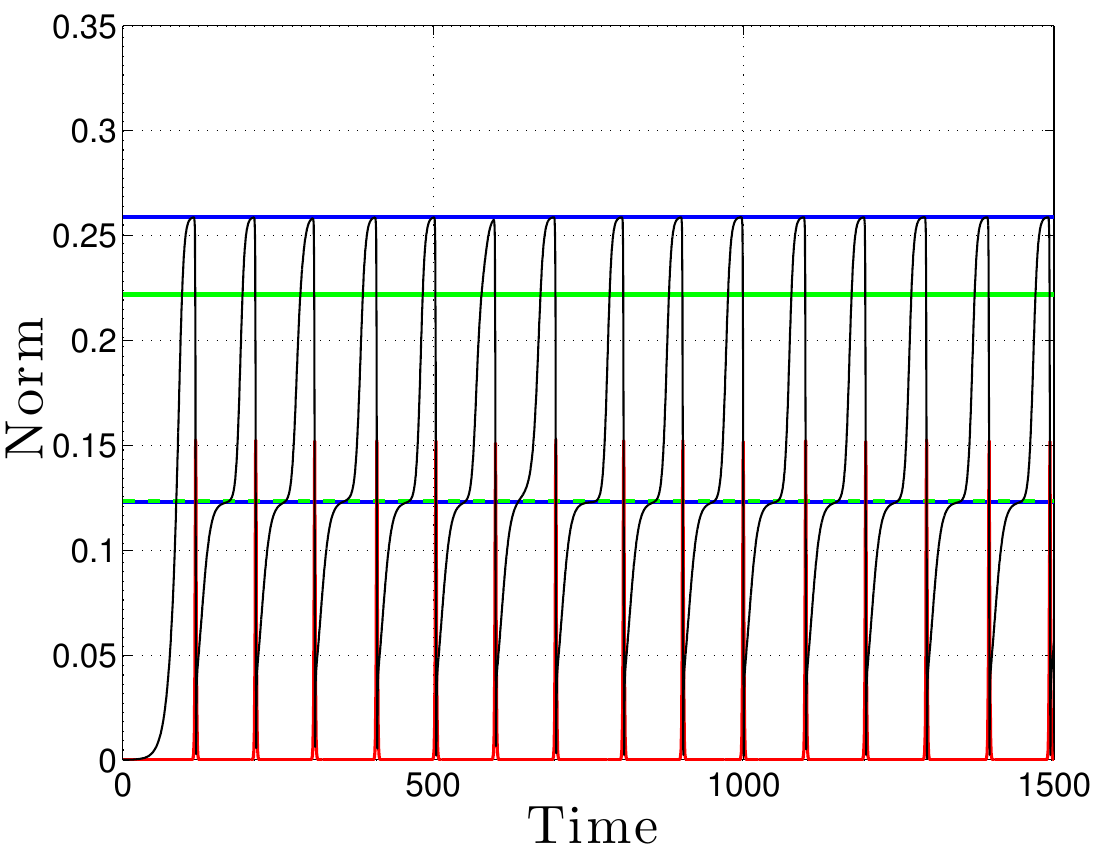}\\
\caption{Time series for $\mu_1=-1,\mu_2=0.1$, $c=0.04$. Same color code as in Fig.  \ref{fig:dpun}.}
\label{fig:hcm20,1}
\end{center}
\end{figure}
\begin{figure}[h]
\begin{center}
\includegraphics[width=0.8\hsize,height=5cm]{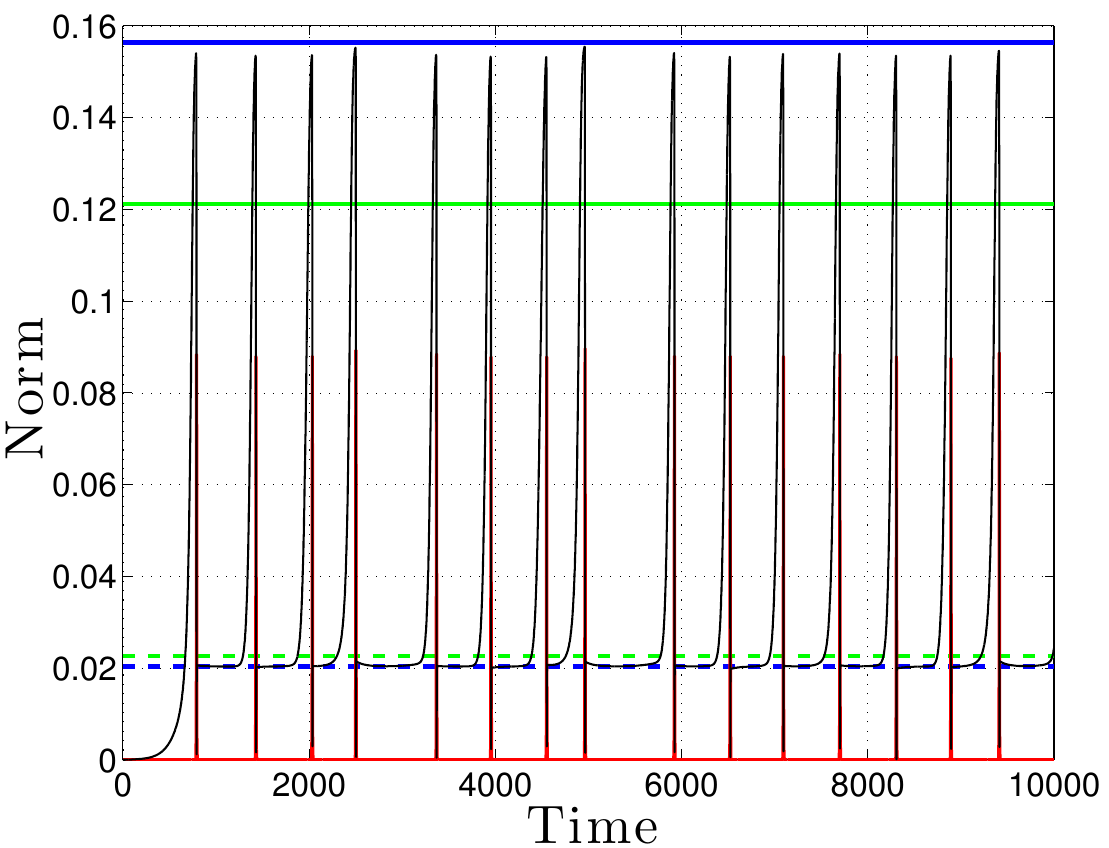}\\
\caption{Time series for $\mu_1=-0.61,\mu_2=0.01$, $c=0.04$. Same color code as in Fig.  \ref{fig:dpun}.}
\label{fig:hcm20,01}
\end{center}
\end{figure}
\begin{figure}[h]
\begin{center}
\includegraphics[width=0.8\hsize,height=5cm]{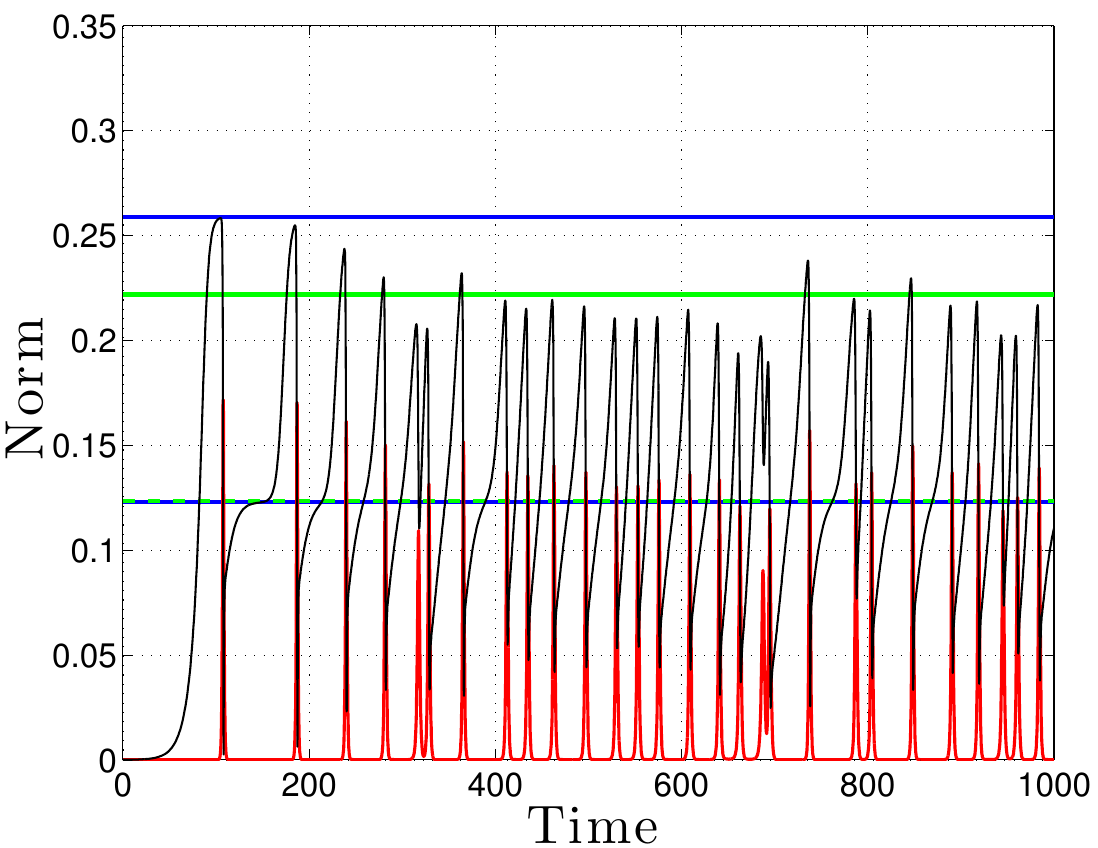}\\
\caption{Time series for$\mu_1=-0.8,\mu_2=0.1$, $c=0.04$. Same color code as in Fig.  \ref{fig:dpun}.}
\label{fig:hcdm20,1}
\end{center}
\end{figure}
\begin{figure}[h]
\begin{center}
\includegraphics[width=0.8\hsize,height=5cm]{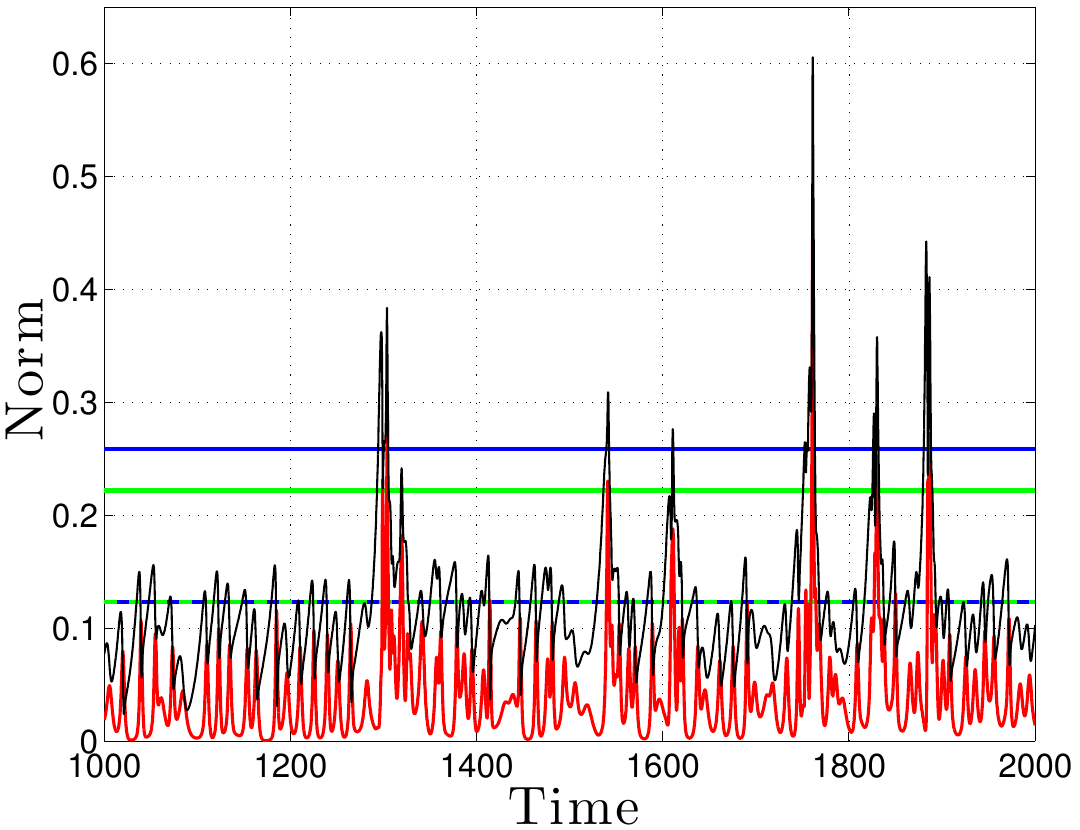}\\
\caption{Time series for $\mu_1=-0.5,\mu_2=0.1$, $c=0.04$. Same color code as in Fig.  \ref{fig:dpun}.}
\label{fig:hcm1-0,5m20,1}
\end{center}
\end{figure}
%

\section{Conclusion}
Intermittent dynamics near onset of convection has been analyzed for Rayleigh-B\'enard type problems in a spherical shell.
Here intermittency is closely linked to the existence of robust heteroclinic connections between bifurcated steady-states, robustness being understood as persistence against perturbative terms in the equations, which keep the spherical symmetry of the problem. When the heteroclinic connections close up in a cycle, nearby dynamics exhibit intermittent behavior with long periods of time passed near the steady-states followed by quick jumps to the next steady-state. A necessary condition for the occurrence of these heteroclinic cycles in problems with spherical symmetry is that spherical modes with degrees $\ell$ and $\ell+1$ compete for destabilization of the initial state of rest. Here we have analyzed the case $\ell=3$, corresponding to an aspect ratio of the order of 0.45. 

We have shown that robust "generalized" heteroclinic cycles do indeed exist in certain range of parameters for this problem. These invariant sets connect steady-states with octahedral patterns, and secondary connections do also exist with axisymmetric patterns. These steady-states are pure $4$-modes. In certain cases, connections with mixed-mode steady-states do also exist.  

We have not rigorously studied the asymptotic stability of the generalized heteroclinic cycles. A complete analysis was performed earlier in the case $\ell=1$, which was already quite involved \cite{ChGuLa99}. Instead, we performed a numerical exploration of the dynamics on the center manifold, for parameter ranges where the heteroclinic cycles are expected to exist near onset of convection. \\
We did not find clearly stable heteroclinic cycles for the cubic approximation of the system reduced to the center manifold. Nevertheless we have numerically shown that dynamics mimicking heteroclinic cycles do indeed exist in small ranges of parameter values. It shows relatively long periods of quasi-static regime with octahedral or axisymmetric patterns, followed by fast switches to other quasi-steady states of the same kind.
Some of these dynamics correspond to trajectories which clearly follow generalized heteroclinic cycles, which therefore are "nearly stable". Others are less clearly driven by the heteroclinic cycles, but nevertheless occur through scenarios which are related to the "heteroclinic" mechanism.

\ttfilr{It should be noted that while the steady-states belong to the invariant subspace $V^4$ and heteroclinic connections exist within this space, the "fast" connections which complete the cycle involve $\ell=3$ modes. Moreover, in the range of parameter values which we have considered, even though the heteroclinic cycles are not stable, a chaotic dynamics is observable, which shows a bursting behavior along the "transverse" modes with $\ell=3$. This is reminiscent of the "in-out" intermittency discussed in \cite{Covasetal}. Further analysis in this direction would be an interesting follow-up of our study.}

The asymptotic stability of these generalized heteroclinic cycles is a pending question. By suitably choosing coefficient values of the normal form (\ref{eq:ode}), which however are not compatible with the physical problem of onset of convection, conditions can be found at which it is numerically observed that the dynamics seemingly converge to generalized heteroclinic cycles. Three videos of the time evolution of such patterns are posted in \url{http://math.unice.fr/~chossat}, corresponding respectively to generalized heteroclinic cycles of type I(i), I(iii) and II. In these cases the eigenvalues at each steady-state in the cycle satisfy strong contraction versus weak expansion. However a rigorous analysis of these cases has not yet been undertaken.

Another question is the range of validity of this analysis. It is probably quite narrow. Direct simulations would help to evaluate this range. \\
Finally, let us mention the problem of adding a small rotation to the domain on this dynamics, which breaks spherical symmetry and can also be analyzed using center manifold reduction.

\newpage
\bibliographystyle{plain}

\clearpage

\appendix

\section{The lattice of isotropies for the $3, 4$ mode interaction} \label{annex:lattice}
Numbers on the left indicate the dimension of corresponding fixed point subspaces. 
\begin{center}
\includegraphics[width=0.95\hsize]{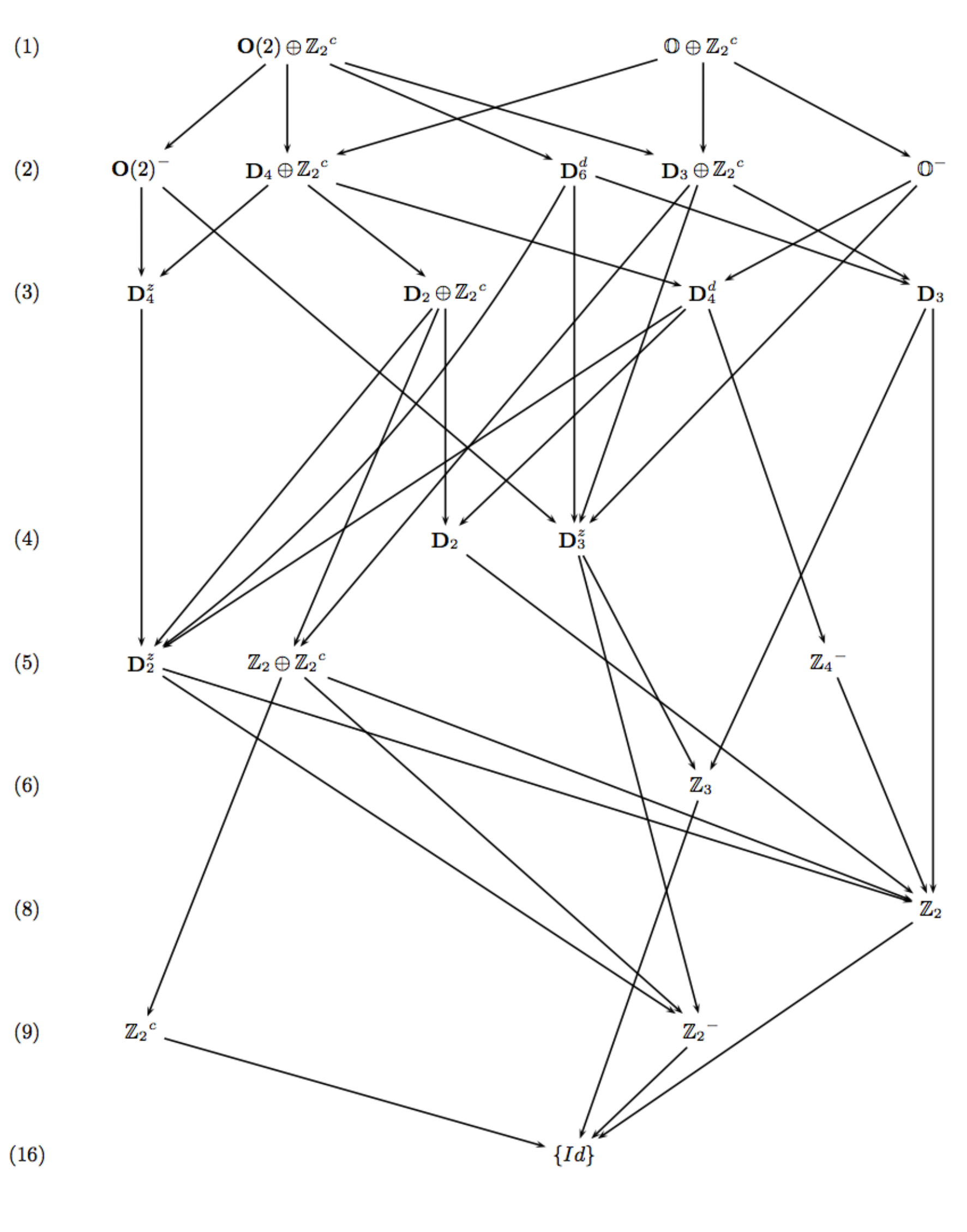}
\end{center}

\label{isotropytypes}

\clearpage

\section{The quadratic and cubic equivariant terms in (\ref{eq:ode})} \label{app:equivariants}
In the tables below we display the coefficients of quadratic and cubic terms in equations (\ref{eq_l=3}) and (\ref{eq_l=4}). Conventions are as follows: Number $m$ in the "Equation" column indicates the index of the component along the spherical harmonic $Y_\ell^m$, with $\ell=3$ or $4$ depending on the map under consideration. We list only the coefficients for the components with $m=0,\cdots, +\ell$, since the components for negative $m$ are obtained from the ones with positive $m$ according to the rule $Y_\ell^{-m}=(-1)^m\overline Y_\ell^m$. The numbers in parenthesis in the column "terms" indicate the indices of the corresponding terms with the following convention: $(i,j)$ for the quadratic terms $x_iy_j$ ($-3\leq i\leq 3$, $-4\leq j\leq 4$), $(ij)$ for the terms $x_ix_j$ ($-3\leq i\leq j\leq 3$) or $y_iy_j$ ($-4\leq i\leq j\leq 4$), and $(ijk)$ for the cubic terms $y_iy_jy_k$ ($-4\leq i\leq j\leq k\leq 4$).

\begin{table}[ht] \footnotesize
\centering
\begin{tabular}{|c|c|c||c|c|c|}
\hline
Equation & Term & Coefficient & Equation & Term & Coefficient \\ \hline
$3$  & $(0,3)$ & $-3\sqrt{7}$ & $2$ & $(-2,4)$ & $\sqrt{70}$ \\ 
    & $(-1,4)$ & $\sqrt{42}$ & & $(-1,3)$ & $-\sqrt{14}$ \\
    & $(1,2)$ & $3\sqrt{6}$ & & $(0,2)$ & $-\sqrt{3}$ \\
    & $(2,1)$ & $-\sqrt{30}$ & & $(1,1)$ & $4\sqrt{2}$ \\
    & $(3,0)$ & $3$ & & $(2,0)$ & $-7$ \\
    & & & & $(3,-1)$ & $\sqrt{30}$ \\
\hline
$1$ & $(-3,4)$ & $\sqrt{42}$ & $0$ & $(-3,3)$ & $\sqrt{7}$ \\
    & $(-2,3)$ & $\sqrt{14}$ & & $(-2,2)$ & $-\sqrt{3}$ \\
    & $(-1,2)$ & $-2\sqrt{10}$ & & $(-1,1)$ & $-\sqrt{15}$ \\
    & $(0,1)$ & $\sqrt{15}$ & & $(0,0)$ & $6$ \\
    & $(1,0)$ & $1$ & & $(1,-1)$ & $-\sqrt{15}$ \\
    & $(2,-1)$ & $-4\sqrt{2}$ & & $(2,-2)$ & $-\sqrt{3}$ \\
    & $(3,-2)$ & $3\sqrt{6}$ & & $(3,-3)$ & $\sqrt{7}$ \\
\hline
\end{tabular}
\caption{\small Coefficients for the quadratic "mixed" $\ell=3$ map $Q^{(1)}$}
\label{mapQ^(1)}
\end{table}

\begin{table}[h] \footnotesize
\centering
\begin{tabular}{|c|c|c||c|c|c|}
\hline
Equation & Term & Coefficient & Equation & Term & Coefficient \\ \hline
$4$  & $(13)$ & $-\sqrt{42}$ & $3$ & $(03)$ & $3\sqrt{7}$ \\ 
    & $(22)$ & $\sqrt{70}/2$ & & $(12)$ & $\sqrt{14}$ \\
\hline
$2$ & $(02)$ & $-\sqrt{3}$ & $1$ & $(01)$ & $\sqrt{15}$ \\
    & $(11)$ & $\sqrt{10}$ & & $(-12)$ & $-4\sqrt{2}$ \\
    & $(-13)$ & $-6\sqrt{3}$ & & $(-23)$ & $-\sqrt{30}$ \\
\hline
$0$ & $(00)$ & $3$ \\
    & $(-11)$ & $-1$ \\
    & $(-22)$ & $-7$ \\
    & $(-33)$ & $-3$ \\ 
\cline{1-3}
\end{tabular}
\caption{\small Coefficients for the quadratic "mixed" $\ell=4$ map $Q^{(2)}$}
\label{mapQ^(2)}
\end{table}

\begin{table}[h] \footnotesize
\centering
\begin{tabular}{|c|c|c||c|c|c|}
\hline
Equation & Term & Coefficient & Equation & Term & Coefficient \\ \hline
$4$  & $(04)$ & $14$ & $3$ & $(-14)$ & $7\sqrt{10}$ \\ 
    & $(13)$ & $-7\sqrt{10}$ & & $(03)$ & $-21$ \\
    & $(22)$ & $3\sqrt{70}/2$ & & $(12)$ & $\sqrt{70}$ \\
\hline
$2$ & $(-24)$ & $3\sqrt{70}$ & $1$ & $(-34)$ & $7\sqrt{10}$ \\
    & $(-13)$ & $-\sqrt{70}$ & & $(-23)$ & $\sqrt{70}$ \\
    & $(02)$ & $-11$ & & $(-12)$ & $-6\sqrt{10}$ \\
    & $(11)$ & $3\sqrt{10}$ & & $(01)$ & $9$ \\ 
\hline
$0$ & $(00)$ & $9$ \\
    & $(-11)$ & $-9$ \\
    & $(-22)$ & $-11$ \\
    & $(-33)$ & $21$ \\ 
    & $(-44)$ & $14$ \\
\cline{1-3}
\end{tabular}
\caption{\small Coefficients for the quadratic "pure" $\ell=4$ map $Q^{(3)}$}
\label{mapQ^(3)}
\end{table}

\begin{table}[h] \footnotesize
\centering
\begin{tabular}{|c|c|c||c|c|c|}
\hline
Equation & Term & Coefficient & Equation & Term & Coefficient \\ \hline
$3$  & $(-333)$ & $-45$ & $2$ & $(-323)$ & $-45$ \\ 
    & $(-223)$ & $45$ & & $(-213)$ & $10\sqrt{15}$ \\
    & $(-113)$ & $-15$ & & $(-222)$ & $20$ \\
    & $(-122)$ & $-5\sqrt{15}$ & & $(-103)$ & $-15\sqrt{2}$ \\
    & $(003)$ & $0$ & & $(-112)$ & $-35$ \\
    & $(012)$ & $15\sqrt{2}$ & & $(002)$ & $30$ \\
    & $(111)$  & $-2\sqrt{15}$ & & $(011)$ & $\sqrt{30}$\\ 
\hline
$1$ & $(-313)$ & $-15$ & $0$ & $(-303)$ & $0$ \\
    & $(-322)$ & $-5\sqrt{15}$ & & $(-312)$ & $-15\sqrt{2}$ \\
    & $(-203)$ & $-15\sqrt{2}$ & & $(-202)$ & $60$ \\
    & $(-212)$ & $35$ & & $(-211)$ & $-5\sqrt{6}$ \\
    & $(-1-13)$ & $-6\sqrt{15}$ & & $(-101)$ & $-36$ \\
    & $(-102)$ & $2\sqrt{30}$ & & $(000)$ & $18$ \\
    & $(-111)$ & $-41$ & & &  \\
    & $(001)$ & $18$ & & &  \\
\hline
\end{tabular}
\caption{\small Coefficients for the cubic "pure" $\ell=3$ map $C^{(1)}$}
\label{mapC^(3)}
\end{table}

\begin{table}[h] \footnotesize
\centering
\begin{tabular}{|c|c|c||c|c|c|}
\hline
Equation & Term & Coefficient & Equation & Term & Coefficient \\ \hline
$4$  & $(-444)$ & $30/7$ & $3$ & $(-434)$ & $30/7$ \\ 
    & $(-334)$ & $-30/7$ & & $(-324)$ & $-5\sqrt{7}/2$ \\
    & $(-224)$ & $-5/7$ & & $(-333)$ & $5/56$ \\
    & $(-233)$ & $5\sqrt{7}/4$ & & $(-214)$ & $15/4$ \\
    & $(-114)$ & $45/14$ & & $(-223)$ & $135/56$ \\
    & $(-123)$ & $-15/4$ & & $(-104)$ & $-\sqrt{10}/4$ \\
    & $(004)$  & $-13/7$ & & $(-113)$ & $-65/56$\\ 
    & $(013)$ & $\sqrt{10}/4$ & & $(-122)$ & $-45\sqrt{7}/56$ \\
    & $(022)$ & $3\sqrt{5/14}$ & & $(003)$ & $-17/28$ \\
    & $(112)$ & $-5\sqrt{7}/14$ & & $(012)$ & $23/4\sqrt{5/14}$ \\
    & & & & $(111)$ & $-15/28\sqrt{7}$ \\ 
\hline
$2$ & $(-424)$ & $-5/7$ & $1$ & $(-414)$ & $-45/14$ \\
    & $(-433)$ & $5\sqrt{7}/4$ & & $(-423)$ & $15/4$ \\
    & $(-314)$ & $-15/4$ & & $(-304)$ & $-\sqrt{10}/4$ \\
    & $(-323)$ & $-135/56$ & & $(-313)$ & $-65/56$ \\
    & $(-204)$ & $3\sqrt{10/7}$ & & $(-322)$ & $-45\sqrt{7}/56$ \\
    & $(-213)$ & $45\sqrt{7}/28$ & & $(-2{-1}4)$ & $5\sqrt{7}/7$ \\
    & $(-222)$ & $0$ & & $(-203)$ & $23/4\sqrt{5/14}$ \\
    & $(-1{-1}4)$ & $-10\sqrt{7}/28$ & & $(-212)$ & $15/8$ \\
    & $(-103)$ & $-23/4\sqrt{5/14}$ & & $(-1{-1}3)$ & $-45\sqrt{7}/28$ \\
    & $(-112)$ & $-15/8$ & & $(-102)$ & $3\sqrt{10}/4$ \\
    & $(002)$ & $7/2$ & & $(-111)$ & $-25/8$ \\
    & $(011)$ & $-3\sqrt{10}/8$ & & $(001)$ & $1$ \\
\hline
$0$ & $(-404)$ & $-26/7$ \\
    & $(-413)$ & $\sqrt{10}/4$ \\
    & $(-422)$ & $3\sqrt{70}/14$ \\
    & $(-3 {-1}4)$ & $\sqrt{10}/4$ \\
    & $(-303)$ & $17/14$ \\
    & $(-312)$ & $-23/4\sqrt{5/14}$ \\
    & $(-2 {-2} 4)$ & $3\sqrt{5/14}$ \\
    & $(-2 {-1} 3)$ & $-23/4\sqrt{5/14}$ \\
    & $(-202)$ & $7$ \\
    & $(-211)$ & $3\sqrt{10}/8$ \\
    & $(-1 {-1} 2)$ & $3\sqrt{10}/8$ \\
    & $(-101)$ & $-2$ \\
    & $(000)$ & $1$ \\
\cline{1-3}
\end{tabular}
\caption{\small Coefficients for the cubic "pure" $\ell=4$ map $C^{(2)}$}
\label{mapC^(2)}
\end{table}
\clearpage
\section{Computation of the coefficients in Equations (\ref{eq:ode})}\label{app:coef}
We give a sketch of the procedure to compute the coefficients of the bifurcation equation (\ref{eq:ode}). Details are found in \cite{ChGu96}. 
Let $Z=(\mathbf{u},\Theta)$ in the system (\ref{eq:pde}), which we assume to belong to a suitable space of square integrable fields. We define the $L^2$ inner product
\begin{equation}
\left< Z_1,Z_2\right>=\int_\Omega \mathbf{u}_1.\overline{\mathbf{u}}_2 dx+\int_\Omega \Theta_1\cdot\overline{\Theta}_2 dx.
\end{equation}
Then we define an orthogonal projection onto the space $V=V^3\oplus V^4$ by
\begin{equation}
X=P_0 Z= \sum_{m=-3}^3 \left<Z,\zeta_m^{*(3)}\right>\zeta_m^{(3)}+\sum_{m=-4}^4 \left<Z,\zeta_m^{*(4)}\right>\zeta_m^{(4)}
\end{equation}
where $(\zeta_m^{(\ell)})_{m=-\ell\dots\ell}$ are an orthonormal basis of $V^\ell$ along the spherical harmonics $Y_l^m$ and $\zeta^{*(\ell)}_m$ are the conjugate vectors. These eigenvectors are obtained from the numerical resolution of the linear system in Section \ref{subsec:linearstability}.

Then, the polynomials $R_{qr}^p$ of the Taylor series expansion (\ref{eq:taylor}) are expressed in terms of operators in the evolution problem (\ref{eq:pde}) by:
\begin{eqnarray}
R_{1,0}^1(X) &=& P_0.L_1X\\
R_{0,1}^1(X) &=& P_0.L_2X\\
R_{0,0}^2(X,X) &=& P_0.M(X,X)\label{eq:Rexpr2}\\
R_{0,0}^3(X,X,X) &=& -2P_0.M(X,S.M(X,X))
\end{eqnarray}
\bffil{where $L_0$ is the linearized part of $F$ at the critical point, $L_1$ and $L_2$ are the linear perturbations of $L_0$ of order respectively $\tilde{\lambda}$ and $\tilde{\eta}$. $M$ is the bilinear and symmetric operator associated with the quadratic terms in the equations (\ref{eq:pde}). 
Finally $S$ designates the {\it pseudo-inverse} of $L_0$: $Y=SZ$ is the unique solution of $L_0Y=(1-P_0)Z$ such that $P_0Y=0$. See \cite{ChGu96} for explicit expressions of these operators.}
Then the bifurcation parameters $\mu_1$ and $\mu_2$ are related to physical parameters by:
\begin{equation} \label{mu1mu2}
\left\{\begin{array}{rl}
\mu_1&=\left<   L_1\zeta_0^{(3)} ,\zeta_0^{*(3)} \right> \tilde\lambda+\left< L_2\zeta_0^{(3)} ,\zeta_0^{*(3)}\right>\tilde\eta \\
\mu_2&=\left<   L_1\zeta_0^{(4)} ,\zeta_0^{*(4)} \right> \tilde\lambda+\left< L_2\zeta_0^{(4)} ,\zeta_0^{*(4)}\right>\tilde\eta
\end{array}\right.
\end{equation}
In order to find the quadratic coefficients, we have to identify one term of the quadratic polynomials (see Tab \ref{mapQ^(3)}) with the terms of the right side of the relation (\ref{eq:Rexpr2}), which gives:
\begin{eqnarray*}
b&=&\frac13\left< M \left( \zeta_0^{(3)},\zeta_0^{(3)} \right) , \zeta_0^{*(4)}\right> \\
\beta&=&\frac16  \left< M \left( \zeta_0^{(3)},\zeta_0^{(4)} \right) , \zeta_0^{*(3)}\right> \\
c&=&\frac19 \left< M \left( \zeta_0^{(4)},\zeta_0^{(4)} \right) , \zeta_0^{*(4)}\right>
\end{eqnarray*}
The same procedure is applied for the cubic coefficients. However, there are two independent cubic polynomials for each representation $\ell$. Therefore, we need to identify two terms of the polynomials in order to get two linear indepedent equations. We chose the terms $(000)$ and $(002)$ of the cubic invariant polynomials.
The coefficient in front of the $(000)$, resp. $(002)$ terms are noted $c^{(i)}_0$, resp. $c^{(i)}_2$ for the $C^{(i)}$ invariant polynomial. We obtain the following expressions:
\begin{eqnarray*}
c^{(1)}_0 & = & -2 \left<
                        M \left(
                                		\zeta_0^{(3)} ,SM \left( \zeta_0^{(3)},\zeta_0^{(3)} \right)
                          \right)
                        , \zeta_0^{*(3)} \right> \\
c^{(1)}_2 & = & -2 \left<
                        M \left(
                                		\zeta_2^{(3)} ,SM \left( \zeta_0^{(3)},\zeta_0^{(3)} \right)
                          \right)
                        , \zeta_2^{*(3)} \right> \\
      &&                -4 \left<
                        M \left(
                                		\zeta_0^{(3)} ,SM \left( \zeta_0^{(3)},\zeta_2^{(3)} \right)
                          \right)
                        , \zeta_2^{*(3)} \right>\\
c^{(2)}_0 & = & -2 \left<
                        M \left(
                                		\zeta_0^{(4)} ,SM \left( \zeta_0^{(4)},\zeta_0^{(4)} \right)
                          \right)
                        , \zeta_0^{*(4)} \right> \\
c^{(2)}_2 & = & -2 \left<
                        M \left(
                                		\zeta_2^{(4)} ,SM \left( \zeta_0^{(4)},\zeta_0^{(4)} \right)
                          \right)
                        , \zeta_2^{*(4)} \right>\\ 
 &&                     -4 \left<
                        M \left(
                                		\zeta_0^{(4)} ,SM \left( \zeta_0^{(4)},\zeta_2^{(4)} \right)
                          \right)
                        , \zeta_2^{*(4)} \right>
\end{eqnarray*}
We then easily obtain:
\begin{eqnarray}
\gamma_1&=&\frac12 \left( 5c^{(1)}_0 -3c^{(1)}_2 \right)\\
\gamma_2&=&\frac1{12} \left( -c^{(1)}_0 +c^{(1)}_2 \right)\\
d_1&=&\frac15 \left( 7c^{(2)}_0 -2c^{(2)}_2 \right)\\
d_2&=&\frac25 \left(-c^{(2)}_0 +c^{(2)}_2 \right)
\end{eqnarray}
\section{Eigenvalues along the heteroclinic connections in generalized heteroclinic cycles of type I}
\label{ap:ev}
In the diagrams of Theorem \ref{thm:generalized hetcycles} we indicated which eigenvalues, at each equilibrium in the generalized heteroclinic cycle, give the rate of expansion or contraction along the connections in the cycle. Expressions for the eigenvalues are given in Tables \ref{vp_O(2)} and \ref{vp_O}. We display in Table \ref{table:evs} these eigenvalues in the different cases where type I generalized heteroclinic cycles appear to drive the dynamics in the numerical simulations of Section \ref{sec:numerics}.  

A criterion for the stability of simple heteroclinic cycles (cycles with heteroclinic saddle-sink connections occuring in invariant planes only) is that the ratio of the product of positive (expanding) eigenvalues by the product of the absolute value of negative (contracting) eigenvalues be less than 1 \cite{krumel}. We observe that this criterion is satisfied in all cases displayed in the table, although the generalized heteroclinic cycles do not seem to be asymptotically stable in the time integrations and with the initial conditions which have been used.
\begin{table}[h] \footnotesize
\centering
\begin{tabular}{|l|c|c|c|}
\hline
 & $c=0.004,\mu_1=0.12$, $\mu_2=0.002$ &  $c=0.04,\mu_1=-1.17,\mu_2=0.1$ & $c=0.04,\mu_1=-0.61,\mu_2=0.01$\\
\hline
 & $\alpha_+$ \hspace{0.5cm} $\alpha_-$ & $\alpha_+$ \hspace{0.5cm} $\alpha_-$ & $\alpha_+$ \hspace{0.5cm} $\alpha_-$\\
\hline
$\sigma_2^{\alpha}$ & $0.6068$ ; ~~~~--~~ & $0.3834$ ; ~~~~--~~  & $0.2381$ ; ~~~~--~~ \\
\hline
$\lambda_2^{\alpha}$ & $-0.09456$ ; $ 0.004189$ & $-0.2301$ ; $0.08246$ & $-0.1126$ ; $0.01753$\\
$\lambda_3^{\alpha}$ & $-0.1172$   ; $ 0.006347$ & $ -0.2325$ ; $0.1585$ & $-0.1353$ ; $0.02747$\\
$\lambda_4^{\alpha}$ & $0.03391$  ; $ -0.001021$ & $ 0.1044$ ; $-0.006122$ & $0.04212$ ; $-0.003897$\\
\hline\hline
& $\beta_+$ \hspace{0.5cm} $\beta_-$ & $\beta_+$ \hspace{0.5cm} $\beta_-$\\
\hline
$\sigma_0^{\beta}$ & $0.1817$ ; $ -0.29706$ & $1.202$ ; $-2.296$  & $0.8229$ ; $-0.7964$\\
$\sigma_2^{\beta}$ & $-0.0697$;  ~~~~--~~~~~~ & $-0.775$ ;  ~~~~--~~~~~~ & $-0.3712$ ; ~~~~--~~~~~~\\
\hline
$\lambda_2^{\beta}$ & $-0.002547$ ; $ 5.944\cdot10^{-5}$ & $-0.1744$ ; $0.01607$ & $-0.08252$ ; $0.005625$\\
$\lambda_3^{\beta}$ & $-0.004408$ ; $ 0.002228$ & $-0.3401$ ; $0.1448$ & $-0.1997$ ; $0.02471$\\
\hline
\end{tabular}
\caption{\small Eigenvalues for the type-I heteroclinic cycles in some cases.}
\label{table:evs}
\end{table}

\end{document}